\documentclass[trackchanges]{aastex701}
\usepackage{amsmath}

\begin{document}

\title{Linear Polarization of Synchrotron Radiation from Very-small-pitch-angle Electrons}

\correspondingauthor{Jirong Mao}
\email{jirongmao@mail.ynao.ac.cn}

\author[orcid=0009-0004-8628-7416, sname='Meng.', gname='Nie']{Meng Nie}
  \affiliation{Yunnan Observatories,
    Chinese Academy of Sciences, 
    Kunming 650216, People's Republic of China}
  \affiliation{University of Chinese Academy of Sciences,
    Beijing 101408, People's Republic of China}
  \affiliation{Center for Astronomical Mega-Science, 
    Chinese Academy of Sciences,
    Beijing 100012, People's Republic of China}
  \email[]{niemeng@ynao.ac.cn}

\author[orcid=0000-0002-7077-7195, sname='J.', gname='Mao']{Jirong Mao}
  \affiliation{Yunnan Observatories,
    Chinese Academy of Sciences, 
    Kunming 650216, People's Republic of China}
  \affiliation{Center for Astronomical Mega-Science, 
    Chinese Academy of Sciences,
    Beijing 100012, People's Republic of China}
  \affiliation{Key Laboratory for the Structure and Evolution of Celestial Objects, 
    Chinese Academy of Sciences, 
    Kunming 650216, People's Republic of China}
  \email[]{jirongmao@mail.ynao.ac.cn}

\begin{abstract}

For synchrotron radiation with pitch angle $\alpha \ll 1/\gamma$ where $\gamma$ is the electron Lorentz factor, referred to as very-small-pitch-angle synchrotron radiation, the polarization properties differ significantly from those of conventional synchrotron radiation ($\alpha \gg 1/\gamma$). In this work, we investigate the linear polarization characteristics of very-small-pitch-angle synchrotron radiation from relativistic thermal and nonthermal electrons. We calculate the linear polarization by applying power-law and Maxwellian electron energy distributions. Moreover, some synchrotron sources exhibit a hybrid distribution of nonthermal and thermal electrons, and we also calculate the polarization degree for a hybrid electron energy distribution. The calculations are performed in the radio, optical, X-ray, and $\gamma$-ray bands. Our results show that the linear polarization degree of very-small-pitch-angle synchrotron radiation is generally lower than that of conventional synchrotron radiation. Very-small-pitch-angle synchrotron radiation from both nonthermal and thermal electrons can produce significant linear polarization. However, the polarization exhibits diverse behaviors across different physical parameters. The general scenario provided in this work can be considered for the application to all kinds of synchrotron sources, as we suggest that the very-small-pitch-angle case is crucial for the study of synchrotron polarization.  

\end{abstract}


\section{Introduction}

Synchrotron radiation is emitted by relativistic electrons accelerated by the Lorentz force while moving in a magnetic field. This radiation mechanism is typically characterized by a significant degree of linear polarization. For a nonthermal electron population with a power-law energy distribution, the linear polarization degree can be expressed as $(3p+3)/(3p+7)$, indicating that the polarization degree is independent of frequency and depends solely on the electron power-law index $p$ \citep{Ginzburg1965, Legg1968,Rybicki1979}. However, this conclusion is modified when the electron Lorentz factor $\gamma$ is confined to a finite range \citep{Westfold1959, Mao2018a}. The conventional synchrotron radiation with isotropic pitch-angle distribution formalism, derived for electrons undergoing helical motion, is valid only under the condition that the pitch angle satisfies $\alpha \gg 1/\gamma$. In the very-small-pitch-angle regime ($\alpha \ll 1/\gamma$), the Airy integral approximation to the Bessel functions commonly employed in standard synchrotron theory breaks down \citep{Melrose1971}. 

The pitch-angle distribution is a key factor influencing the radiation power and polarization of synchrotron radiation. In many cases, electrons are assumed to have an isotropic pitch-angle distribution, which simplifies the analysis of polarization properties. However, when electrons exhibit an anisotropic pitch-angle distribution, synchrotron emission and polarization can differ substantially from those in the isotropic case. Some recent researches on the anisotropic electron distribution have been presented. The formation of small pitch angles may originate from multiple physical mechanisms, such as rapid cooling and inefficient diffusion transverse to the field \citep{Lloyd2002,Baring2004}. In emission regions with ordered magnetic fields, or in magnetic reconnection regions, the electron pitch-angle distribution may be anisotropic. Numerical simulations further suggest that when relativistic electrons propagate along magnetic flux tubes with ordered field configurations, an anisotropic pitch-angle distribution can naturally arise \citep{Sun2026}. The very small pitch angle represents an extreme case of an anisotropic pitch-angle distribution. In this limit, not only do the standard approximations used in the conventional synchrotron theory break down, but radiation spectrum and polarization properties can also differ fundamentally from those presented in conventional synchrotron theory. 

The synchrotron polarization in the very-small-pitch-angle case is seldom investigated. \citet{Epstein1973a} derived the total power spectrum and polarization properties of synchrotron radiation emitted by a single electron in the regime of very small pitch angles. Because $\alpha \ll 1/\gamma$, the radiation emitted by a single electron is highly collimated along the magnetic field direction. In any given direction, the emission is effectively monochromatic, with a frequency given by $2\nu_L\gamma/(1 + \gamma^2 \theta^2)$, where $\nu_L$ is the nonrelativistic electron cyclotron frequency and $\theta$ is the angle between the direction of emission and the magnetic field \citep{Epstein1973a, Goundyrev1995, Vedantham2022}. At high frequencies, the spectrum follows a power law. For conventional synchrotron radiation, the spectral slope is $(1 - p)/2$, whereas in the very-small-pitch-angle synchrotron, the spectral slope becomes $2 - p$. Moreover, in the case of anisotropic pitch-angle distribution, electron acceleration with a small pitch angle suppresses synchrotron cooling, producing a hard synchrotron spectrum \citep{Comisso2021,Goto2022}. In addition, the polarization properties under different electron energy distributions are strongly affected by the anisotropic pitch-angle distribution \citep{Comisso2023,Lai2025}.

Very-small-pitch-angle synchrotron radiation has limited astrophysical applications so far, and its polarization properties have not been precisely described in recent years. Nevertheless, very-small-pitch-angle synchrotron radiation has been proposed as a possible radiation mechanism for a few astrophysical cases. The Crab pulsar spectrum and the luminosity from the infrared up to soft X-rays can be reproduced by the very-small-pitch-angle synchrotron radiation \citep{Crusius2001}. In pulsar magnetospheres, electrons are expected to possess very small pitch angles \citep{ODell1970}. If the electrons responsible for X-ray emission initially have relativistic perpendicular momenta, the subsequent very-small-pitch-angle synchrotron radiation becomes negligible compared to the conventional synchrotron radiation. In contrast, if the initial perpendicular momentum is nonrelativistic, the very-small-pitch-angle synchrotron radiation can make a nonnegligible contribution to the X-ray luminosity \citep{Rudak1999,Kisaka2014}. Considering $\gamma$-ray emission from pulsars, synchrotron radiation with a very small pitch angle only operates within the inner magnetosphere, where the magnetic field strength exceeds $1.0 \times 10^9\ \mathrm{G}$ \citep{Grenier1993}. In the gamma-ray burst (GRB) research field, for an isotropic power-law electron distribution, the low-energy spectral index of optically thin synchrotron radiation is constrained to be $\leq -2/3$ \citep{Preece1998}. However, a nonnegligible fraction of detected GRB prompt emission spectra exhibit a low-energy spectral index exceeding $-2/3$ \citep{Crider1997,Preece2002,Ghirlanda2003}. Very-small-pitch-angle synchrotron radiation is one of the likely radiation mechanisms to explain this phenomenon \citep{Lloyd2000,Ryde2006,Savchenko2009,Fouka2011AdS,Fouka2011ApJ,Ryde2011}. Furthermore, very-small-pitch-angle synchrotron radiation provides an important dissipation mechanism in the model of relativistic magnetized GRB outflows \citep{Neronov2010}. In addition, the very-small-pitch-angle synchrotron radiation model has also been proposed as a mechanism to avoid catastrophic cooling of relativistic electrons due to inverse Compton upscattering of their own synchrotron photons in extragalactic radio sources \citep{Callahan1975,Begelman1984}. From the above reviews, we believe that a full investigation of very-small-pitch-angle synchrotron radiation is necessary. In this paper, we focus primarily on the linear polarization properties of the very-small-pitch-angle synchrotron.

The electron energy distribution is also a key factor in determining the polarization properties of synchrotron radiation. In particular, power-law and relativistic Maxwellian distributions are widely used in astrophysical processes. When modeling the properties of synchrotron sources, a power-law electron distribution is commonly assumed. This is partly due to its mathematical simplicity, and partly because power-law radio spectra are widely observed in many extended radio sources \citep{Jones1979}. 
One may consider a distribution that has a thermal core at low energies and transits into a power-law tail at high energies \citep{Pandya2016}. Moreover, the second-order Fermi acceleration can generate a Maxwellian electron energy distribution \citep{Stawarz2018}. In some synchrotron-emitting sources, which are characterized by strong relativistic shocks and turbulence, both first- and second-order Fermi acceleration processes play important roles in shaping the polarization properties of the radiation. Motivated by this, \citet{Giannios2009} proposed a hybrid electron energy distribution consisting of a power-law component combined with a Maxwellian component. Subsequently, \citet{Mao2018a} performed a detailed calculation of the polarization properties of conventional synchrotron radiation for a hybrid electron energy distribution. The results of \citet{Mao2018a} show that relativistic thermal electrons can produce highly polarized photons. In this paper we investigate the small-pitch-angle synchrotron polarization properties for power-law, Maxwellian, and hybrid electron energy distributions.

We present the formalism for the linear polarization of conventional and very-small-pitch-angle synchrotron radiation in Section \ref{2}. In Section \ref{3}, we introduce a hybrid electron energy distribution consisting of both thermal and nonthermal electrons. The polarization results corresponding to different electron energy distributions are presented in Section \ref{3}. We briefly compare the application conditions of very-small-pitch-angle synchrotron radiation to those of the curvature radiation in Section \ref{sec:4}. Coherent radiation is also simply discussed. The conclusions are presented in Section \ref{sec:5}.

\section{Thermal-Nonthermal electrons in the very-small-pitch-angle case}\label{2}

The spectrum and polarization properties of synchrotron radiation are strongly dependent on the pitch angle $\alpha$. Radiation from relativistic electrons exhibits a characteristic angular collimation, with most of the radiation confined within a conical region of opening angle $\sim 1/\gamma$, commonly referred to as the beaming angle. When the pitch angle satisfies $\alpha \gg 1/\gamma $, the radiation corresponds to conventional synchrotron radiation \citep{Ginzburg1969,Rybicki1979}. In contrast, when the pitch angle satisfies $\alpha \ll 1/\gamma$, the radiation corresponds to very-small-pitch-angle synchrotron radiation \citep{ODell1970,Melrose1971,Epstein1973a}. 

\subsection{Conventional Synchrotron Polarization}
In order to derive the synchrotron polarization in the very-small-pitch-angle case, we first simply introduce the well-known conventional synchrotron polarization in general in this subsection. The power of synchrotron radiation emitted by a relativistic electron with Lorentz factor $\gamma$ can be decomposed into two orthogonal components. Following \citet{Rybicki1979}, the perpendicular and parallel components of the emitted power per frequency can be expressed as

\begin{equation}
    \frac{dP_\perp(\nu)}{d\nu}=\frac{\sqrt{3}\pi e^2\nu_L\sin\alpha}{c}\left[F(x)+G(x)\right]  \label{power-perp}  
\end{equation}

\noindent and

\begin{equation}
	\frac{dP_\parallel(\nu)}{d\nu}=\frac{\sqrt{3}\pi e^2\nu_L\sin\alpha}{c}\left[F(x)-G(x)\right] \label{power-parallel}
\end{equation}

\noindent respectively, with $F(x) = x\int_x^\infty K_{5/3}(\xi)d\xi$ and $G(x) = xK_{2/3}(x)$, where $K_{2/3}$ and $K_{5/3}$ are the modified Bessel functions of the second kind. Here, $x=\nu/\nu_c$, $\nu$ is the observational frequency, $\nu_c=\frac{3}{2}\gamma^2\nu_L\mathrm{sin}\alpha$ is the critical frequency, and $\nu_L=eB/2\pi m_ec$ is the gyrofrequency. From Equation (\ref{power-perp}) and (\ref{power-parallel}) we can obtain the total power and the linearly polarized power of synchrotron radiation as

\begin{equation}
	\frac{dP(\nu)}{d\nu} = \frac{dP_\perp(\nu)}{d\nu} + \frac{dP_\parallel(\nu)}{d\nu} 
    = \frac{2\sqrt{3}\pi e^2\nu_L\sin\alpha}{c}x\int_x^\infty K_{5/3}(\xi)d\xi \label{power}
\end{equation}

\noindent and

\begin{equation}
	\frac{dP_L(\nu)}{d\nu} = \frac{dP_\perp(\nu)}{d\nu} - \frac{dP_\parallel(\nu)}{d\nu} 
    = \frac{2\sqrt{3}\pi e^2\nu_L\sin\alpha}{c}xK_{2/3}(x)\label{power-linear}
\end{equation}

\noindent respectively. Using Equation (\ref{power}) and (\ref{power-linear}), the linear polarization degree of synchrotron radiation from a single electron is

\begin{equation} \label{single-pl}
	\pi_L = \frac{dP_L(\nu)/d\nu}{dP(\nu)/d\nu} 
    = \frac{K_{2/3}(x)}{\int_x^\infty K_{5/3}(\xi)d\xi}
\end{equation}

For an electron energy distribution $N_e(\gamma)$, and $d\Omega = 2\pi\sin \alpha d\alpha $, the total linear polarization degree of conventional synchrotron radiation is given by

\begin{equation} \label{HYPL}
\Pi_L =\frac{\int N_{e}(\gamma, \alpha) (dP_L(\nu)/d\nu) d\gamma d\Omega}{\int N_{e}(\gamma, \alpha)(dP(\nu)/d\nu) d\gamma d\Omega } 
=\frac{\int_{0}^{\pi/2}\int_{\gamma_{\min}}^{\gamma_{\max}}N_{e}(\gamma, \alpha)\sin\alpha [xK_{2/3}(x)]d\alpha d\gamma}{\int_{0}^{\pi/2}\int_{\gamma_{\min}}^{\gamma_{\max}}N_{e}(\gamma, \alpha)\sin\alpha[x\int_{x}^{\infty}K_{5/3}(t)dt]d\alpha d\gamma}
\end{equation}

\subsection{Very-small-pitch-angle Synchrotron Polarization}

When $\gamma \alpha \ll 1$, the emission is dominated by the fundamental harmonic, and the standard Airy integral approximation to the Bessel functions employed in conventional synchrotron theory breaks down \citep{Melrose1971}. Therefore, Equation (\ref{power-perp}) to (\ref{HYPL}) are valid only under the condition that the pitch angle satisfies $\gamma\alpha \gg 1$. According to \citet{Epstein1973a}, one can designate the inertial frame of the observer as $K$, and $K_0$ is for the inertial frame in which the electron motion is circular. When $\gamma\alpha \ll 1$, the speed of the electron in the $K_0$ frame is $\beta_0 \approx \gamma \alpha \ll 1$ (quantities measured in $K_0$ are denoted by the subscript 0). Therefore, the electron energy is nonrelativistic and the emission appears predominantly as monochromatic cyclotron radiation in the $K_0$ frame. The emissivity and linear polarization degree of cyclotron radiation from a single electron are given by \citep{Bekefi1966,Epstein1973a}

\begin{equation}
	\eta_0 = \frac{\pi e^2\beta_0^2\nu_L^2}{2c}(1+\cos^2\theta_0)\delta(\nu_0-\nu_B)
\end{equation}

\noindent and

\begin{equation}
	\pi_{L,0} = \frac{1-\cos^2\theta_0}{1+\cos^2\theta_0},
\end{equation}

\noindent respectively, where $\theta_0$ is the angle between the direction of emission and the magnetic field in the $K_0$ frame. Through an appropriate transformation, we can obtain the emissivity and linear polarization degree of very-small-pitch-angle synchrotron radiation from a single electron in the $K$ frame as

\begin{equation}
	\eta =\frac{\pi e^2\gamma\alpha^2\nu^3}{\nu_L c}\left[1-\frac{\nu}{\gamma\nu_L}+\frac{\nu^2}{2\gamma^2\nu_L^2}\right]\delta\left(\nu-\frac{2\gamma\nu_L}{1+\theta^2\gamma^2}\right) \label{emiss}
\end{equation}

\noindent and

\begin{equation}
	\pi_{L,\rm VSPA} =\frac{\nu(2\gamma\nu_L-\nu)}{\nu^2-2\gamma\nu\nu_L+2\gamma^2\nu_L^2}, \label{VSPA-SPLC}
\end{equation}

\noindent respectively. We note that Equations (\ref{emiss}) and (\ref{VSPA-SPLC}) are derived by retaining the lowest-order terms in $\gamma\alpha$ and $\theta$. The emission in any given direction is monochromatic with a frequency $2\gamma \nu_L / (1+\theta^2\gamma^2)$. By performing a straightforward transformation and differentiation of Equation (\ref{VSPA-SPLC}), we can find that the linear polarization degree of very-small-pitch-angle synchrotron radiation from a single electron reaches a peak of 100\% at frequency around $\nu \sim \gamma \nu_L$. The power of very-small-pitch-angle synchrotron radiation from a single electron can be obtained by integrating $\eta$ over $\theta$. The results are as follows:

\begin{equation} \label{VSPA-PS}
\frac{dP_{\rm VSPA}(\nu)}{d\nu}=
	\begin{cases}\frac{2\pi^2e^2\alpha^2\nu}{c}\left[1-\left(\frac{\nu}{\gamma \nu_L}\right)+\frac{1}{2}\left(\frac{\nu}{\gamma \nu_L}\right)^2\right], &\left(\frac{\nu_L}{\gamma}\ll\nu \leq 2\gamma\nu_L \right)\\
		0, &(\nu>2\gamma\nu_L)
	\end{cases}
\end{equation}

The $\delta$-function in $\eta$ constrains the power to a finite frequency interval $(\nu_L/\gamma,\, 2\gamma \nu_L)$, implying that the emission is restricted to a narrow bandwidth. The linearly polarized power of very-small-pitch-angle synchrotron radiation is $dP_{L,\rm VSPA}(\nu)/d\nu = \pi_{L,\rm VSPA} dP_{\rm VSPA}(\nu)/d\nu$. Following the same method as in Equation (\ref{HYPL}), the total linear polarization degrees of very-small-pitch-angle synchrotron radiation can be written as

\begin{equation}\label{VSPA-PL}
	\Pi_{L,{\rm VSPA}}  = \frac{\int N_{e}(\gamma, \alpha) (dP_{L,\rm VSPA}(\nu)/d\nu) d\gamma d\Omega}{\int N_{e}(\gamma, \alpha)(dP_{\rm VSPA}(\nu)/d\nu) d\gamma d\Omega }.
\end{equation}

\begin{figure*}
\centering
\includegraphics[scale=0.7]{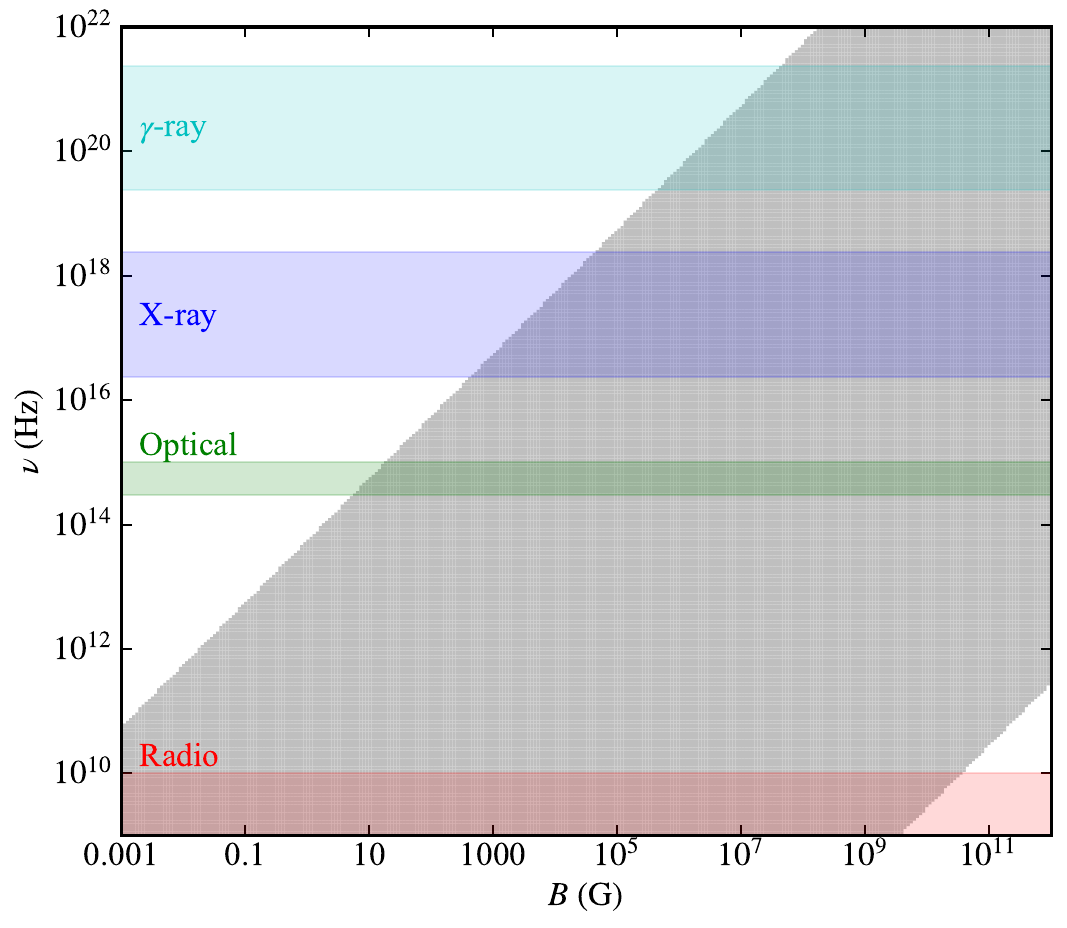}
\caption{Allowed regions in the parameter space for very-small-pitch-angle synchrotron radiation. Variation of the cutoff frequency with magnetic field strength in the radio, optical, X-ray, and $\gamma$-ray bands. The gray shaded region indicates the parameter space where $\max (\nu_L/\nu,\; \nu/2\nu_L) \leq 10^{7}$.
\label{VSPArange}}
\end{figure*}

Both Equations (\ref{VSPA-SPLC}) and (\ref{VSPA-PL}) need to satisfy the condition in Equation (\ref{VSPA-PS}). This condition can also be expressed as a requirement on the Lorentz factor 	$\gamma > \max (\nu_L/\nu, \nu/2\nu_L )$. Figure \ref{VSPArange} presents the valid conditions of the very-small-pitch-angle case for different frequency bands and magnetic field strengths. The white regions correspond to $\max(\nu_L/\nu,\, \nu/2\nu_L) > 10^{7}$, indicating that within the Lorentz factor range $[1,\,10^{7}]$ there are no electrons energetic enough to produce significant radiation and polarization. The conditions shown in Figure 1 provide the guidance for the choice of parameter numbers in our calculations of the polarization degree for synchrotron radiation from very-small-pitch-angle electrons.

\section{Results} \label{3}

\subsection{Electron Populations and Parameter choices}

We consider an electron distribution that can be decomposed into an electron energy distribution function $f_e(\gamma)$ and a pitch-angle distribution function $g_e(\alpha)$, i.e.,
\begin{equation}
    \frac{dN_e(\gamma, \alpha)}{d\gamma d\Omega} = N_0 f_e(\gamma)g_e(\alpha)
\end{equation}
where $N_0$ is a normalization constant.

\subsubsection{Hybrid Electron Energy Distribution Function}
Electron energy is usually assumed as having either a thermal or a nonthermal distribution. In this subsection, we suggest a hybrid electron energy distribution. Particle-in-cell simulations of relativistic shocks indicate that the downstream electron distribution develops a thermal component, which transitions smoothly to the low-energy end of a nonthermal power-law tail \citep{Spitkovsky2008b}. \citet{Giannios2009} proposed a hybrid electron distribution of the Maxwellian and power-law forms. 
Similarly, the kappa distributions provide a natural description of plasmas with both thermal and nonthermal components, consisting of a thermal core at small $\gamma$, which transitions to a nonthermal power-law tail at large $\gamma$ \citep{Pandya2016}.
Thus, the electron distribution is a combination of thermal and nonthermal electron components. In this paper, we use this hybrid distribution to calculate the linear polarization degree of synchrotron radiation. One can define a conjunctive Lorentz factor $\gamma_{\rm th}$, such that electrons with $\gamma$ below $\gamma_{\rm th}$ follow a Maxwellian distribution while those above $\gamma_{\rm th}$ follow a power-law distribution. The electron distribution can be written as \citep{Giannios2009, Mao2018a}

\begin{equation} \label{hydis}
f_{e} (\gamma) =
\begin{cases}
\gamma^{2} \exp(-\gamma/\Theta)/2\Theta^{3}   & (\rm{for \; \gamma \leq \gamma_{\rm th}}) \\
\gamma_{\rm th}^{2} \exp(-\gamma_{\rm th}/\Theta)(\gamma/\gamma_{\rm th})^{-p}/2\Theta^{3}  & (\rm{for \; \gamma >\gamma_{\rm th}})
\end{cases}
\end{equation}

\noindent where $\Theta = kT_e/m_ec^2$ is the characteristic temperature, $T_e$ is the thermal electrons temperature, and $p$ is the power-law index. If $\Theta \ll 1$, the thermal electrons are nonrelativistic. The synchrotron polarization of thermal electrons strongly depends on $\Theta$. This hybrid electron distribution reduces to a pure power-law distribution for $\gamma_{\rm th} =\gamma_{\min}$, and to a pure Maxwellian distribution when $\gamma_{\rm th} = \gamma_{\max}$. 

\subsubsection{Pitch-angle Distribution Function}

For the conventional synchrotron radiation, we assume an isotropic pitch-angle distribution. The pitch-angle distribution can be written as

\begin{equation}
    g_e(\alpha) = \frac{1}{4\pi}.
\end{equation}

For the very-small-pitch-angle synchrotron radiation, the pitch-angle distribution is highly anisotropic. Therefore, following \citet{Epstein1973a}, the pitch-angle distribution can be written as
\begin{equation} \label{andis}
    g_e(\alpha) = \exp(-\frac{\alpha^{2}}{\tilde{\alpha}^{2}}),
\end{equation}
where $\tilde{\alpha}$ is the characteristic angular width of the pitch-angle distribution. We consider an electron distribution function that depends only on the electron energy and is independent of the pitch angle in this work. Therefore, the Lorentz factor and pitch angle are two independent parameters when we calculate the power and polarization of very-small-pitch-angle synchrotron radiation. For the polarization degree of very-small-pitch-angle synchrotron radiation, the common pitch-angle factor cancels out between the polarized and total powers under the separability assumption. However, an explicit pitch-angle distribution is required for the spectra shown in Appendix \ref{sec:power}. When calculating the specific flux of very-small-pitch-angle synchrotron radiation, the pitch angle can significantly affect the flux. In the Appendix \ref{sec:power}, we consider the Equation (\ref{andis}) and set the characteristic angular width $\tilde{\alpha} = 10^{-3}$.

\subsubsection{Parameter Choices}

The adopted parameters are chosen primarily to explore the general polarization properties of very-small-pitch-angle synchrotron radiation rather than to model a specific source. In all calculations, the electron Lorentz factor is integrated over the common interval $\gamma_{\min} = 1$ and $\gamma_{\max} = 10^7$. This broad interval includes mildly relativistic electrons and extends sufficiently far into the ultrarelativistic regime to cover the frequency and magnetic field ranges considered. The Lorentz factor and magnetic field ranges are selected for electrons satisfying the condition $\gamma>\max(\nu_L/\nu,\nu/2\nu_L)$. For the power-law energy distribution, we consider $p=1.5$, $2.5$, and $3.5$ in order to cover relatively hard, representative, and soft electron spectra, respectively. For the thermal energy distribution, the electron temperature is varied over $T_e=10^9$--$10^{13}\ {\rm K}$ to investigate the transition from mildly relativistic to ultrarelativistic electron populations. Unless otherwise stated, we use $p=2.5$ and $T_e=10^{12}\ {\rm K}$ as the fiducial values. For the hybrid energy distribution, the conjunctive Lorentz factor is varied over $1\leq\gamma_{\rm th}\leq10^7$, so that both the power-law and thermal limits are included. For the illustrative hybrid spectra in Figure 10 of Appendix \ref{sec:power}, we adopt $\gamma_{\rm th}=300$ as an intermediate value for which both the thermal core and the nonthermal tail make visible contributions.

The anisotropic distribution of electrons has been mentioned in some references \citep{Comisso2020,Comisso2023,Comisso2024}. The results are obtained from numerical simulations. These results provide the hints for our work, as a very small pitch angle is an extreme case of the anisotropic distribution.
Further work can be performed for astrophysical applications in the future \citep{Cheng2026}.

\subsection{Polarization of Power-law Electron Energy Distribution} \label{3.1}

\begin{figure*}
\centering
\includegraphics[scale=0.45]{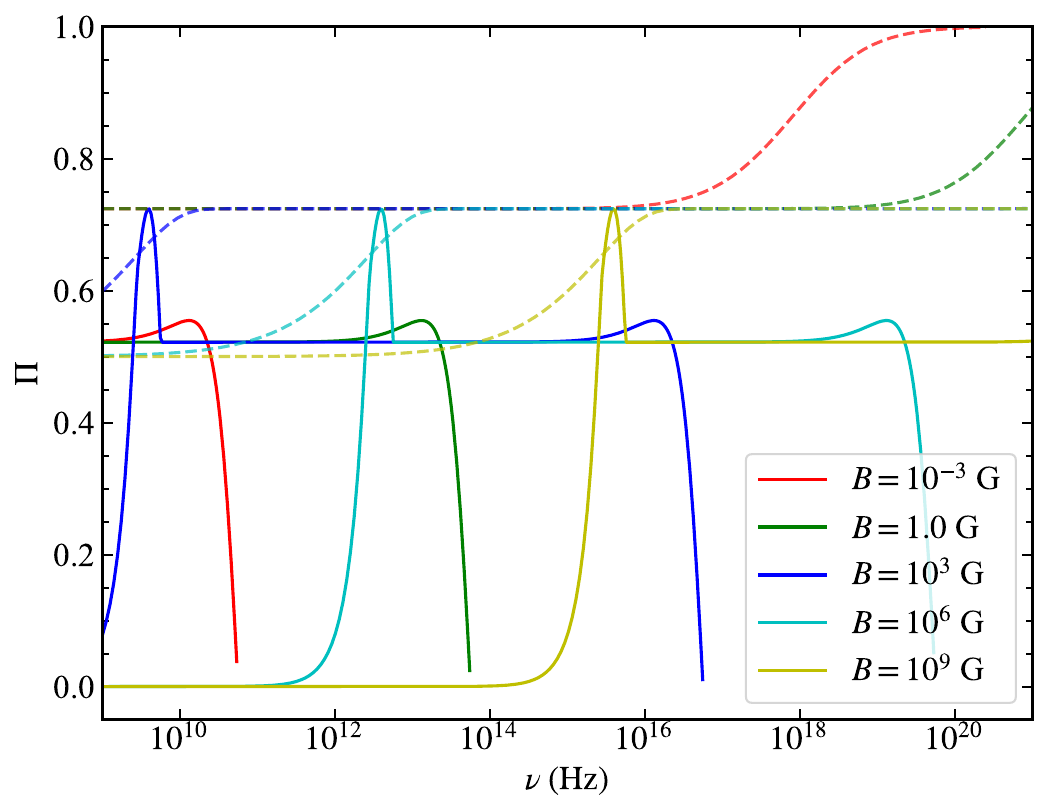}
\includegraphics[scale=0.45]{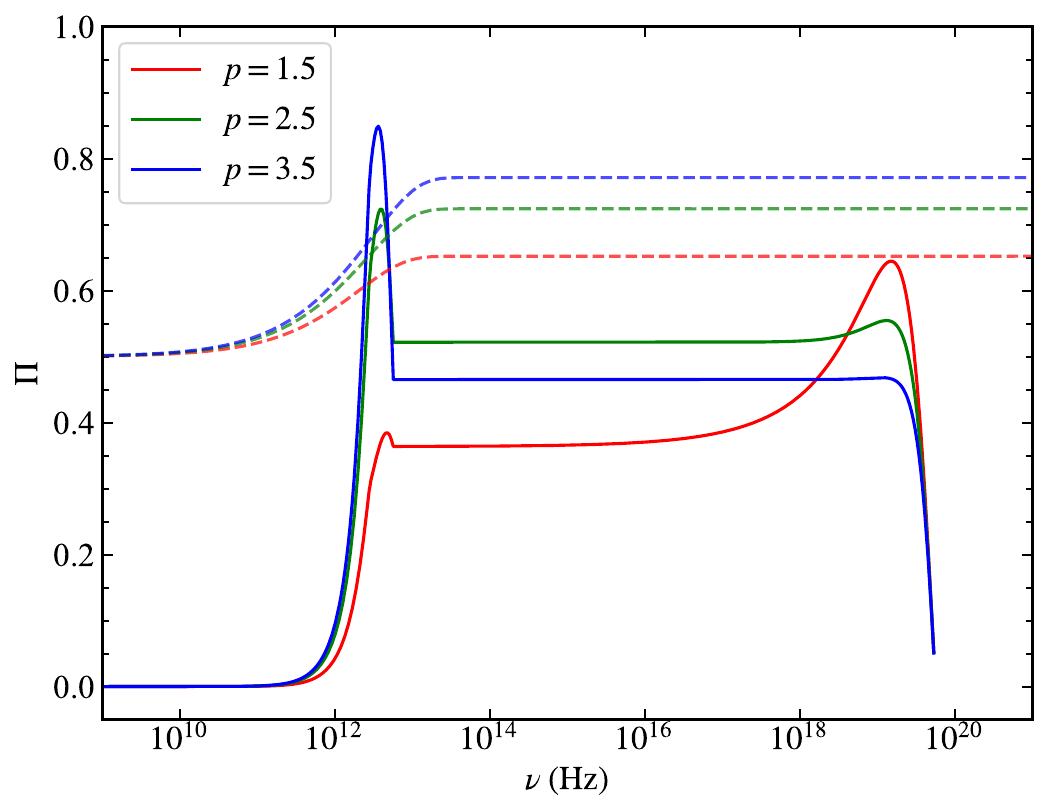}
\caption{Linear polarization degree for power-law electron energy distributions. The solid and dashed lines represent the very-small-pitch-angle and conventional synchrotron radiation, respectively. Left panel: we adopt power-law index $p = 2.5$. Right panel: we adopt magnetic field strength $B = 1.0 \times 10^{6}\, \mathrm{G}$.
\label{CONV_VSPA_PL}}
\end{figure*}

We first calculate the linear polarization degree of very-small-pitch-angle and conventional synchrotron radiation for a power-law electron energy distribution. The results of the polarization degree as a function of observational frequency are shown in Figure \ref{CONV_VSPA_PL}. The observational frequency range covers from the radio to the $\gamma$-ray band. 

For a power-law electron energy distribution, the polarization degree of very-small-pitch-angle synchrotron radiation approaches zero at low frequency and exhibits a cutoff at high frequency. In addition, the polarization profile exhibits two distinct peaks. In the region between the two peaks, the polarization degree remains nearly independent of observational frequency. In contrast, the polarization degree of conventional synchrotron radiation does not exhibit a high-frequency cutoff. It approaches $50\%$ at low frequencies and tends toward $100\%$ at high frequencies. However, over a broad frequency range, the polarization degree of the conventional synchrotron radiation remains nearly unchanged with observational frequency and follows the usual polarization behavior $(3p+3)/(3p+7)$. Overall, the polarization degree of very-small-pitch-angle and conventional synchrotron radiation remains nearly constant over a broad frequency range. The linear polarization degree of very-small-pitch-angle synchrotron radiation for a power-law electron energy distribution is generally lower than that of conventional synchrotron radiation. At the low-frequency peak, the polarization degree of very-small-pitch-angle synchrotron radiation can exceed that of the conventional synchrotron radiation.

We investigate the effects of different magnetic field strengths and power-law indices on the very-small-pitch-angle synchrotron polarization as a function of observational frequency for a power-law electron energy distribution. The left panel of Figure \ref{CONV_VSPA_PL} shows the polarization degree for different magnetic field strengths at $p=2.5$. We note that the cutoff frequency is strongly correlated with the magnetic field strength, such that the stronger magnetic field leads to a higher cutoff frequency. Moreover, as the magnetic field increases, the polarization profile remains unchanged but shifts as a whole toward higher frequencies. The polarization degree for different power-law indices at $B=1.0\times 10^6\,\mathrm{G}$ is shown in the right panel of Figure \ref{CONV_VSPA_PL}. When we adopt power-law indices $p = 1.5,\; 2.5$, and $3.5$, the polarization degree changes significantly. For $p = 2.5$, the polarization degree is approximately $52.2\%$. As the power-law index increases to $p = 3.5$, the polarization degree decreases to $46.8\%$, while decreasing the index to $p = 1.5$ reduces the polarization degree to $36.4\%$. 

\subsection{Polarization of Maxwellian Electron Energy Distribution} \label{3.2}

\begin{figure*}
\centering
\includegraphics[scale=0.45]{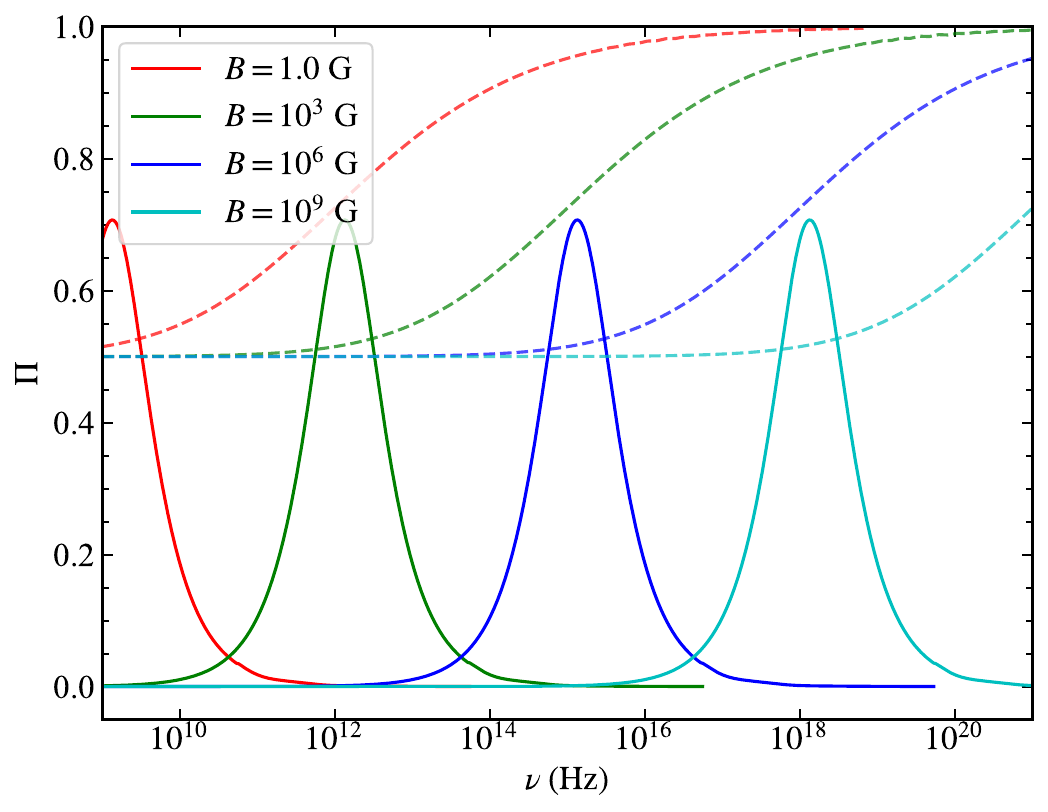}
\includegraphics[scale=0.45]{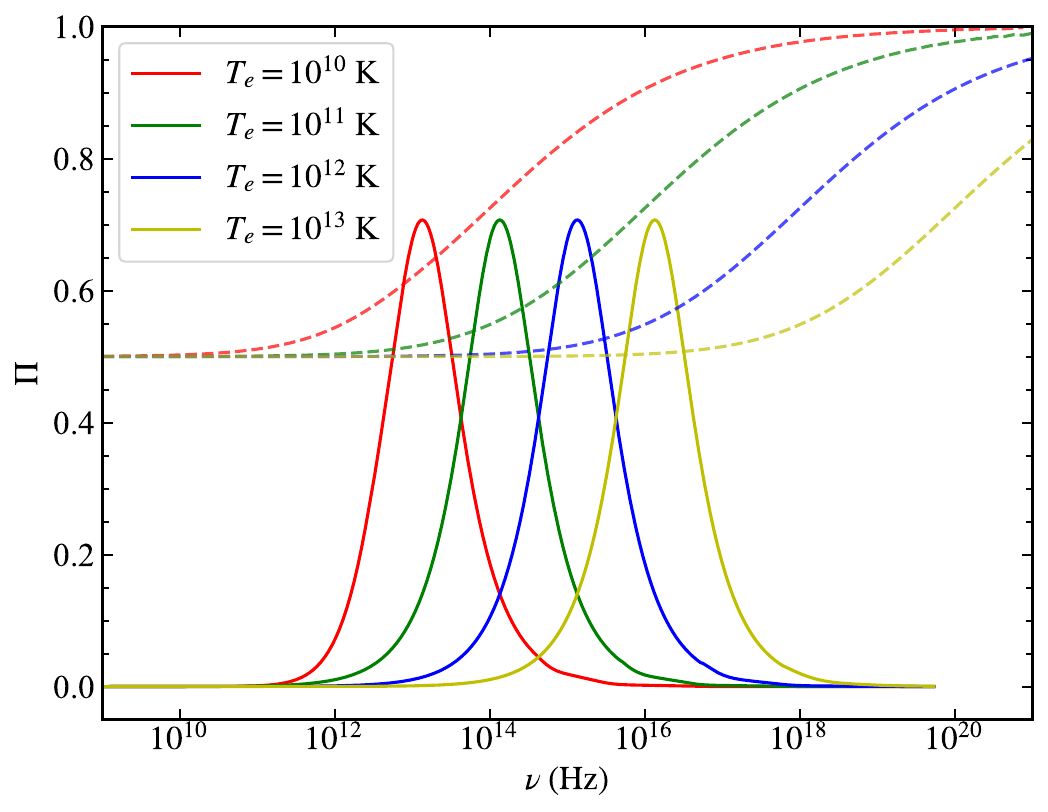}
\caption{Linear polarization degree for Maxwellian electron energy distributions. The solid and dashed lines represent the very-small-pitch-angle and conventional synchrotron radiation, respectively. Left panel: we adopt an electron temperature $T_e = 1.0 \times 10^{12}\,\mathrm{K}$. Right panel: we adopt magnetic field strength $B = 1.0 \times 10^{6}\, \mathrm{G}$.
\label{CONV_VSPA_TL}}
\end{figure*}

We calculate the linear polarization degree of very-small-pitch-angle and conventional synchrotron radiation for a Maxwellian electron energy distribution. The results of the polarization degree as a function of observational frequency are shown in Figure \ref{CONV_VSPA_TL}.

For a Maxwellian electron energy distribution, the polarization profile of very-small-pitch-angle synchrotron radiation exhibits an approximately single Gaussian-shaped peak structure. The peak of the polarization degree is approximately $70.7\%$, occurring at an observational frequency of about $\nu \approx 4.8\,\nu_L T_e / (10^{10}\, \rm K)$. Moreover, the polarization degree rapidly approaches zero at both the low- and high-frequency ends. The frequency range over which significant polarization can be produced spans approximately 4 orders of magnitude. In contrast, the conventional synchrotron polarization increases monotonically with observational frequency and exhibits a lower polarization limit of $50\%$. Overall, the linear polarization degree of very-small-pitch-angle synchrotron radiation for a Maxwellian electron energy distribution is generally lower than that of conventional synchrotron radiation. However, the peak of the polarization degree of very-small-pitch-angle synchrotron radiation can exceed the polarization degree of the conventional synchrotron radiation.

We investigate the effects of different magnetic field strengths and electron temperatures on the very-small-pitch-angle synchrotron polarization as a function of observational frequency for a Maxwellian electron energy distribution. The left panel of Figure \ref{CONV_VSPA_TL} shows the polarization degree for different magnetic field strengths at $T_e =10^{12}\,\mathrm{K}$. The polarization profile does not change its shape with magnetic field strength. It shifts as a whole toward higher frequencies as the magnetic field strength increases. When $B = 1.0\,\mathrm{G}$, the peak of the polarization degree occurs at $\nu \approx 1.34 \times 10^{9}\,\mathrm{Hz}$, whereas when the magnetic field increases to $1.0 \times 10^{3}\,\mathrm{G}$, the peak frequency shifts to $1.34 \times 10^{12}\,\mathrm{Hz}$. Therefore, at a fixed electron temperature, the frequency corresponding to the peak is proportional to the magnetic field strength. The right panel of Figure \ref{CONV_VSPA_TL} shows the polarization degree for different electron temperatures at $B =1.0\times 10^6\,\mathrm{G}$. When $T_e = 10^{10}\,\mathrm{K}$, the peak of the polarization degree occurs at $\nu \approx 1.34 \times 10^{10}\,\mathrm{Hz}$, whereas when the electron temperature increases to $10^{11}\,\mathrm{K}$, the peak frequency shifts to $1.34 \times 10^{11}\,\mathrm{Hz}$. Analogously, at a fixed magnetic field, the frequency corresponding to the peak is proportional to the electron temperature. 

\subsection{Polarization of Hybrid Electron Energy Distribution} \label{3.3}

In this section, we calculate the linear polarization degree of very-small-pitch-angle and conventional synchrotron radiation as a function of the conjunctive Lorentz factor $\gamma_{\rm th}$ for a hybrid electron energy distribution. The radiation frequency covers from the radio band to the $\gamma$-ray band. In higher-frequency bands, stronger magnetic fields are required to satisfy the conditions for very-small-pitch-angle synchrotron radiation. In lower-frequency bands, very-small-pitch-angle synchrotron radiation requires weaker magnetic fields to produce significant polarization. Therefore, in this section, we calculate the polarization in the radio, optical, X-ray, and $\gamma$-ray bands.

\subsubsection{Polarization in the Radio Band}

\begin{figure*}
\centering
\includegraphics[scale=0.45]{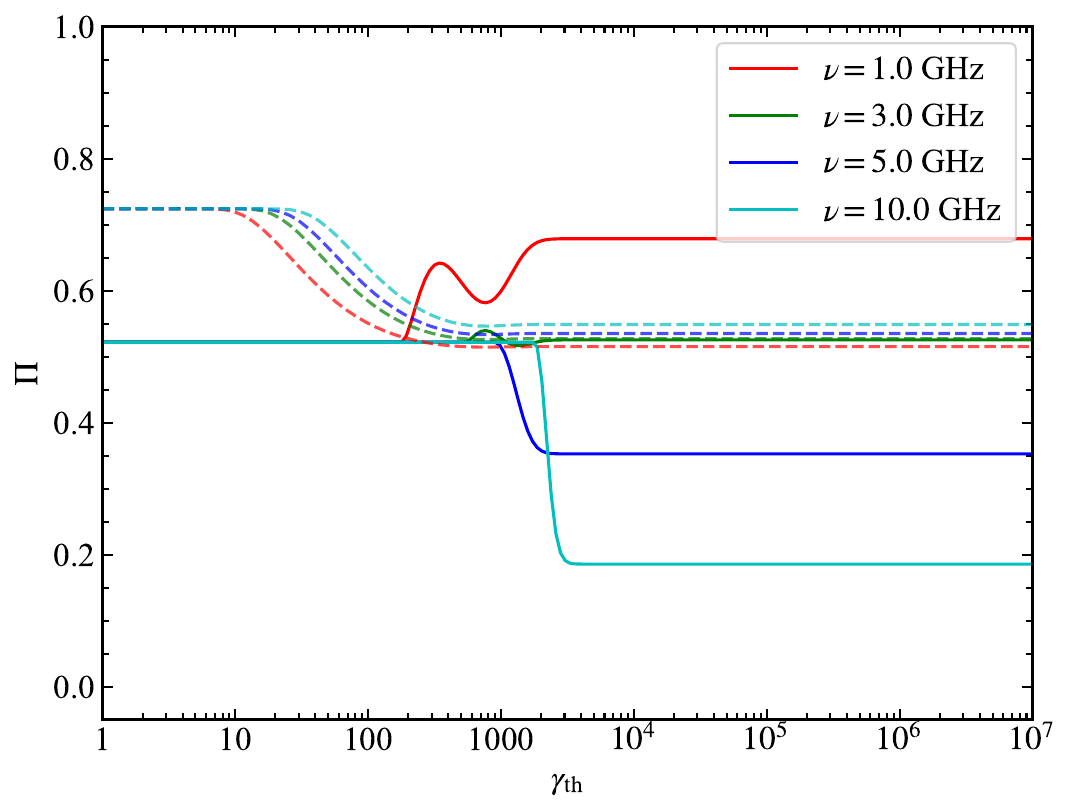}
\includegraphics[scale=0.45]{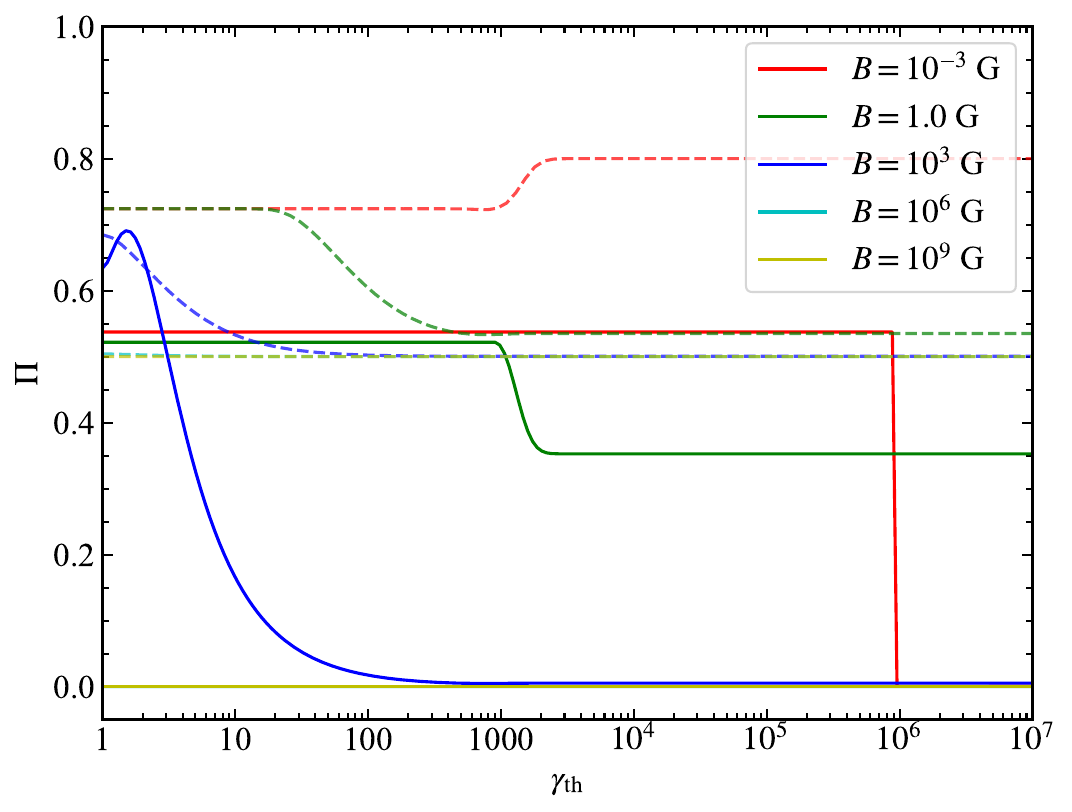}
\includegraphics[scale=0.45]{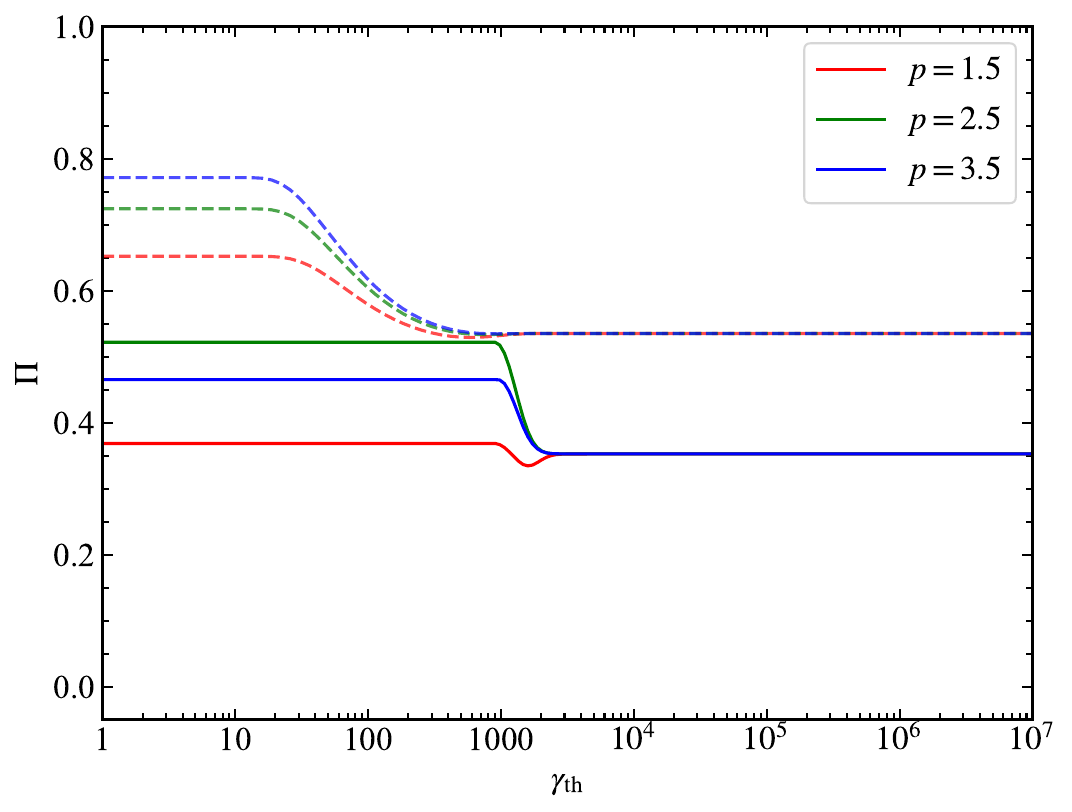}
\includegraphics[scale=0.45]{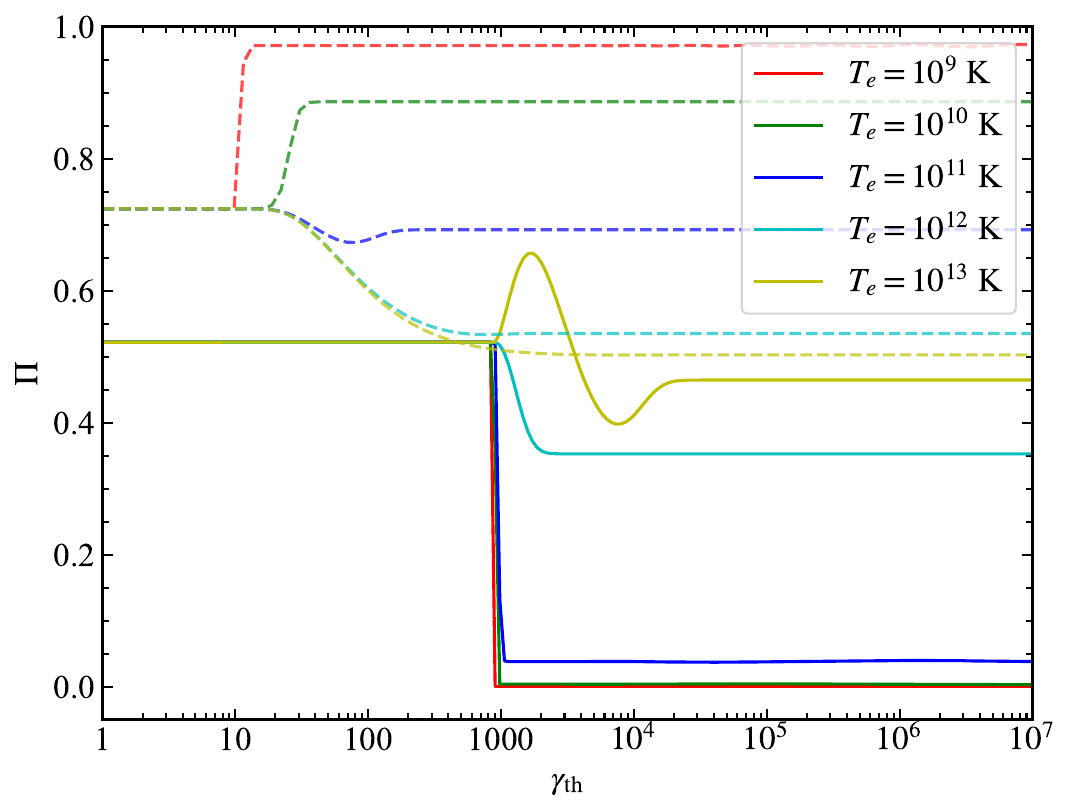}
\caption{Linear polarization degree in the radio band. The solid and dashed lines represent the very-small-pitch-angle and conventional synchrotron radiation, respectively. Top left panel: we adopt $B = 1.0 \, \mathrm{G}$, $p = 2.5$, and $T_e = 1.0 \times 10^{12}\,\mathrm{K}$. It is noted that at $\nu = 3.0\,\mathrm{GHz}$, very-small-pitch-angle and conventional synchrotron radiation exhibit the same polarization degree for $\gamma_{\rm th} \gtrsim 2 \times 10^{3}$. Top right panel: we adopt $\nu = 5.0 \, \mathrm{GHz}$, $p = 2.5$, and $T_e = 1.0 \times 10^{12}\,\mathrm{K}$. The solid lines corresponding to $B = 1.0 \times 10^{6}\,\mathrm{G}$ and $B = 1.0 \times 10^{9}\,\mathrm{G}$ overlap. Bottom left panel: we adopt $\nu = 5.0 \, \mathrm{GHz}$, $B = 1.0 \, \mathrm{G}$, and $T_e = 1.0 \times 10^{12}\,\mathrm{K}$. Bottom right panel: we adopt $\nu = 5.0 \, \mathrm{GHz}$, $B = 1.0 \, \mathrm{G}$, and $p = 2.5$.
\label{Radio-HY}}
\end{figure*}

Figure \ref{Radio-HY} shows the linear polarization degrees of very-small-pitch-angle and conventional synchrotron radiation in the radio band as a function of $\gamma_{\rm th}$ for a hybrid electron energy distribution. In the radio band, the linear polarization degree of the very-small-pitch-angle synchrotron radiation is generally lower than that of conventional synchrotron radiation. 

We investigate the effects of different observational frequencies, magnetic field strengths, power-law indices, and electron temperatures on the very-small-pitch-angle synchrotron polarization in the radio band as a function of $\gamma_{\rm th}$.
The polarization degree related to the observational frequency is shown in the top left panel of Figure \ref{Radio-HY}. We take the observational frequencies at the $L$ (1 GHz), $S$ (3 GHz), $C$ (5 GHz) and $X$ (10 GHz) bands. In the $L$ band, the polarization degree remains constant at $52.2\%$ for $\gamma_{\rm th} \lesssim 200$ and remains constant at $67.9\%$ for $\gamma_{\rm th} \gtrsim 2.0 \times 10^{3}$. The polarization profile of $L$ band exhibits a transition in the range $200 \lesssim \gamma_{\rm th} \lesssim 2.0 \times 10^{3}$. In the $S$ band, the polarization degree remains constant at $52.2\%$ for $\gamma_{\rm th} \lesssim 500$ and remains constant at $52.5\%$ for $\gamma_{\rm th} \gtrsim 2.0 \times 10^{3}$. The polarization profile of the $S$ band exhibits a transition in the range $500 \lesssim \gamma_{\rm th} \lesssim 2.0 \times 10^{3}$. In the $C$ band, the polarization degree remains constant at $52.2\%$ for $\gamma_{\rm th} \lesssim 1.0 \times 10^{3}$ and remains constant at $35.3\%$ for $\gamma_{\rm th} \gtrsim 3.0 \times 10^{3}$. The polarization profile of the $C$ band exhibits a transition in the range $1.0 \times 10^{3} \lesssim \gamma_{\rm th} \lesssim 2.0 \times 10^{3}$. In the $X$ band, the polarization degree remains constant at $52.2\%$ for $\gamma_{\rm th} \lesssim 2.0 \times 10^{3}$ and remains constant at $18.6\%$ for $\gamma_{\rm th} \gtrsim 3.0 \times 10^{3}$. The polarization profile of the $X$ band exhibits a transition in the range $2.0 \times 10^{3} \lesssim \gamma_{\rm th} \lesssim 3.0 \times 10^{3}$. Overall, the transition exhibits a very sharp change at higher observational frequencies. It seems that, in the radio band, the observational frequency does not affect the polarization degree in the low-$\gamma_{\rm th}$ range, whereas it has a strong influence on the polarization degree in the high-$\gamma_{\rm th}$ range. 
The polarization degree related to the magnetic field is shown in the top right panel of Figure \ref{Radio-HY}. When $B = 1.0 \times 10^{-3}\,\mathrm{G}$, the polarization degree remains constant at $53.7\%$ for $\gamma_{\rm th} \lesssim 1.0 \times 10^{6}$ and approaches zero for $\gamma_{\rm th} \gtrsim 1.0 \times 10^{6}$. When $B = 1.0\,\mathrm{G}$, the polarization degree remains constant at $52.2\%$ for $\gamma_{\rm th} \lesssim 1.0 \times 10^{3}$ and remains constant at $35.3\%$ for $\gamma_{\rm th} \gtrsim 3.0 \times 10^{3}$. The polarization profile exhibits a transition in the range $1.0 \times 10^{3} \lesssim \gamma_{\rm th} \lesssim 3.0 \times 10^{3}$. When $B = 1.0\times 10^3\,\mathrm{G}$, the polarization degree first increases and then decreases with increasing $\gamma_{\rm th}$, reaching a maximum value of approximately $69.1\%$ at $\gamma_{\rm th} \approx 1.5$, and approaches zero for $\gamma_{\rm th} \gtrsim 1.0 \times 10^{3}$. When $B = 1.0\times 10^{6}\,\mathrm{G}$ or $1.0\times 10^{9}\,\mathrm{G}$, the polarization degree remains zero over the entire range of $\gamma_{\rm th}$. Overall, the transition exhibits a very sharp change at lower magnetic field strengths. We see from the above results that the magnetic field has a significant influence on the polarization in the radio band. 
The polarization degree related to the power-law index of the nonthermal electrons is shown in the bottom left panel of Figure \ref{Radio-HY}. When $p = 1.5$, the polarization degree remains constant at $36.8\%$ for $\gamma_{\rm th} \lesssim 1.0 \times 10^{3}$ and remains constant at $35.3\%$ for $\gamma_{\rm th} \gtrsim 3.0 \times 10^{3}$. When $p = 2.5$, the polarization degree remains constant at $52.2\%$ for $\gamma_{\rm th} \lesssim 1.0 \times 10^{3}$ and remains constant at $35.3\%$ for $\gamma_{\rm th} \gtrsim 3.0 \times 10^{3}$. When $p = 3.5$, the polarization degree remains constant at $46.5\%$ for $\gamma_{\rm th} \lesssim 1.0 \times 10^{3}$ and remains constant at $35.3\%$ for $\gamma_{\rm th} \gtrsim 3.0 \times 10^{3}$. 
Overall, for different power-law indices, the polarization profile exhibits the same transition range of $1.0 \times 10^{3} \lesssim \gamma_{\rm th} \lesssim 3.0 \times 10^{3}$. In the radio band, the power-law index has a strong influence on the polarization degree in the low-$\gamma_{\rm th}$ range, whereas it does not affect the polarization degree in the high-$\gamma_{\rm th}$ range. 
The polarization degree related to the electron temperature is shown in the bottom right panel of Figure \ref{Radio-HY}. When $T_e = 1.0 \times 10^{9}\,\mathrm{K}$, the polarization degree remains constant at $52.2\%$ for $\gamma_{\rm th} \lesssim 900 $ and approaches zero for $\gamma_{\rm th} \gtrsim 900$. When $T_e = 1.0 \times 10^{10}\,\mathrm{K}$, the polarization degree remains constant at $52.2\%$ for $\gamma_{\rm th} \lesssim 1.0 \times 10^{3}$ and remains constant at $0.4\%$ for $\gamma_{\rm th} \gtrsim 1.0 \times 10^{3}$. When $T_e = 1.0 \times 10^{11}\,\mathrm{K}$, the polarization degree remains constant at $52.2\%$ for $\gamma_{\rm th} \lesssim 1.0 \times 10^{3}$ and remains constant at $3.7\%$ for $\gamma_{\rm th} \gtrsim 1.0 \times 10^{3}$. When $T_e = 1.0 \times 10^{12}\,\mathrm{K}$, the polarization degree remains constant at $52.2\%$ for $\gamma_{\rm th} \lesssim 1.0 \times 10^{3}$ and remains constant at $35.3\%$ for $\gamma_{\rm th} \gtrsim 3.0 \times 10^{3}$. The polarization profile exhibits a transition in the range $1.0 \times 10^{3} \lesssim \gamma_{\rm th} \lesssim 3.0 \times 10^{3}$. When $T_e = 1.0 \times 10^{13}\,\mathrm{K}$, the polarization degree remains constant at $52.2\%$ for $\gamma_{\rm th} \lesssim 1.0 \times 10^{3}$ and remains constant at $46.4\%$ for $\gamma_{\rm th} \gtrsim 2.0 \times 10^{4}$. The polarization profile exhibits a transition in the range $1.0 \times 10^{3} \lesssim \gamma_{\rm th} \lesssim 2.0 \times 10^{4}$. Overall, the transition exhibits a very sharp change at lower electron temperatures. In the radio band, the electron temperature has a strong influence on the polarization degree in the high-$\gamma_{\rm th}$ range, whereas it does not affect the polarization degree in the low-$\gamma_{\rm th}$ range.

\subsubsection{Polarization in the Optical Band}

\begin{figure*}
\centering
\includegraphics[scale=0.45]{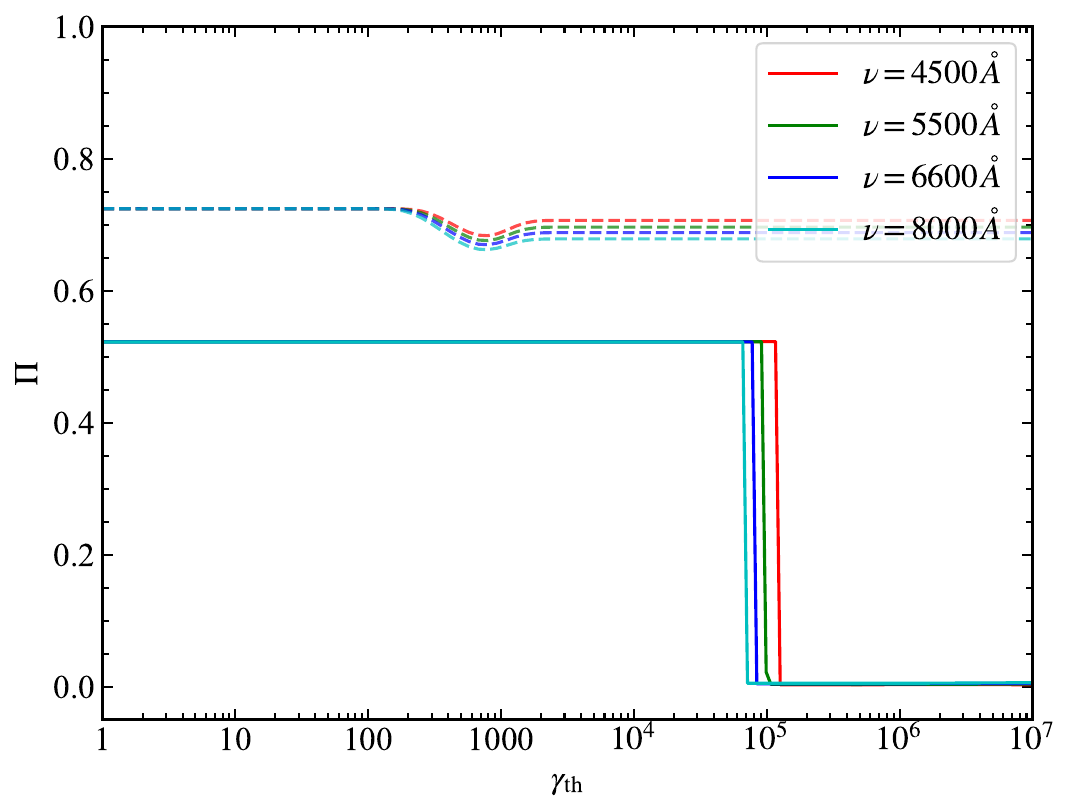}
\includegraphics[scale=0.45]{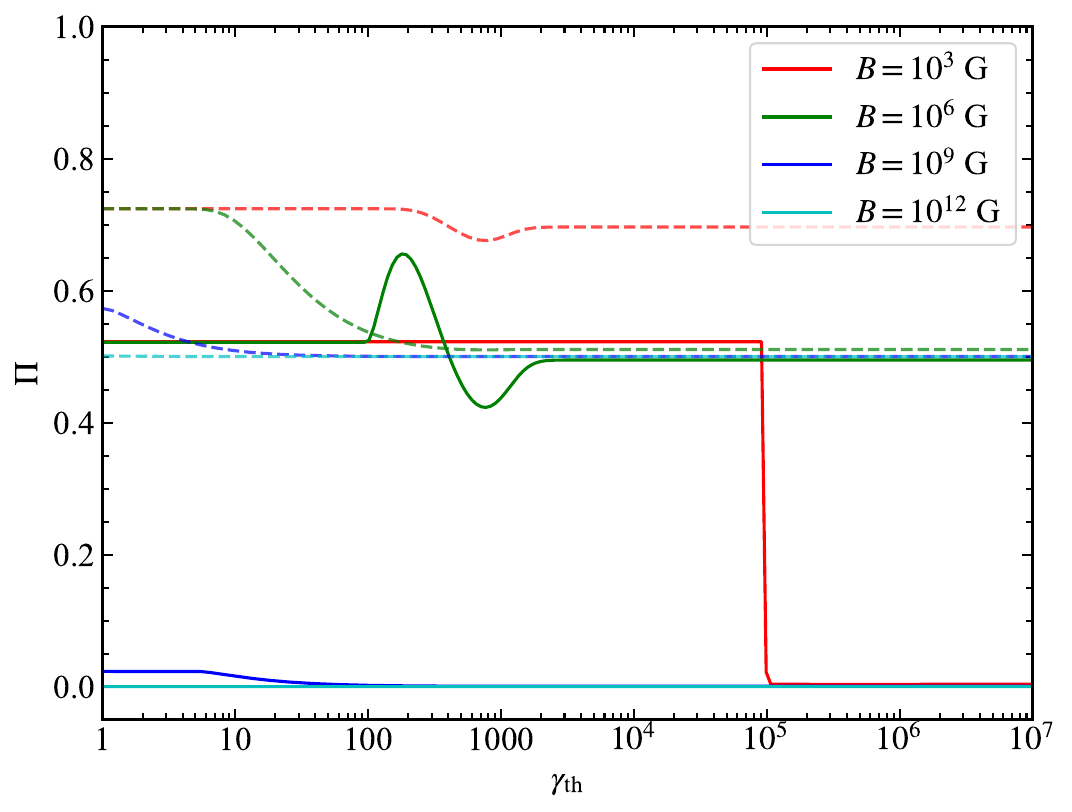}
\includegraphics[scale=0.45]{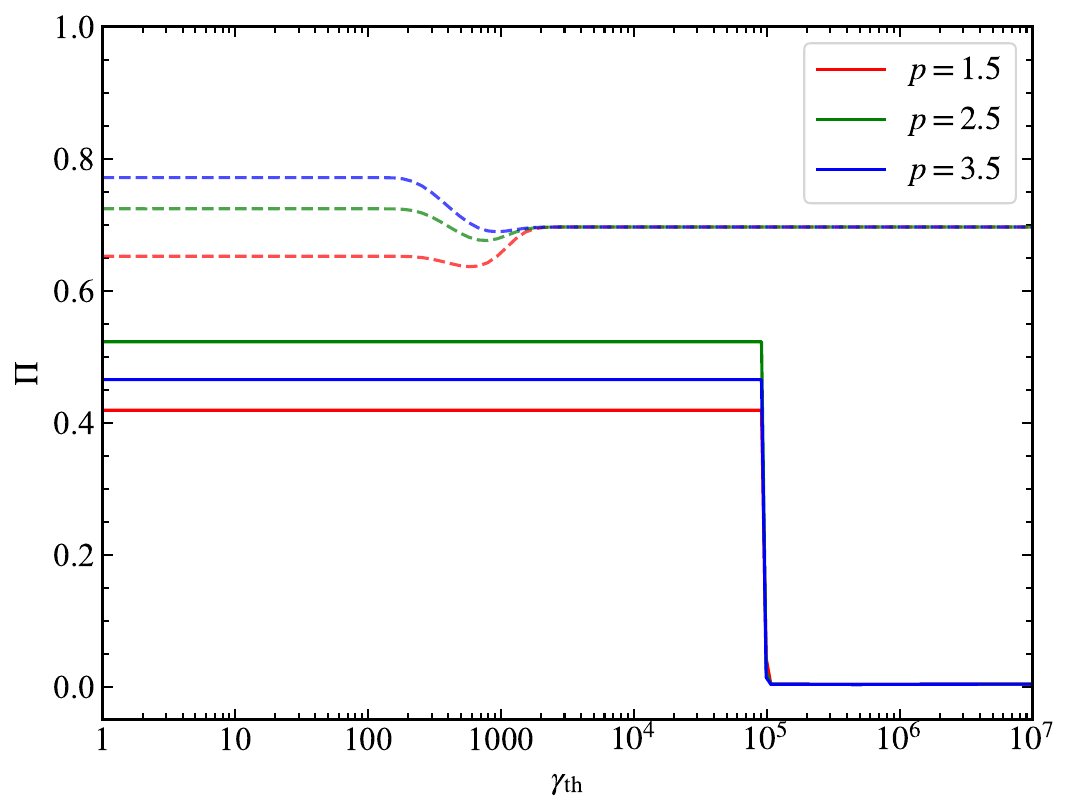}
\includegraphics[scale=0.45]{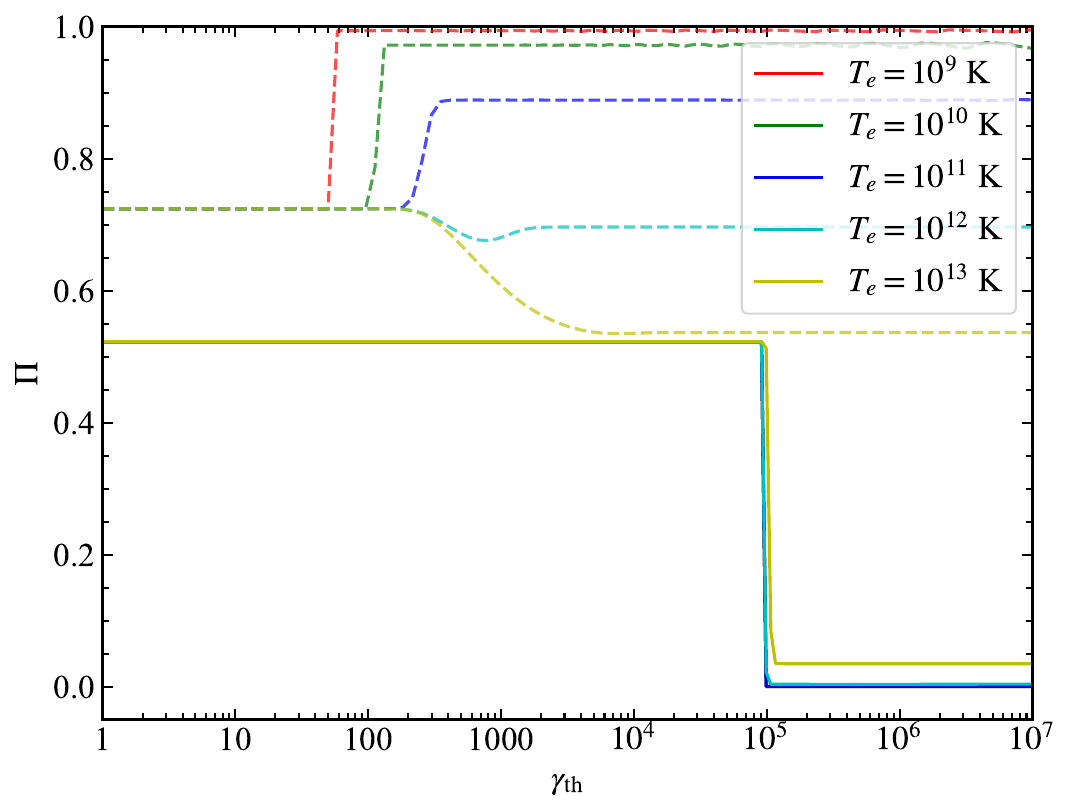}
\caption{Linear polarization degree in the optical band. The solid and dashed lines represent the very-small-pitch-angle and conventional synchrotron radiation, respectively. Top left panel: we adopt $B = 1.0 \times 10^{3} \, \mathrm{G}$, $p = 2.5$, and $T_e = 1.0 \times 10^{12}\,\mathrm{K}$. Top right panel: we adopt $\nu = 5500 \, \mathrm{\AA}$, $p = 2.5$, and $T_e = 1.0 \times 10^{12}\,\mathrm{K}$. Bottom left panel: we adopt $\nu = 5500 \, \mathrm{\AA}$, $B = 1.0 \times 10^{3} \, \mathrm{G}$, and $T_e = 1.0 \times 10^{12}\,\mathrm{K}$. Bottom right panel: we adopt $\nu = 5500 \, \mathrm{\AA}$, $B = 1.0 \times 10^{3} \, \mathrm{G}$, and $p = 2.5$. When $T_e \leq 1.0 \times 10^{12}\,\mathrm{K}$, very-small-pitch-angle synchrotron radiation exhibits the same polarization degree; therefore the corresponding solid lines overlap.
\label{Optical-HY}}
\end{figure*}

Figure \ref{Optical-HY} shows the linear polarization degree of very-small-pitch-angle and conventional synchrotron radiation in the optical band as a function of $\gamma_{\rm th}$ for a hybrid electron energy distribution. In the optical band, the linear polarization degree of the very-small-pitch-angle synchrotron radiation is generally lower than that of conventional synchrotron radiation. 

We investigate the effects of different observational frequencies, magnetic field strengths, power-law indices, and electron temperatures on the very-small-pitch-angle synchrotron polarization in the optical band as a function of $\gamma_{\rm th}$. 
The polarization degree related to the observational frequency is shown in the top left panel of Figure \ref{Optical-HY}. We take the observational frequencies at the $B$ ($4500 \mathrm{~\AA}$), $V$ ($5500 \mathrm{~\AA}$), $R$ ($6600 \mathrm{~\AA}$) and $I$ ($8000 \mathrm{~\AA}$) bands. In the $B$ band, the polarization degree remains constant at $52.2\%$ for $\gamma_{\rm th} \lesssim 7.0 \times 10^{4}$ and approaches zero for $\gamma_{\rm th} \gtrsim 7.0 \times 10^{4}$. In the $V$ band, the polarization degree remains constant at $52.2\%$ for $\gamma_{\rm th} \lesssim 8.0 \times 10^{4}$ and approaches zero for $\gamma_{\rm th} \gtrsim 8.0 \times 10^{4}$. In the $R$ band, the polarization degree remains constant at $52.2\%$ for $\gamma_{\rm th} \lesssim 1.0 \times 10^{5}$ and approaches zero for $\gamma_{\rm th} \gtrsim 1.0 \times 10^{5}$. In the $I$ band, the polarization degree remains constant at $52.2\%$ for $\gamma_{\rm th} \lesssim 1.4 \times 10^{5}$ and approaches zero for $\gamma_{\rm th} \gtrsim 1.4 \times 10^{5}$. Overall, the polarization profiles for different optical bands all show a very sharp transition. In the optical band, the polarization profiles are similar at different observational frequencies. 
The polarization degree related to the magnetic field is shown in the top right panel of Figure \ref{Optical-HY}. When $B = 1.0 \times 10^{3}\,\mathrm{G}$, the polarization degree remains constant at $52.2\%$ for $\gamma_{\rm th} \lesssim 1.0 \times 10^{5}$ and approaches zero for $\gamma_{\rm th} \gtrsim 1.0 \times 10^{5}$. When $B = 1.0 \times 10^{6}\,\mathrm{G}$, the polarization degree remains constant at $52.2\%$ for $\gamma_{\rm th} \lesssim 100 $ and remains constant at $49.5\%$ for $\gamma_{\rm th} \gtrsim 2.0 \times 10^{3}$. The polarization profile exhibits a transition in the range $100 \lesssim \gamma_{\rm th} \lesssim 2.0 \times 10^{3}$. When $B = 1.0 \times 10^{9}\,\mathrm{G}$, the polarization degree remains constant at $2.3\%$ for $\gamma_{\rm th} \lesssim 5 $ and approaches zero for $\gamma_{\rm th} \gtrsim 100$. The polarization profile exhibits a transition in the range $5 \lesssim \gamma_{\rm th} \lesssim 100$. When $B = 1.0\times 10^{9}\,\mathrm{G}$, the polarization degree remains zero over the entire range of $\gamma_{\rm th}$. Overall, the transition exhibits a very sharp change at lower magnetic field strengths. In the optical band, the magnetic field has a significant influence on the polarization. 
The polarization degree related to the power-law index is shown in the bottom left panel of Figure \ref{Optical-HY}. When $p = 1.5$, the polarization degree remains constant at $41.9\%$ for $\gamma_{\rm th} \lesssim 1.0 \times 10^{5}$ and approaches zero for $\gamma_{\rm th} \gtrsim 1.0 \times 10^{5}$. When $p = 2.5$, the polarization degree remains constant at $52.2\%$ for $\gamma_{\rm th} \lesssim 1.0 \times 10^{5}$ and approaches zero for $\gamma_{\rm th} \gtrsim 1.0 \times 10^{5}$. When $p = 3.5$, the polarization degree remains constant at $46.5\%$ for $\gamma_{\rm th} \lesssim 1.0 \times 10^{5}$ and approaches zero for $\gamma_{\rm th} \gtrsim 1.0 \times 10^{5}$. Overall, the polarization profiles for different power-law indices all show a very sharp transition. In the optical band, the power-law index has a strong influence on the polarization degree in the low-$\gamma_{\rm th}$ range, whereas it does not affect the polarization degree in the high-$\gamma_{\rm th}$ range. 
The polarization degree related to the electron temperature is shown in the bottom right panel of Figure \ref{Optical-HY}. When $T_e \leq 1.0 \times 10^{12}\,\mathrm{K}$, the polarization degree remains constant at $52.2\%$ for $\gamma_{\rm th} \lesssim 1.0 \times 10^{5} $ and approaches zero for $\gamma_{\rm th} \gtrsim 1.0 \times 10^{5}$. When $T_e = 1.0 \times 10^{13}\,\mathrm{K}$, the polarization degree remains constant at $52.2\%$ for $\gamma_{\rm th} \lesssim 1.2 \times 10^{5}$ and remains constant at $3.5\%$ for $\gamma_{\rm th} \gtrsim 1.2 \times 10^{5}$. Overall, the polarization profiles for different electron temperatures all show a very sharp transition. In the optical band, the electron temperature has a weak influence on the polarization degree in the high-$\gamma_{\rm th}$ range, whereas it does not affect the polarization degree in the low-$\gamma_{\rm th}$ range.

\subsubsection{Polarization in the X-Ray Band}

\begin{figure*}
\centering
\includegraphics[scale=0.45]{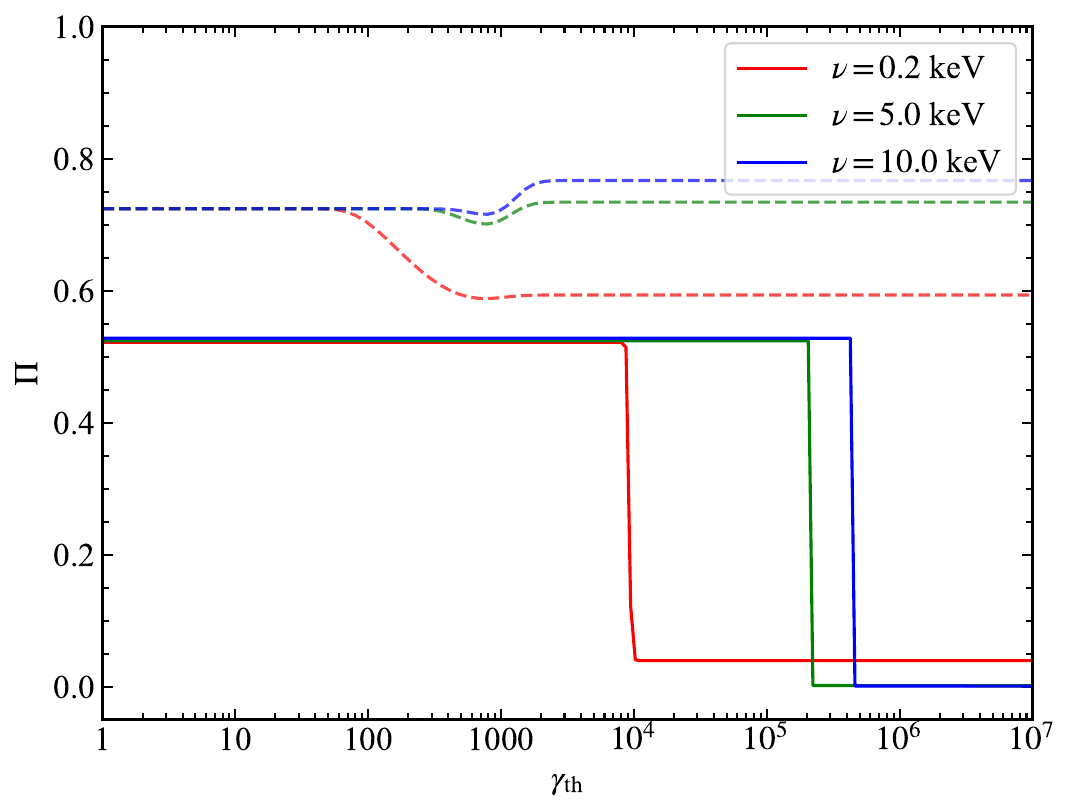}
\includegraphics[scale=0.45]{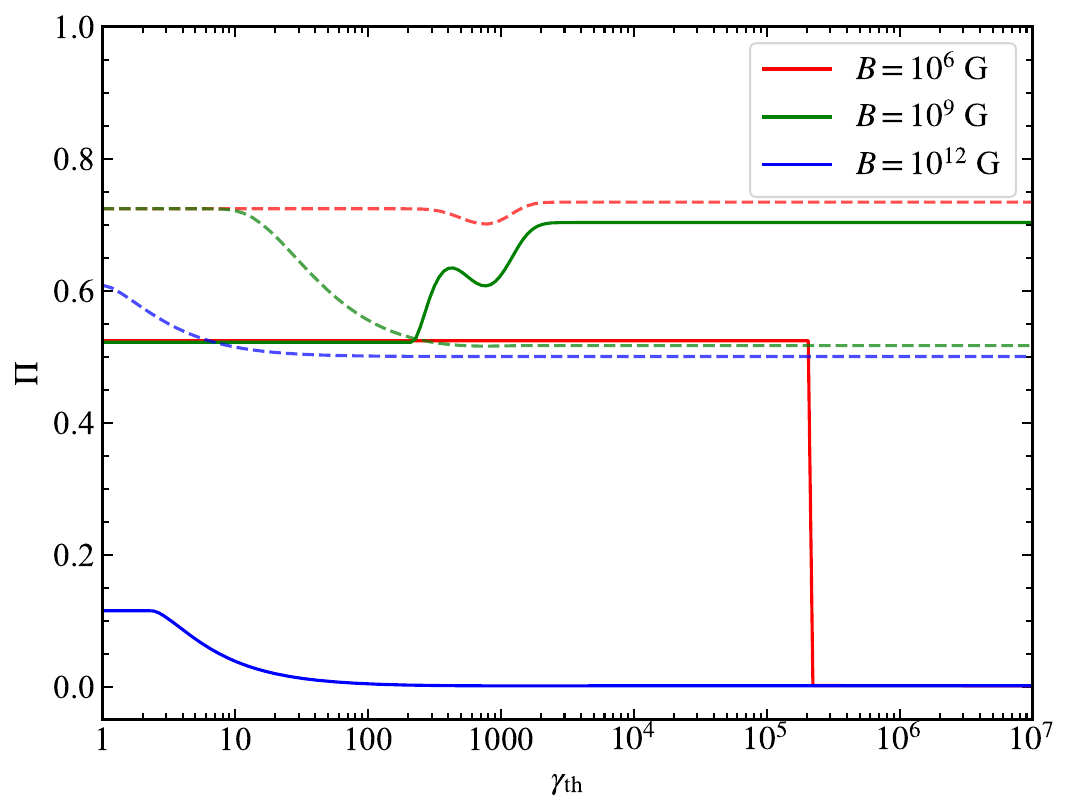}
\includegraphics[scale=0.45]{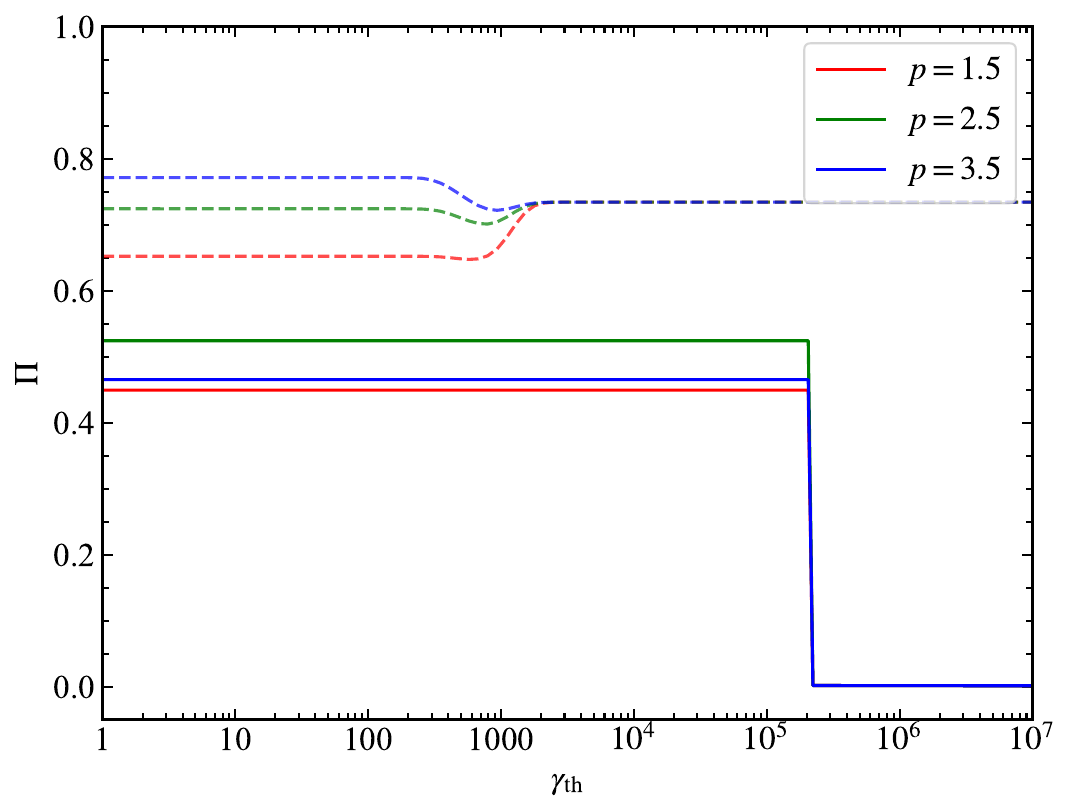}
\includegraphics[scale=0.45]{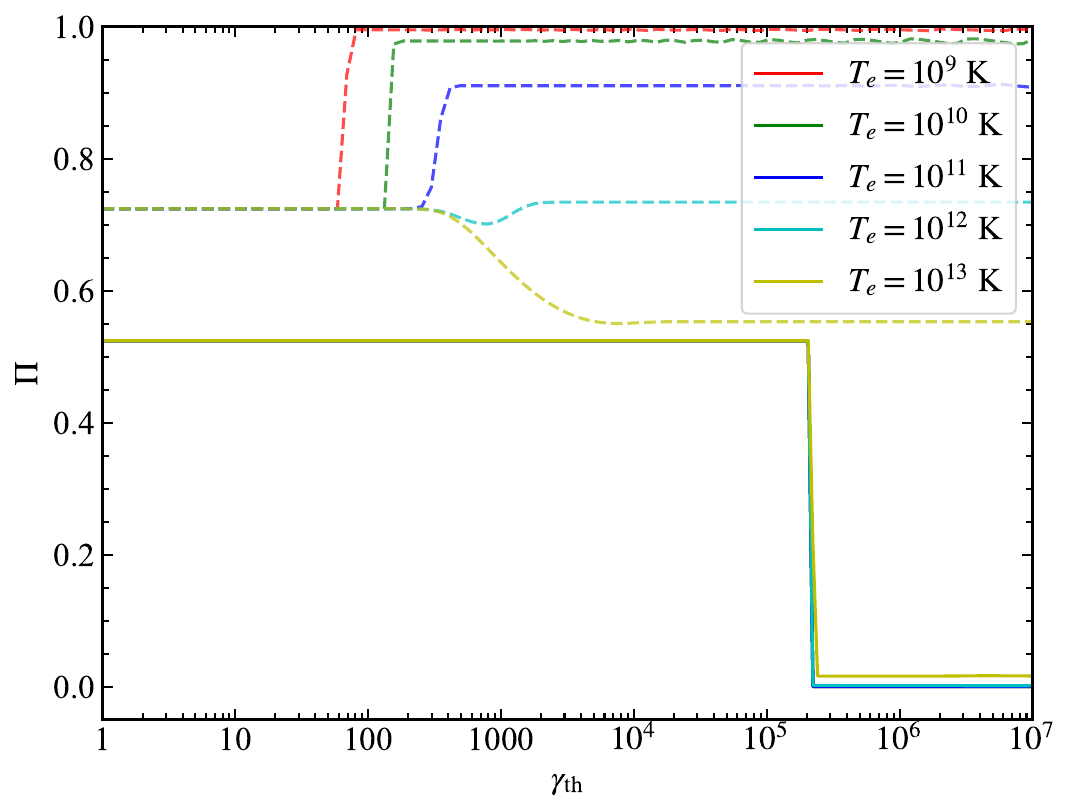}
\caption{Linear polarization degree in the X-ray band. The solid and dashed lines represent the very-small-pitch-angle and conventional synchrotron radiation, respectively. Top left panel: we adopt $B = 1.0 \times 10^{6} \, \mathrm{G}$, $p = 2.5$, and $T_e = 1.0 \times 10^{12}\,\mathrm{K}$. Top right panel: we adopt $\nu = 5.0 \, \mathrm{keV}$, $p = 2.5$, and $T_e = 1.0 \times 10^{12}\,\mathrm{K}$. Bottom left panel: we adopt $\nu = 5.0 \, \mathrm{keV}$, $B = 1.0 \times 10^{6} \, \mathrm{G}$, and $T_e = 1.0 \times 10^{12}\,\mathrm{K}$. Bottom right panel: we adopt $\nu = 5.0 \, \mathrm{keV}$, $B = 1.0 \times 10^{6} \, \mathrm{G}$, and $p = 2.5$. When $T_e \leq 1.0 \times 10^{12}\,\mathrm{K}$, very-small-pitch-angle synchrotron radiation exhibits the same polarization degree; therefore the corresponding solid lines overlap. 
\label{X-ray-HY}}
\end{figure*}

Figure \ref{X-ray-HY} shows the  linear polarization degree of very-small-pitch-angle and conventional synchrotron radiation in the X-ray band as a function of $\gamma_{\rm th}$ for a hybrid electron energy distribution. 
In the X-ray band, the linear polarization degree of the very-small-pitch-angle synchrotron radiation is generally lower than that of conventional synchrotron radiation.

We investigate the effects of different observational frequencies, magnetic field strengths, power-law indices, and electron temperatures on the very-small-pitch-angle synchrotron polarization in the X-ray band as a function of $\gamma_{\rm th}$. 
The polarization degree related to the observational frequency is shown in the top left panel of Figure \ref{X-ray-HY}. When $\nu = 0.2 \,\mathrm{keV}$, the polarization degree remains constant at $52.2\%$ for $\gamma_{\rm th} \lesssim 1.0 \times 10^{4}$ and remains constant at $3.9\%$ for $\gamma_{\rm th} \gtrsim 1.0 \times 10^{4}$. When $\nu = 5.0 \,\mathrm{keV}$, the polarization degree remains constant at $52.4\%$ for $\gamma_{\rm th} \lesssim 2.0 \times 10^{5}$ and approaches zero for $\gamma_{\rm th} \gtrsim 2.0 \times 10^{5}$. When $\nu = 10.0 \,\mathrm{keV}$, the polarization degree remains constant at $52.4\%$ for $\gamma_{\rm th} \lesssim 5.0 \times 10^{5}$ and approaches zero for $\gamma_{\rm th} \gtrsim 5.0 \times 10^{5}$. Overall, the polarization profiles for different X-ray bands all show a very sharp transition. In the X-ray band, the observational frequency has a weak effect on the polarization degree in the low-$\gamma_{\rm th}$ range, whereas it has a strong influence on the polarization degree in the high-$\gamma_{\rm th}$ range.
The polarization degree related to the magnetic field is shown in the top right panel of Figure \ref{X-ray-HY}.  When $B = 1.0 \times 10^{6}\,\mathrm{G}$, the polarization degree remains constant at $52.2\%$ for $\gamma_{\rm th} \lesssim 2.2 \times 10^{5}$ and approaches zero for $\gamma_{\rm th} \gtrsim 2.2 \times 10^{5}$. When $B = 1.0 \times 10^{9}\,\mathrm{G}$, the polarization degree remains constant at $52.2\%$ for $\gamma_{\rm th} \lesssim 200 $ and remains constant at $70.3\%$ for $\gamma_{\rm th} \gtrsim 2.0 \times 10^{3}$. The polarization profile exhibits a transition in the range $200 \lesssim \gamma_{\rm th} \lesssim 2.0 \times 10^{3}$. When $B = 1.0 \times 10^{12}\,\mathrm{G}$, the polarization degree remains constant at $11.5\%$ for $\gamma_{\rm th} \lesssim 2.5 $ and approaches zero for $\gamma_{\rm th} \gtrsim 100$. The polarization profile exhibits a transition in the range $2.5 \lesssim \gamma_{\rm th} \lesssim 100$. Overall, the transition exhibits a very sharp change at lower magnetic field strengths. In the X-ray band, the magnetic field has a significant influence on the polarization.
The polarization degree related to the power-law index is shown in the bottom left panel of Figure \ref{X-ray-HY}. When $p = 1.5$, the polarization degree remains constant at $44.9\%$ for $\gamma_{\rm th} \lesssim 2.2 \times 10^{5}$ and approaches zero for $\gamma_{\rm th} \gtrsim 2.2 \times 10^{5}$. When $p = 2.5$, the polarization degree remains constant at $52.4\%$ for $\gamma_{\rm th} \lesssim 2.2 \times 10^{5}$ and approaches zero for $\gamma_{\rm th} \gtrsim 2.2 \times 10^{5}$. When $p = 3.5$, the polarization degree remains constant at $46.5\%$ for $\gamma_{\rm th} \lesssim 2.2 \times 10^{5}$ and approaches zero for $\gamma_{\rm th} \gtrsim 2.2 \times 10^{5}$.  Overall, the polarization profiles for different power-law indices all show a very sharp transition. In the X-ray band, the power-law index has a strong influence on the polarization degree in the low-$\gamma_{\rm th}$ range, whereas it does not affect the polarization degree in the high-$\gamma_{\rm th}$ range. 
The polarization degree related to the electron temperature is shown in the bottom right panel of Figure \ref{X-ray-HY}. When $T_e \leq 1.0 \times 10^{12}\,\mathrm{K}$, the polarization degree remains constant at $52.4\%$ for $\gamma_{\rm th} \lesssim 2.2 \times 10^{5} $ and approaches zero for $\gamma_{\rm th} \gtrsim 2.2 \times 10^{5}$. When $T_e = 1.0 \times 10^{13}\,\mathrm{K}$, the polarization degree remains constant at $52.4\%$ for $\gamma_{\rm th} \lesssim 2.5 \times 10^{5}$ and remains constant at $1.5\%$ for $\gamma_{\rm th} \gtrsim 2.5 \times 10^{5}$. Overall, the polarization profiles for different electron temperatures all show a very sharp transition. In the X-ray band, the electron temperature has a weak influence on the polarization degree in the high-$\gamma_{\rm th}$ range, whereas it does not affect the polarization degree in the low-$\gamma_{\rm th}$ range.

\subsubsection{Polarization in the $\gamma$-Ray Band}

\begin{figure*}
\centering
\includegraphics[scale=0.45]{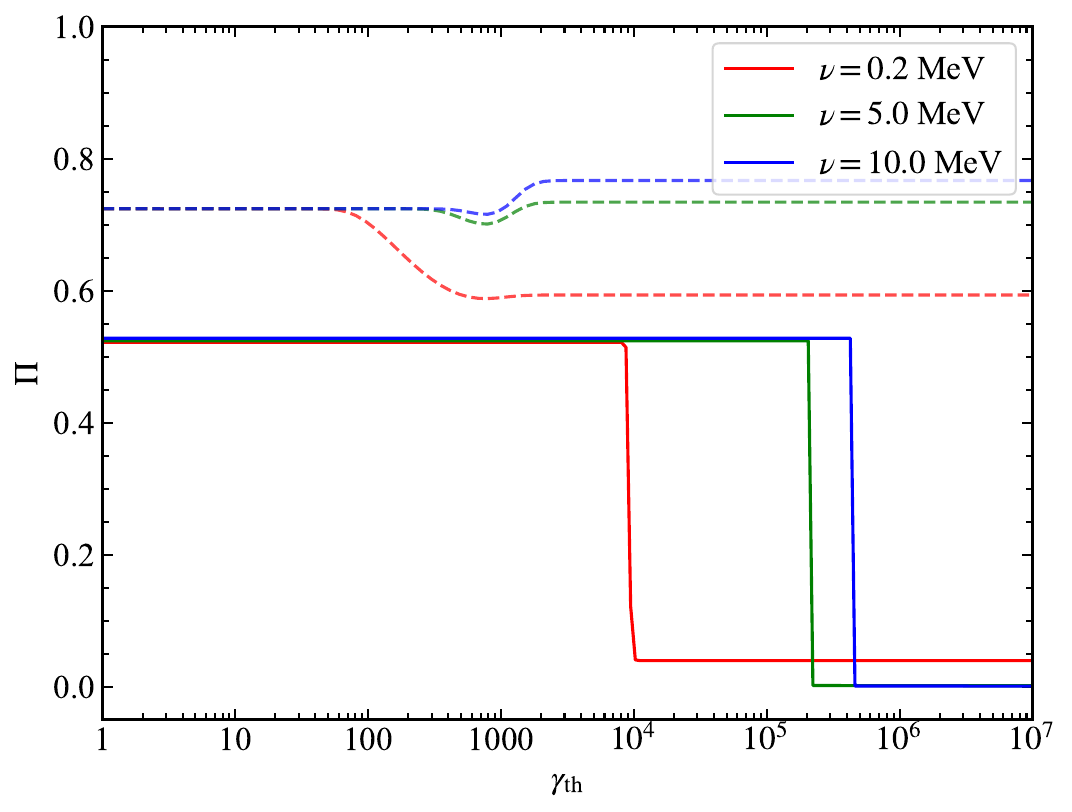}
\includegraphics[scale=0.45]{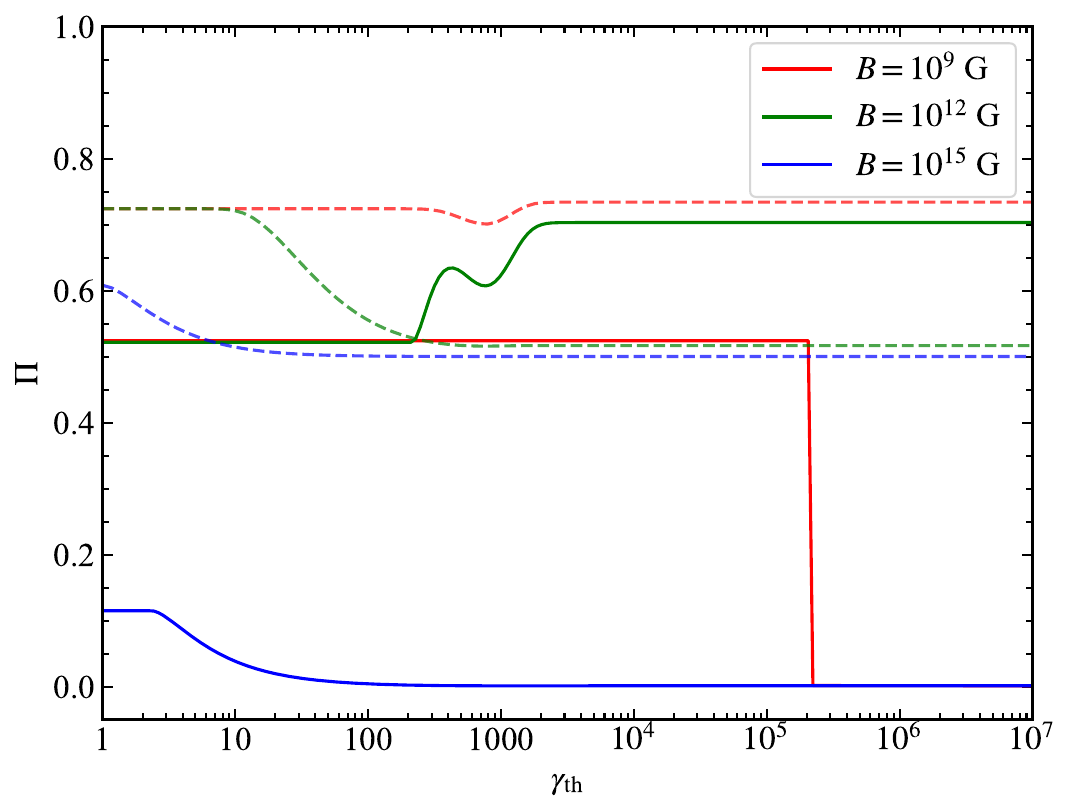}
\includegraphics[scale=0.45]{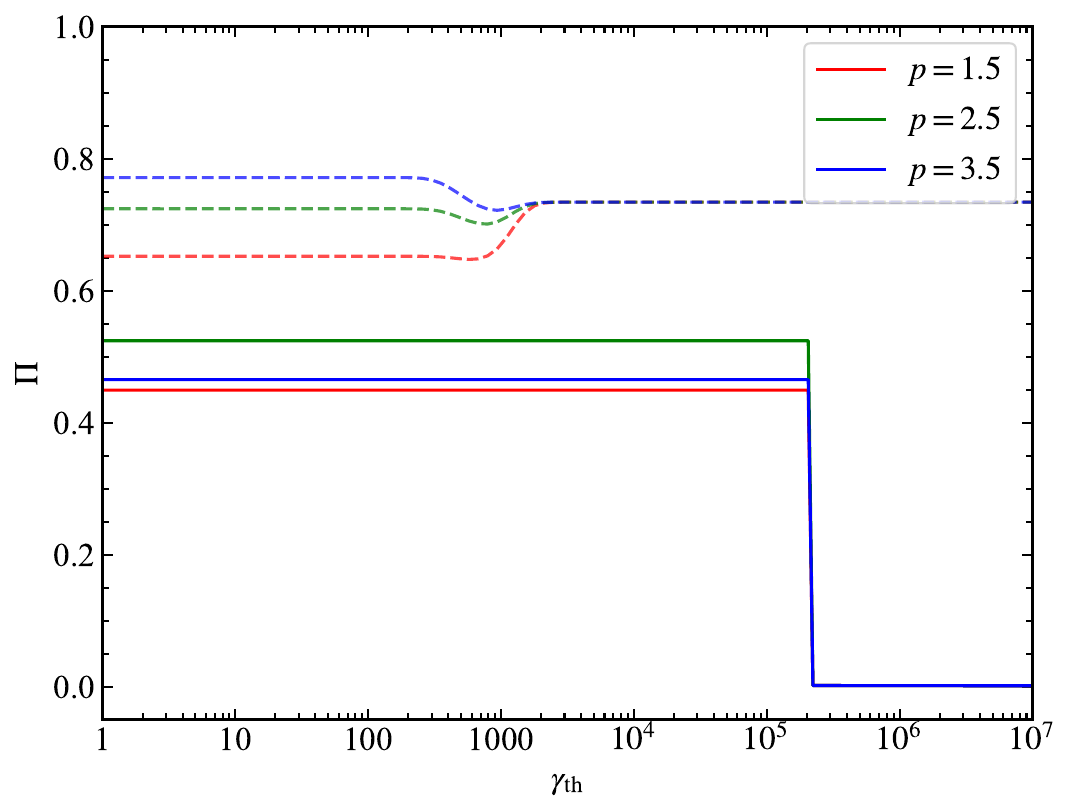}
\includegraphics[scale=0.45]{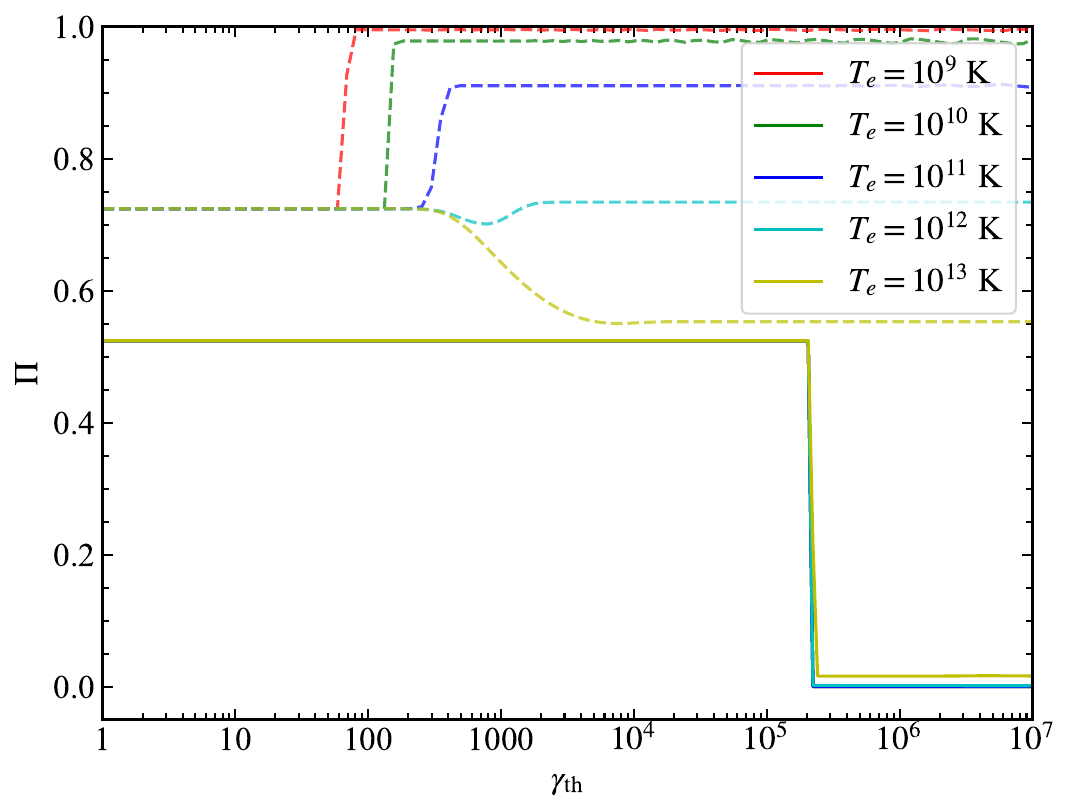}
\caption{Linear polarization degree in the $\gamma$-ray band. The solid and dashed lines represent the very-small-pitch-angle and conventional synchrotron radiation, respectively. Top left panel: we adopt $B = 1.0 \times 10^{9} \, \mathrm{G}$, $p = 2.5$, and $T_e = 1.0 \times 10^{12}\,\mathrm{K}$. Top right panel: we adopt $\nu = 5.0 \, \mathrm{MeV}$, $p = 2.5$, and $T_e = 1.0 \times 10^{12}\,\mathrm{K}$. Bottom left panel: we adopt $\nu = 5.0 \, \mathrm{MeV}$, $B = 1.0 \times 10^{9} \, \mathrm{G}$, and $T_e = 1.0 \times 10^{12}\,\mathrm{K}$. Bottom right panel: we adopt $\nu = 5.0 \, \mathrm{MeV}$, $B = 1.0 \times 10^{9} \, \mathrm{G}$, and $p = 2.5$. When $T_e \leq 1.0 \times 10^{12}\,\mathrm{K}$, very-small-pitch-angle synchrotron radiation exhibits the same polarization degree; therefore the corresponding solid lines overlap. 
\label{Gamma-ray-HY}}
\end{figure*}

Figure \ref{Gamma-ray-HY} shows the linear polarization degree of very-small-pitch-angle and conventional synchrotron radiation in the $\gamma$-ray band as a function of $\gamma_{\rm th}$ for a hybrid electron energy distribution. We note that the shape of the polarization profile in the $\gamma$-ray band is nearly consistent with the shape of the polarization profile in the X-ray band. This is because, according to Equations (\ref{HYPL}) and (\ref{VSPA-PL}), the linear polarization of conventional and very-small-pitch-angle synchrotron radiation depends on $\nu / \nu_L$. In the $\gamma$-ray band, both the observational frequency and the magnetic field strength are increased by 3 orders of magnitude relative to the X-ray band. Therefore, the ratio $\nu / \nu_L$ is identical in the two bands, leading to the polarization profiles in the X-ray and $\gamma$-ray bands having the same shape.

We investigate the effects of different observational frequencies, magnetic field strengths, power-law indices, and electron temperatures on the very-small-pitch-angle synchrotron polarization in the $\gamma$-ray band as a function of $\gamma_{\rm th}$. 
The polarization degree related to the observational frequency is shown in the top left panel of Figure \ref{Gamma-ray-HY}. When $\nu = 0.2 \,\mathrm{MeV}$, the polarization degree remains constant at $52.2\%$ for $\gamma_{\rm th} \lesssim 1.0 \times 10^{4}$ and remains constant at $3.9\%$ for $\gamma_{\rm th} \gtrsim 1.0 \times 10^{4}$. When $\nu = 5.0 \,\mathrm{MeV}$, the polarization degree remains constant at $52.4\%$ for $\gamma_{\rm th} \lesssim 2.0 \times 10^{5}$ and approaches zero for $\gamma_{\rm th} \gtrsim 2.0 \times 10^{5}$. When $\nu = 10.0 \,\mathrm{MeV}$, the polarization degree remains constant at $52.4\%$ for $\gamma_{\rm th} \lesssim 5.0 \times 10^{5}$ and approaches zero for $\gamma_{\rm th} \gtrsim 5.0 \times 10^{5}$. Overall, the polarization profiles for different $\gamma$-ray bands all show a very sharp transition. In the $\gamma$-ray band, the observational frequency has a weak effect on the polarization degree in the low-$\gamma_{\rm th}$ range, whereas it has a strong influence on the polarization degree in the high-$\gamma_{\rm th}$ range.
The polarization degree related to the magnetic field is shown in the top right panel of Figure \ref{Gamma-ray-HY}. When $B = 1.0 \times 10^{9}\,\mathrm{G}$, the polarization degree remains constant at $52.2\%$ for $\gamma_{\rm th} \lesssim 2.2 \times 10^{5}$ and approaches zero for $\gamma_{\rm th} \gtrsim 2.2 \times 10^{5}$. When $B = 1.0 \times 10^{12}\,\mathrm{G}$, the polarization degree remains constant at $52.2\%$ for $\gamma_{\rm th} \lesssim 200 $ and remains constant at $70.3\%$ for $\gamma_{\rm th} \gtrsim 2.0 \times 10^{3}$. The polarization profile exhibits a transition in the range $200 \lesssim \gamma_{\rm th} \lesssim 2.0 \times 10^{3}$. When $B = 1.0 \times 10^{15}\,\mathrm{G}$, the polarization degree remains constant at $11.5\%$ for $\gamma_{\rm th} \lesssim 2.5 $ and approaches zero for $\gamma_{\rm th} \gtrsim 100$. The polarization profile exhibits a transition in the range $2.5 \lesssim \gamma_{\rm th} \lesssim 100$. Overall, the transition exhibits a very sharp change at lower magnetic field strengths. In the $\gamma$-ray band, the magnetic field has a significant influence on the polarization.
The polarization degree related to the power-law index is shown in the bottom left panel of Figure \ref{Gamma-ray-HY}. When $p = 1.5$, the polarization degree remains constant at $44.9\%$ for $\gamma_{\rm th} \lesssim 2.2 \times 10^{5}$ and approaches zero for $\gamma_{\rm th} \gtrsim 2.2 \times 10^{5}$. When $p = 2.5$, the polarization degree remains constant at $52.4\%$ for $\gamma_{\rm th} \lesssim 2.2 \times 10^{5}$ and approaches zero for $\gamma_{\rm th} \gtrsim 2.2 \times 10^{5}$. When $p = 3.5$, the polarization degree remains constant at $46.5\%$ for $\gamma_{\rm th} \lesssim 2.2 \times 10^{5}$ and approaches zero for $\gamma_{\rm th} \gtrsim 2.2 \times 10^{5}$. Overall, the polarization profiles for different power-law indices all show a very sharp transition. In the $\gamma$-ray band, the power-law index has a strong influence on the polarization degree in the low-$\gamma_{\rm th}$ range, whereas it does not affect the polarization degree in the high-$\gamma_{\rm th}$ range. 
The polarization degree related to the electron temperature is shown in the bottom right panel of Figure \ref{Gamma-ray-HY}. When $T_e \leq 1.0 \times 10^{12}\,\mathrm{K}$, the polarization degree remains constant at $52.4\%$ for $\gamma_{\rm th} \lesssim 2.2 \times 10^{5} $ and approaches zero for $\gamma_{\rm th} \gtrsim 2.2 \times 10^{5}$. When $T_e = 1.0 \times 10^{13}\,\mathrm{K}$, the polarization degree remains constant at $52.4\%$ for $\gamma_{\rm th} \lesssim 2.5 \times 10^{5}$ and remains constant at $1.5\%$ for $\gamma_{\rm th} \gtrsim 2.5 \times 10^{5}$. Overall, the polarization profiles for different electron temperatures all show a very sharp transition. In the $\gamma$-ray band, the electron temperature has a weak influence on the polarization degree in the high-$\gamma_{\rm th}$ range, whereas it does not affect the polarization degree in the low-$\gamma_{\rm th}$ range.

\subsubsection{Polarization of the Nonthermal and Thermal Electrons}

\begin{figure*}
\centering
\includegraphics[scale=0.45]{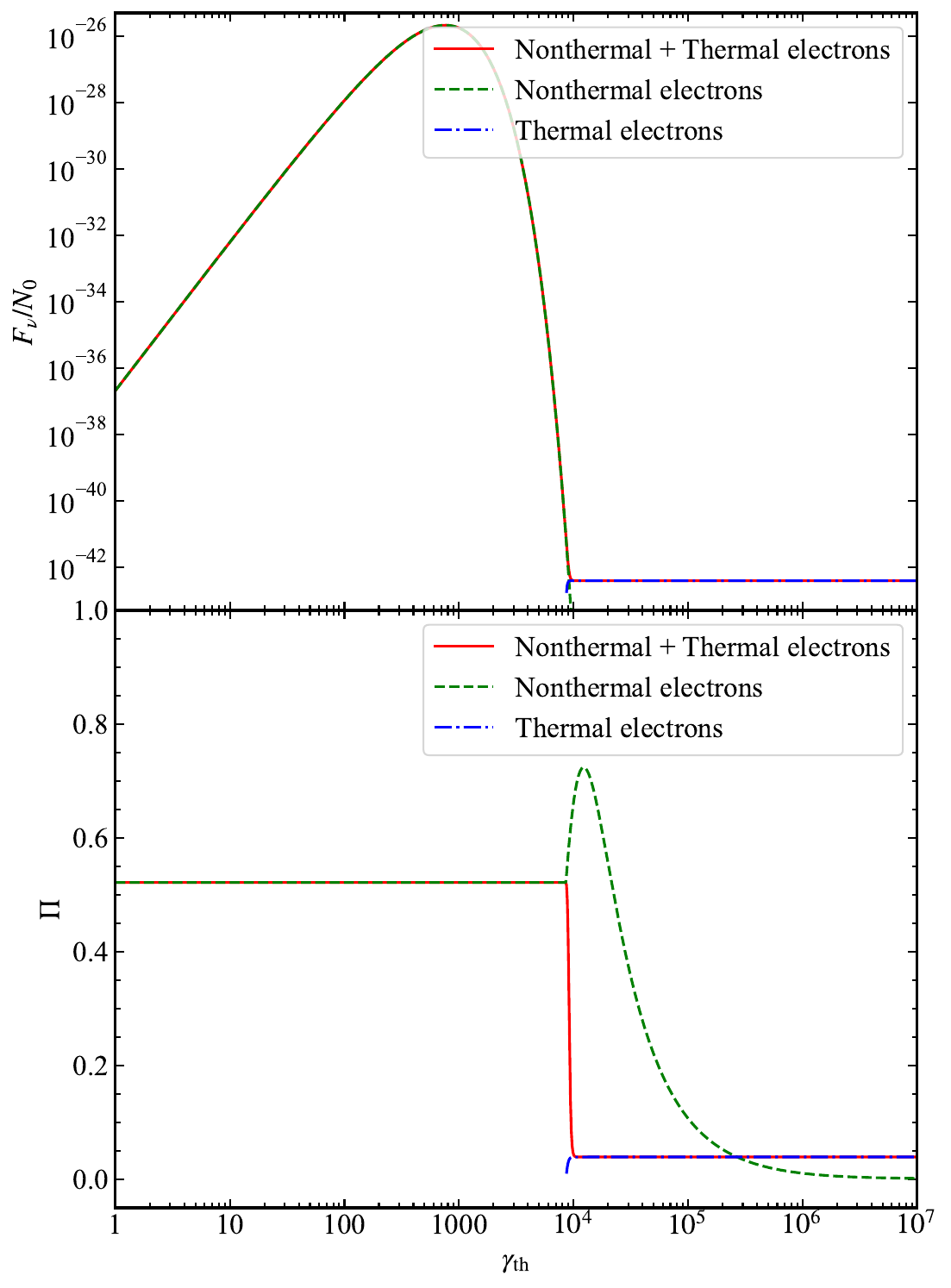}
\includegraphics[scale=0.45]{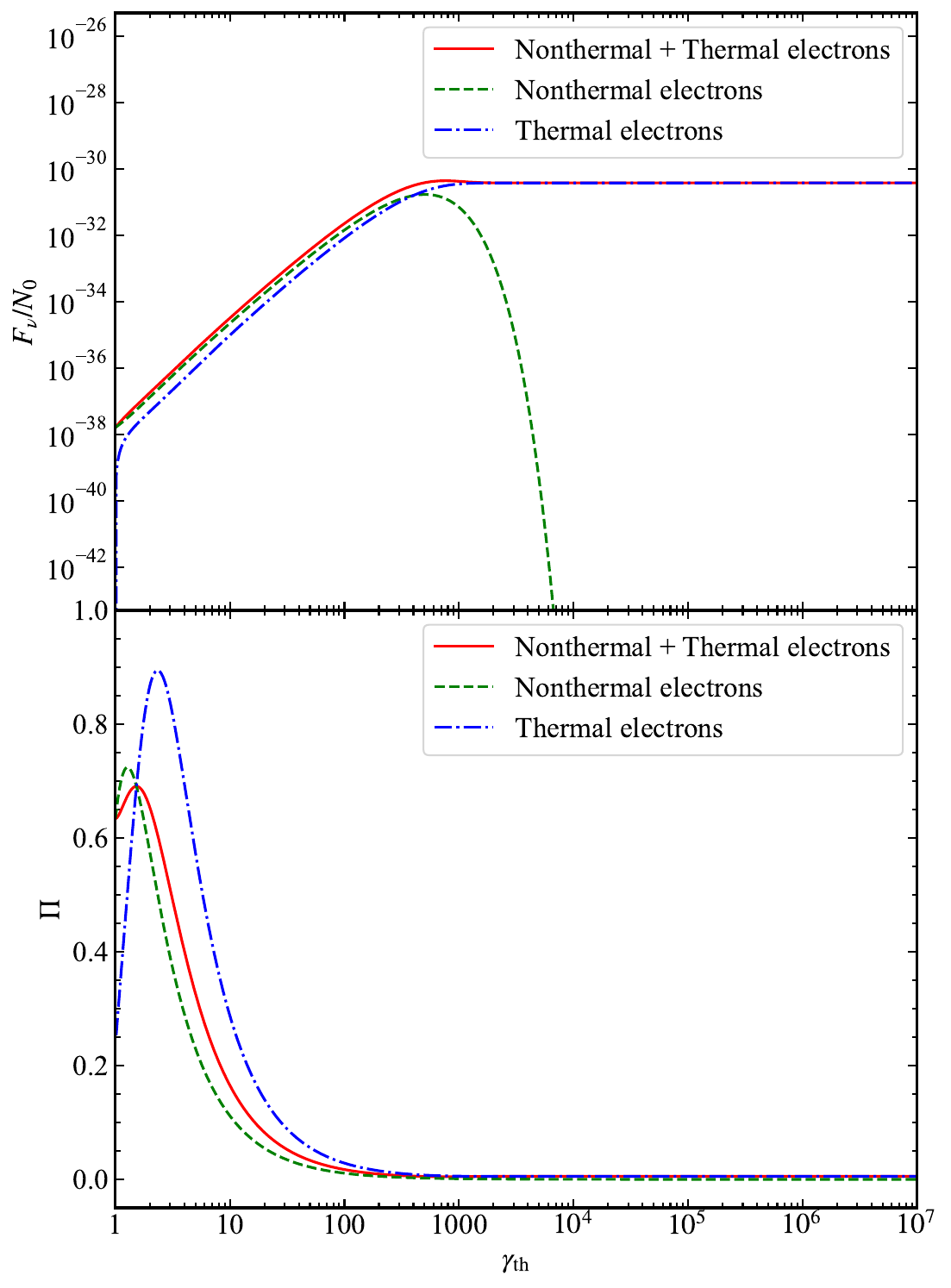}
\caption{The spectra and polarization degree of the very-small-pitch-angle synchrotron radiation for a hybrid electron energy distribution. The red, green, and blue lines represent the total, nonthermal electron, and thermal electron spectra and polarization, respectively. Left panel: we adopt $\nu = 0.2\, \mathrm{keV}$, $B = 1.0\times 10^6\, \mathrm{G}$, $p=2.5$, $T_e = 1.0 \times 10^{12}\, \mathrm{K}$, and $\tilde{\alpha}=10^{-3}$. Right panel: we adopt $\nu = 5.0\, \mathrm{GHz}$, $B = 1.0\times 10^3 \, \mathrm{G}$, $p=2.5$, $T_e = 1.0 \times 10^{12}\, \mathrm{K}$, and $\tilde{\alpha}=10^{-3}$.
\label{vspa-contributions}}
\end{figure*}

According to Equation (\ref{hydis}), $\gamma_{\rm th}$ separates the electron energy distribution into nonthermal and thermal components. When $\gamma_{\rm th} = 1$, the electron distribution is entirely nonthermal, while the fraction of thermal electrons in the distribution increases with increasing $\gamma_{\rm th}$. The polarization can be divided into the nonthermal–electron–dominated phase, the nonthermal–thermal electron conjunctive phase, and the thermal–electron–dominated phase, according to the value of $\gamma_{\rm th}$. As $\gamma_{\rm th}$ increases, the polarization transitions from the nonthermal–electron–dominated phase to the thermal–electron–dominated phase.  We set $\gamma_{\rm th}$ as the upper integration limit for thermal electrons and the lower integration limit for nonthermal electrons to obtain the spectra and polarization of very-small-pitch-angle synchrotron radiation from thermal and nonthermal electrons (the details of the synchrotron spectra can be found in Appendix \ref{sec:power}). In order to further explore the contributions of nonthermal and thermal electrons to the total spectra and polarization of very-small-pitch-angle synchrotron radiation, we consider two examples. We adopt the parameters $\nu = 0.2\,\mathrm{keV}$, $B = 1.0 \times 10^{6}\,\mathrm{G}$, $p = 2.5$, and $T_e = 1.0 \times 10^{12}\,\mathrm{K}$ in the first example. For the second example, we adopt parameters $\nu = 5.0\,\mathrm{GHz}$, $B = 1.0 \times 10^{3}\,\mathrm{G}$, $p = 2.5$, and $T_e = 1.0 \times 10^{12}\,\mathrm{K}$. 
The first example is shown in the left panel of Figure \ref{vspa-contributions}. We can see that in the range of small $\gamma_{\rm th}$, where the radiation is dominated by nonthermal electrons, the thermal electrons do not produce radiation. This is because, for the adopted parameters, thermal electrons can produce very-small-pitch-angle synchrotron radiation only when $\gamma_{\rm th} > \max(\nu_L/\nu,\ \nu/2\nu_L) \approx 8.6 \times 10^{3}$. In the range of large $\gamma_{\rm th}$, where the radiation is dominated by thermal electrons, the polarization degree of the nonthermal electrons can exceed that of the thermal electrons. However, because the radiation of nonthermal electrons is much weaker, the total polarization degree is dominated by the thermal electrons. 
The second example is shown in the right panel of Figure \ref{vspa-contributions}. For the adopted parameters, the condition for thermal electrons to produce very-small-pitch-angle synchrotron radiation is $\gamma_{\rm th} > \max(\nu_L/\nu,\ \nu/2\nu_L) \approx 0.89$. Therefore, in the range of small $\gamma_{\rm th}$, thermal electrons can contribute to the total radiation. In the range of small $\gamma_{\rm th}$, the radiation from nonthermal electrons exceeds that from thermal electrons, although the polarization degree of the thermal electrons can be higher than that of the nonthermal electrons. In the range of large $\gamma_{\rm th}$, where the radiation is dominated by thermal electrons, the polarization degrees of both thermal and nonthermal electrons are zero. From the two examples mentioned above, we see the complicated diversities of synchrotron polarization due to different electron energy distributions.

\subsubsection{Effects of Magnetic Field, Power-law Index, and Electron Temperature on the Polarization Degree}

\begin{figure*}
\centering
\includegraphics[scale=0.33]{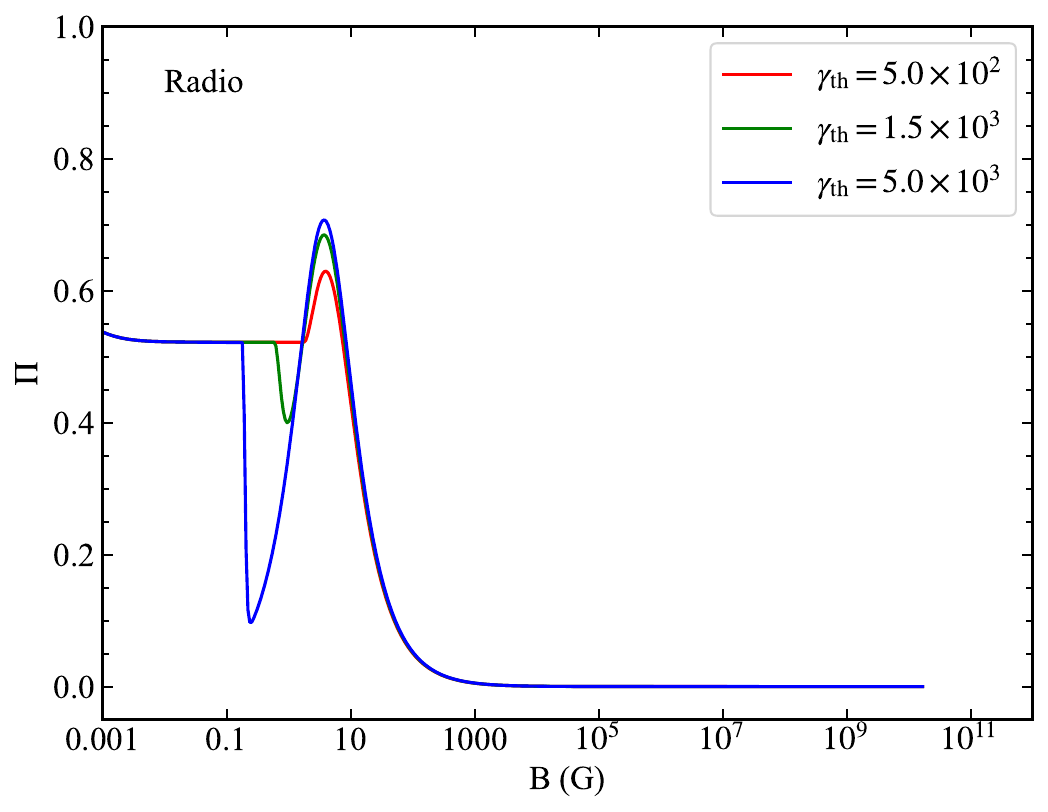}
\includegraphics[scale=0.33]{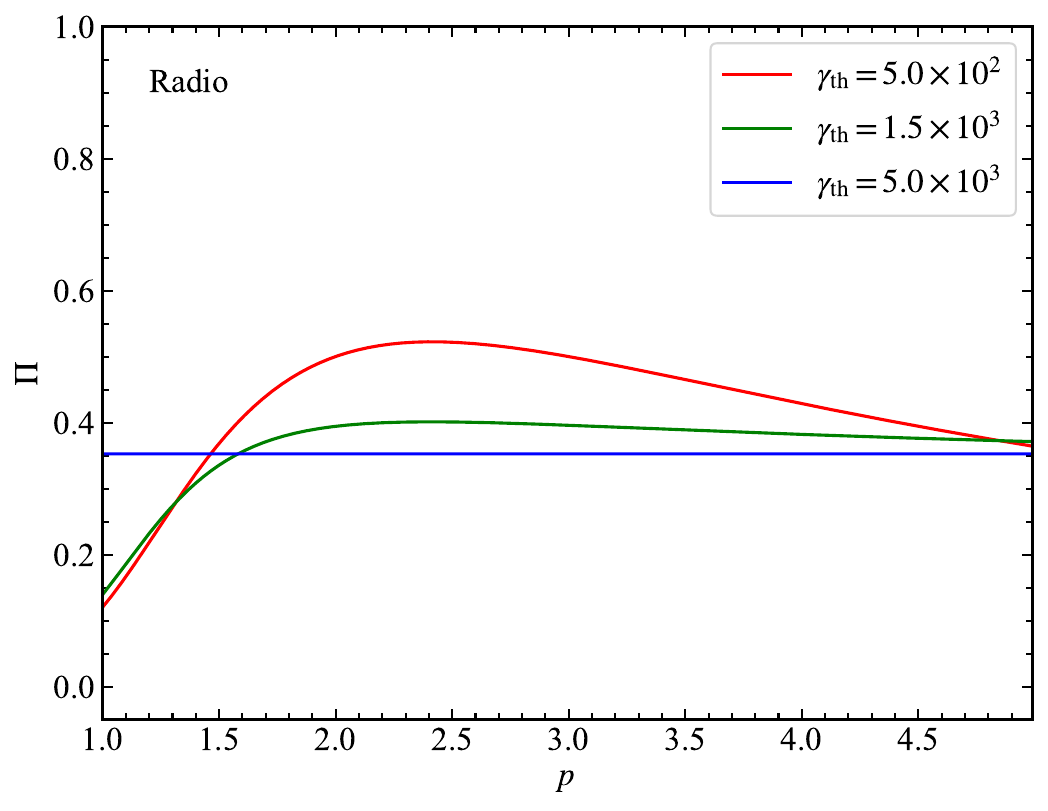}
\includegraphics[scale=0.33]{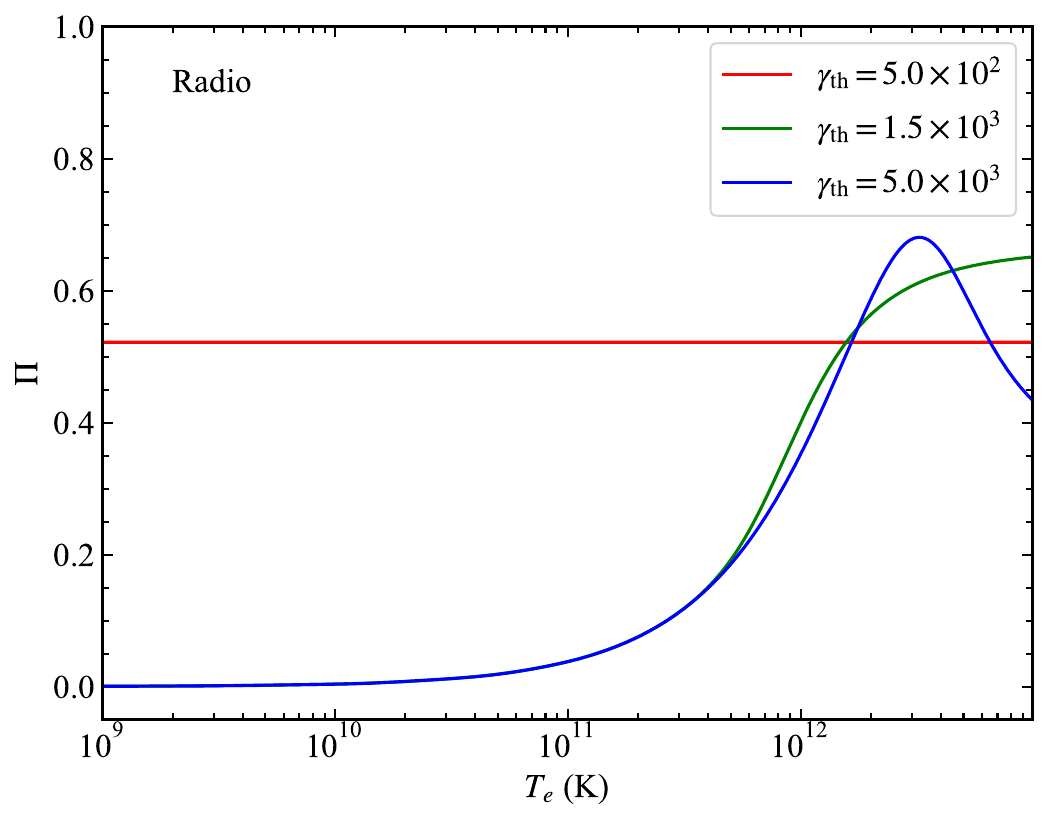}
\includegraphics[scale=0.33]{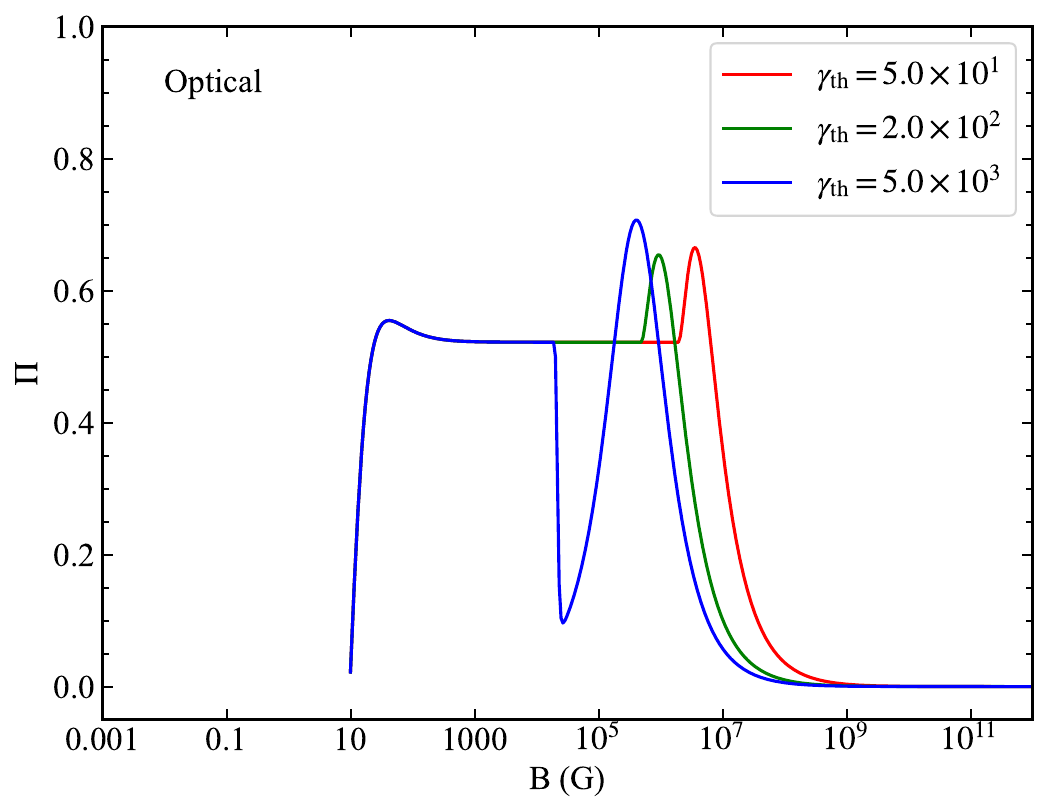}
\includegraphics[scale=0.33]{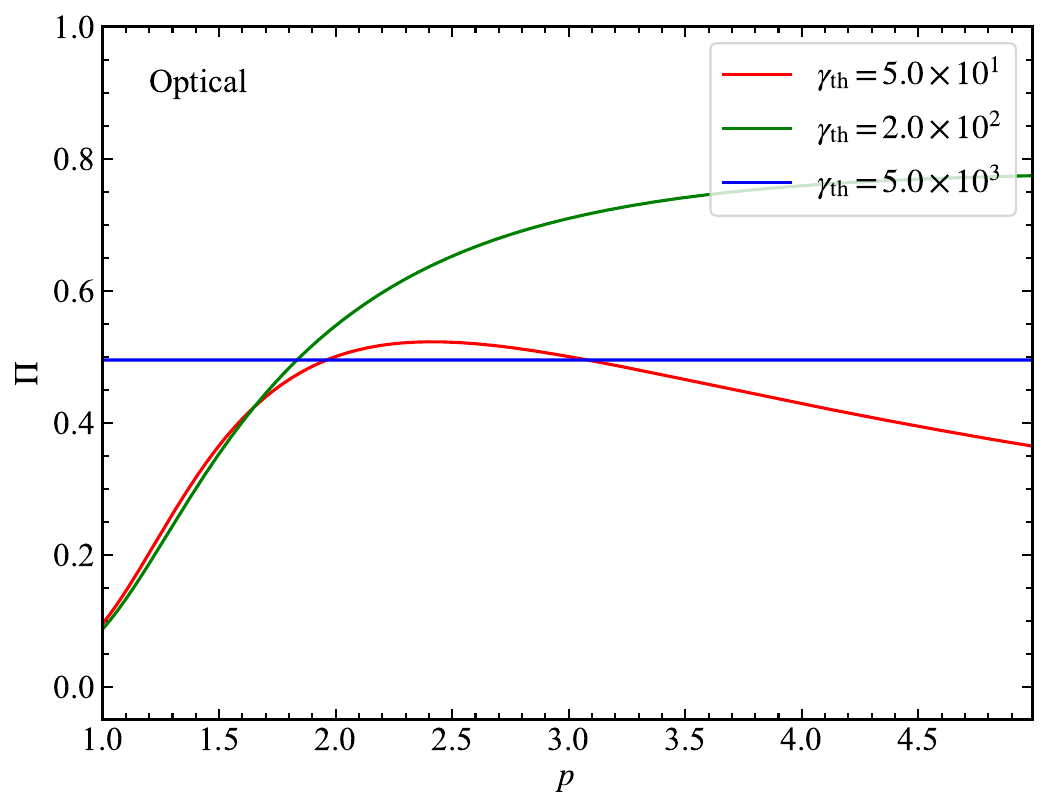}
\includegraphics[scale=0.33]{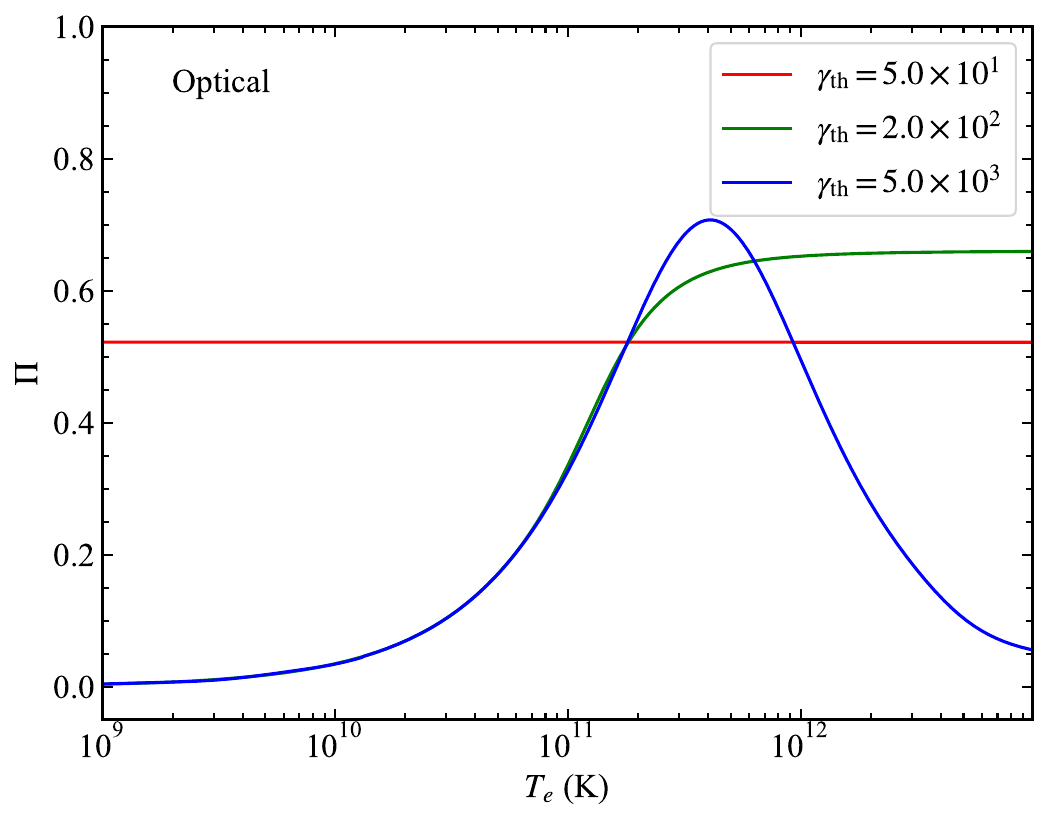}
\includegraphics[scale=0.33]{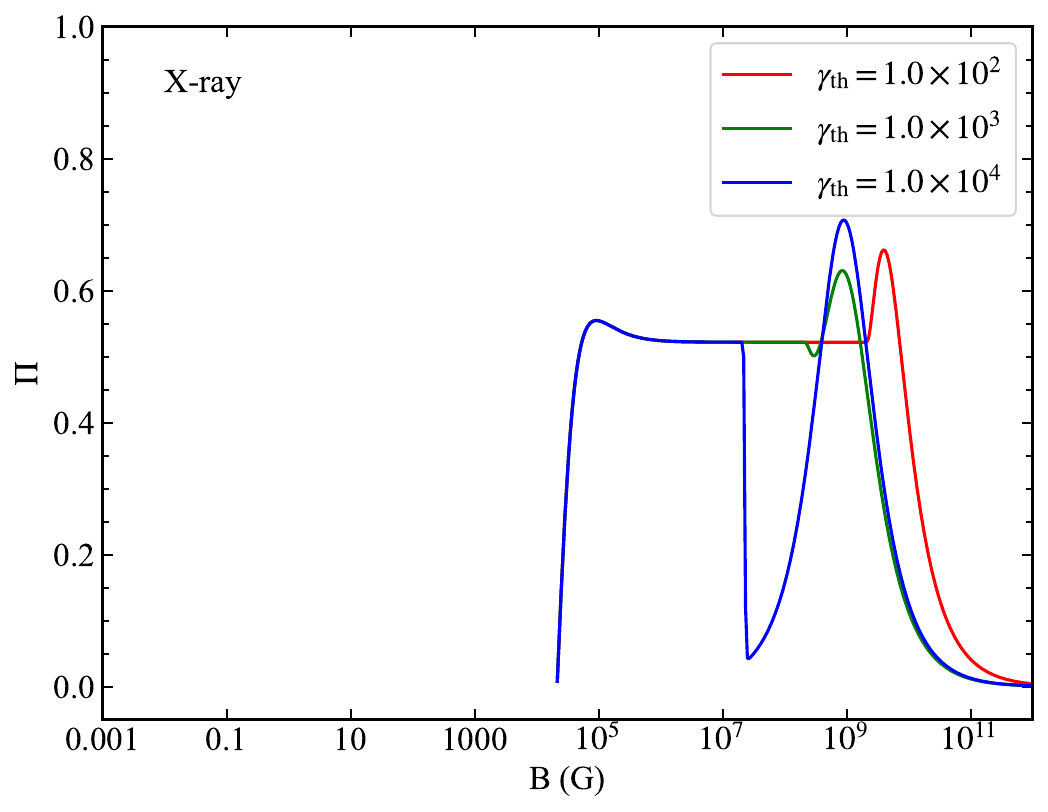}
\includegraphics[scale=0.33]{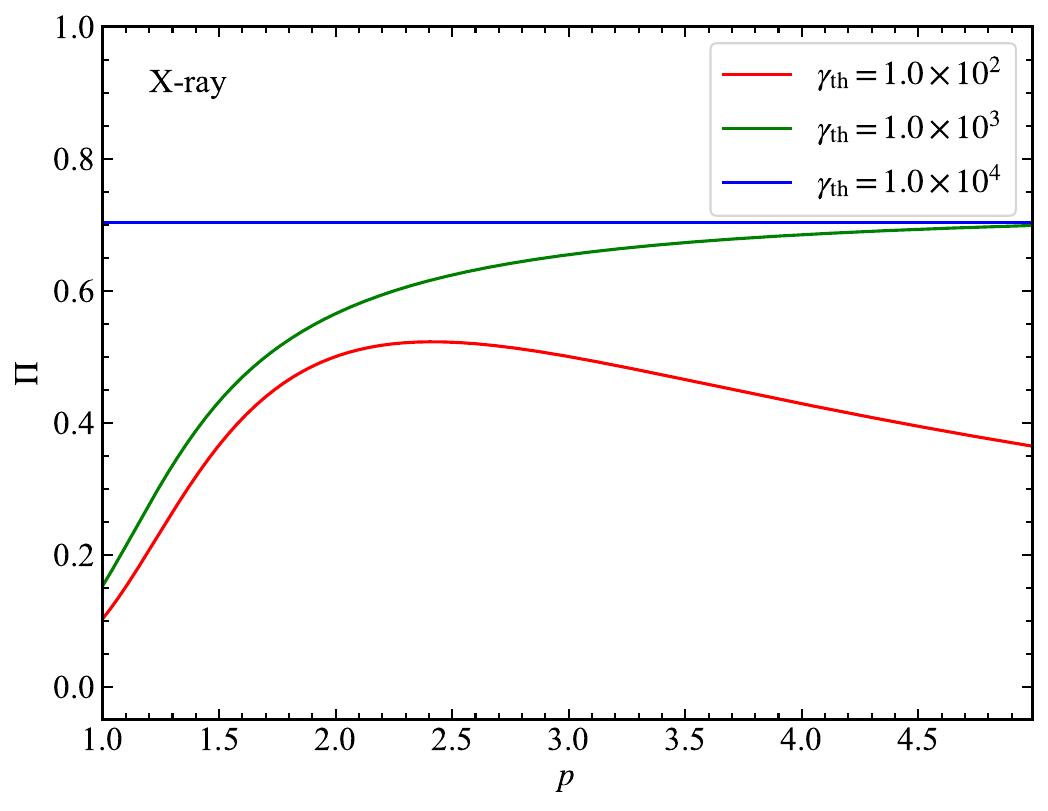}
\includegraphics[scale=0.33]{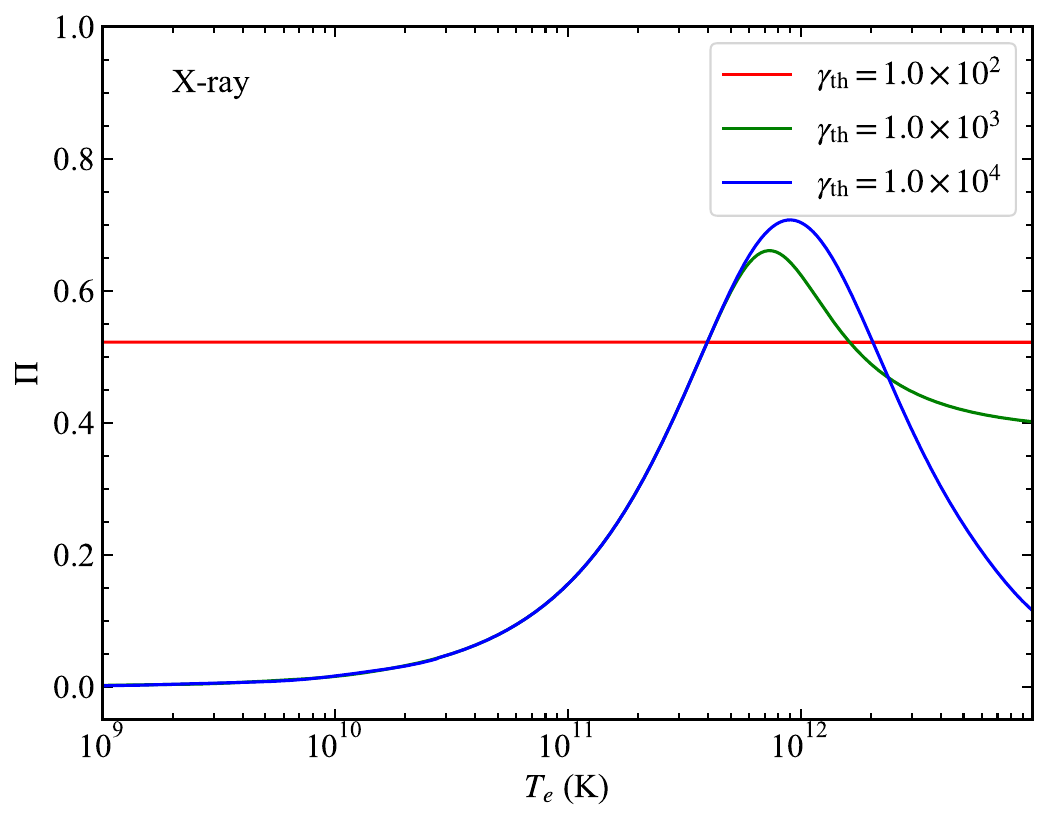}
\includegraphics[scale=0.33]{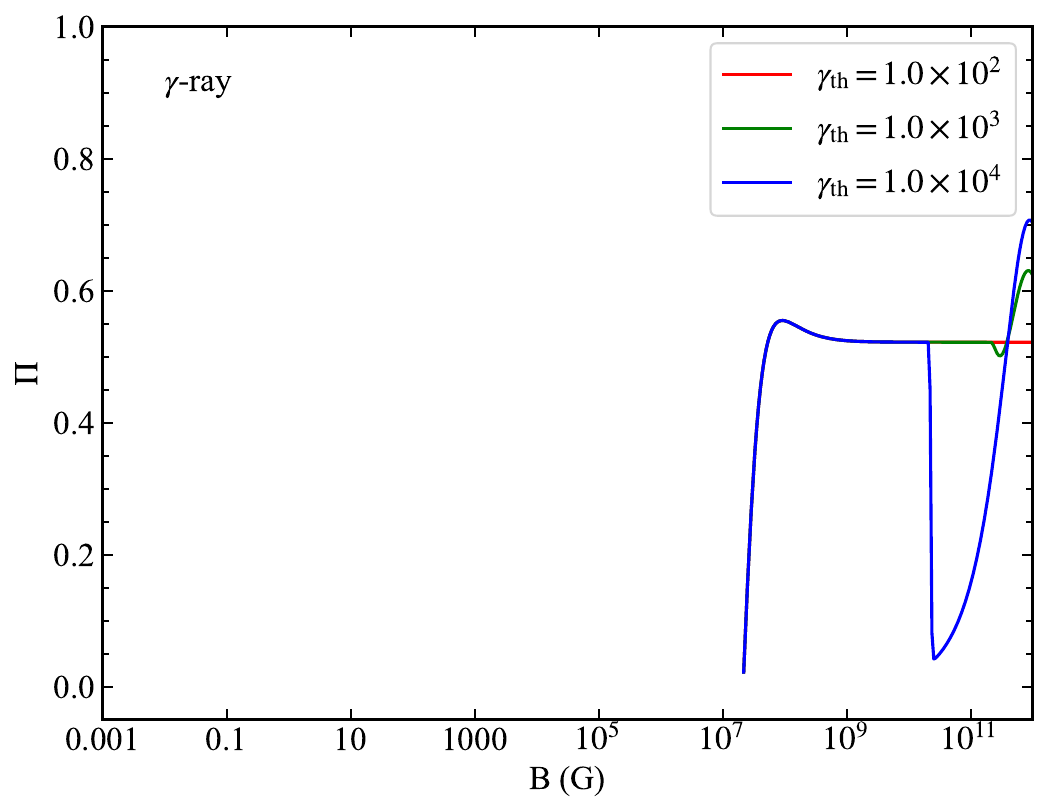}
\includegraphics[scale=0.33]{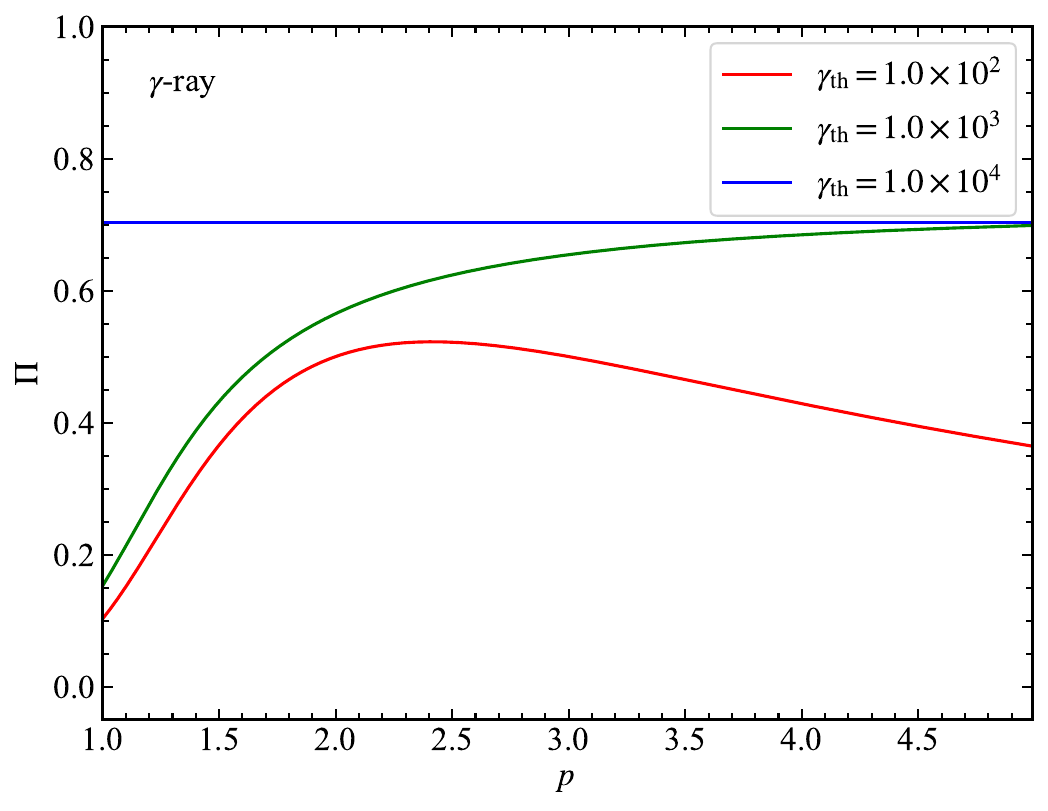}
\includegraphics[scale=0.33]{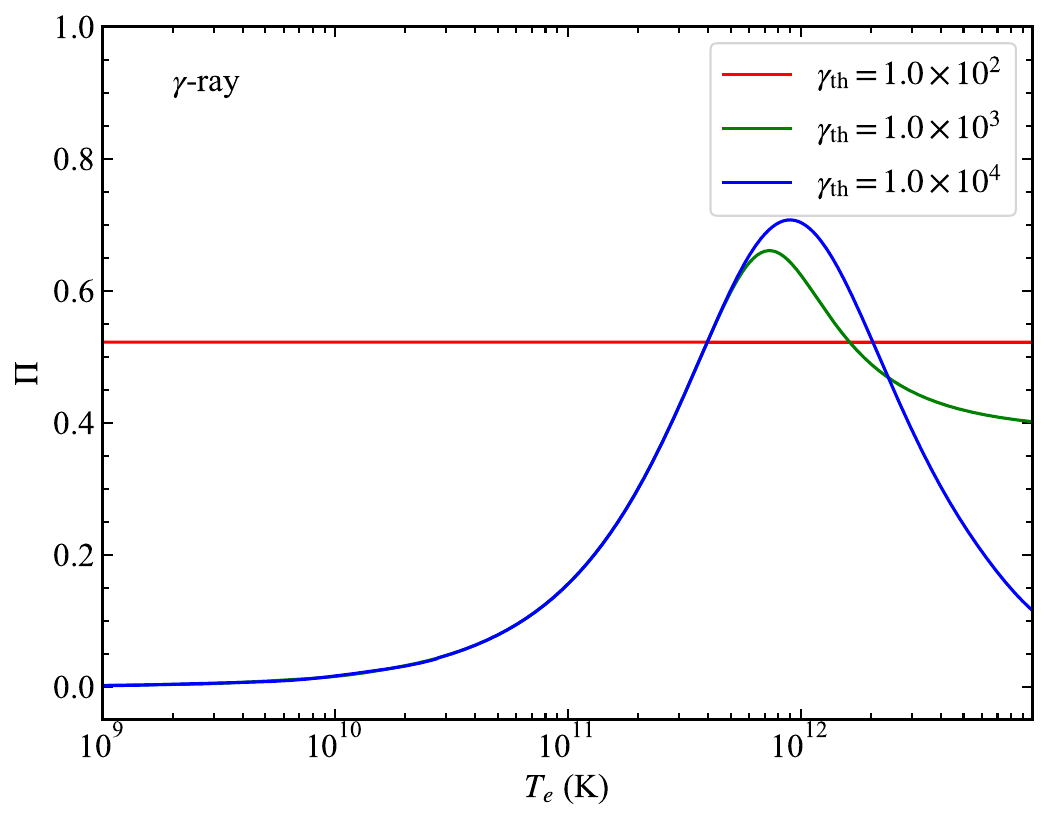}
\caption{Polarization degree of the very-small-pitch-angle synchrotron radiation for a hybrid electron energy distribution as a function of the magnetic field strength, power-law index, and electron temperature for the radio (5 GHz), optical (5500 $\mathrm{\AA}$), X-ray (5 keV), and $\gamma$-ray (5 MeV) bands. The values of $\gamma_{\rm th}$ corresponding to the red, green, and blue curves represent the nonthermal-electron-dominated phase, the nonthermal-thermal electron conjunctive phase, and the thermal-electron-dominated phase, respectively. Left panels: we adopt $p=2.5$, $T_e=1.0\times10^{12}\,\mathrm{K}$. 
Middle panels: we adopt $T_e=1.0\times10^{12}\,\mathrm{K}$ with $B=1.0\,\mathrm{G}$ (radio), $1.0 \times 10^6\,\mathrm{G}$ (optical), $1.0 \times 10^9\,\mathrm{G}$ (X-ray), or $1.0 \times 10^{12}\,\mathrm{G}$ ($\gamma$-ray). Right panels: we adopt $p=2.5$ and vary the magnetic field strength in the same manner as in the middle panels.
\label{VSPA-B-p-Te}}
\end{figure*}

Figure \ref{VSPA-B-p-Te} shows the polarization degree of very-small-pitch-angle synchrotron radiation for a hybrid energy distribution as a function of magnetic field strength, power-law index, and electron temperature across different frequency bands. 
In Figure \ref{VSPA-B-p-Te}, the different values of $\gamma_{\rm th}$ corresponding to the red, green, and blue lines represent the nonthermal-electron-dominated phase, nonthermal-thermal electron conjunctive phase, and thermal-electron-dominated phase, respectively. 
The polarization degree related to the magnetic field is shown in the left panels of Figure \ref{VSPA-B-p-Te}. The polarization profiles for different bands all show a significant peak structure, which is primarily attributed to thermal electron contributions. At low magnetic field strengths, the polarization degree is dominated by nonthermal electrons. Therefore, the overall polarization degree does not change with the magnetic field strength. Moreover, the polarization degree is the same for different values of $\gamma_{\rm th}$. At high magnetic field strengths, the polarization corresponding to smaller $\gamma_{\rm th}$ transitions to the thermal-electron-dominated phase at higher magnetic field strengths. In addition, from the radio to the $\gamma$-ray bands, the polarization profile shifts toward higher magnetic field strengths. 
The polarization degree related to the power-law index is shown in the middle panels of Figure \ref{VSPA-B-p-Te}. The power-law index does not affect the polarization degree for all frequency bands in the thermal-electron-dominated phase. When the power-law index increases, the polarization degree increases in the nonthermal-thermal electron conjunctive phases, while in the nonthermal-electron-dominated phase it first increases and then decreases. Across all frequency bands, the polarization shows the same dependence on the power-law index in the nonthermal–electron–dominated phase, but shows significant differences in the other phases. 
The polarization degree related to the electron temperature is shown in the right panels of Figure \ref{VSPA-B-p-Te}. The electron temperature does not affect the polarization degree for all frequency bands in the nonthermal-electron-dominated phase. In the thermal-electron-dominated phase, the polarization for all bands shows similar dependence on the electron temperature. Moreover, the polarization profile exhibits a peak and the polarization degree approaches to zero at low electron temperatures. In the nonthermal–thermal electron conjunctive phase, the polarization degree in the radio and optical bands increases with increasing temperature, whereas in the X-ray and $\gamma$-ray bands, the polarization degree first increases with increasing temperature and then decreases. 

From the detailed investigations mentioned in this subsection, we see that polarization of very-small-pitch-angle synchrotron radiation for a hybrid electron energy distribution shows diverse behaviors across different physical parameters.

\section{Discussion}\label{sec:4}

For an energy distribution with an injected power law, synchrotron energy losses of the relativistic electrons increase the spectral index, causing the electron energy distribution to evolve into a broken power law \citep{Kardashev1962}. The very-small-pitch-angle synchrotron cooling timescale can be estimated as $P_{\rm VSPA} t_{\mathrm{cool}}\sim\gamma m_e c^2$, where $P_{\rm VSPA}$ is the radiation power of the very-small-pitch-angle synchrotron radiation obtained by integrating $dP_{\rm VSPA}(\nu)/d\nu$ over all frequencies:

\begin{equation}
    P_{\rm VSPA} \simeq \frac{2e^{4}B^{2}\gamma^{2}\alpha^{2}}{3m_e^2c^3}.
\end{equation}

The cooling timescale is given by

\begin{equation}
    t_{\mathrm{cool}}\sim\frac{3m_e^3c^5}{2e^4\gamma B^2\alpha^2}.
\end{equation}
\noindent Since $\alpha\ll1$, synchrotron cooling is strongly suppressed by the factor $\alpha^2$, even in strong magnetic fields. This suppression substantially increases the cooling timescale and shifts the corresponding cooling break toward higher electron energies. Therefore, synchrotron cooling is expected to have a much weaker influence on the electron energy distribution in the very-small-pitch-angle synchrotron radiation than in the conventional synchrotron radiation.

Curvature radiation refers to the radiation produced by relativistic electrons moving along curved magnetic field lines in strong magnetic fields. Very-small-pitch-angle synchrotron radiation likewise involves the motion primarily along the magnetic field lines, although the electrons still possess a small velocity component perpendicular to the field lines. Very-small-pitch-angle synchrotron radiation is similar to the curvature radiation, but the two radiation mechanisms apply to different magnetic field environments. Curvature radiation is typically applicable to astrophysical objects with strong magnetic fields, whereas very-small-pitch-angle synchrotron radiation can be applied not only in strong magnetic field objects but also in weak magnetic field regimes. 

Coherent curvature radiation is widely regarded as the most likely mechanism for pulsars and fast radio bursts \citep{Gangadhara2010,Wang2012,Lu2018}. In coherent curvature radiation, the charge bunches must have a size smaller than the emission wavelength, i.e., the antenna mechanism \citep{Ruderman1975}. In contrast, the very-small-pitch-angle synchrotron radiation discussed in this work assumes that the electron emissions are independent, with random phases between different electrons, representing a typical incoherent radiation process. However, the establishment of a very small pitch angle must occur at first instance, implying that electron acceleration should maintain ordered motion along the magnetic field lines. This suggests that such a configuration is more likely to arise from coherent electrodynamic acceleration mechanisms \citep{Baring2004}. In the present work, we consider only incoherent very-small-pitch-angle synchrotron radiation. The polarization properties of coherent very-small-pitch-angle synchrotron radiation will be investigated in future studies.

\section{Conclusions}\label{sec:5}

In this work, we calculate the linear polarization degree of very-small-pitch-angle and conventional synchrotron radiation 
from power-law, Maxwellian, and hybrid electron energy distributions. It should be noted that the linear polarization degree derived in this work is intrinsic. We do not consider the effects of polarized radiative transfer, depolarization, or Faraday rotation. The linear polarization degree of very-small-pitch-angle synchrotron radiation is generally lower than that of conventional synchrotron radiation, with significantly different polarization behaviors. Regarding the linear polarization properties of very-small-pitch-angle synchrotron radiation, we obtain the following conclusions:
\begin{enumerate}
\item Within a finite range of electron Lorentz factors, the allowed parameter range for very-small-pitch-angle synchrotron radiation is constrained by $\nu_{\rm L}/\gamma \ll \nu \ll 2\gamma \nu_{\rm L}$. The allowed parameter range is strongly correlated with the magnetic field strength.

\item For very-small-pitch-angle synchrotron radiation with a power-law electron distribution, the polarization degree typically depends only on the power-law index and is generally lower than the usual polarization behavior $(3p+3)/(3p+7)$. For a Maxwellian electron energy distribution, the polarization profile exhibits a single Gaussian-like peak structure, with the peak of the polarization degree of about $70.7\%$ occurring at $\nu \approx 4.8\,\nu_L T_e / (10^{10}\, \rm K)$. Therefore, very-small-pitch-angle synchrotron radiation from thermal electrons produces significant linear polarization only within a relatively narrow frequency range.

\item For very-small-pitch-angle synchrotron radiation with a hybrid electron energy distribution, the conjunctive Lorentz factor $\gamma_{\rm th}$ separates the electrons into nonthermal and thermal components. As $\gamma_{\rm th}$ increases, the radiation gradually transitions from nonthermal electron domination to thermal electron domination. When the radiation is dominated by nonthermal electrons, we can obtain a significant polarization degree. In contrast, significant polarization can be produced only within a narrow parameter range when thermal electrons dominate the radiation. Overall, the polarization properties exhibit diverse behaviors depending on different physical parameters.

\end{enumerate}

\begin{acknowledgments}

We thank the anonymous referee  for the careful reading of our manuscript and for the constructive comments.  This work is supported by the Natural Science Foundation of China 12393813, the National Key R\&D Program of China (2023YFE0101200),  and the Yunnan Revitalization Talent Support Program (YunLing Scholar Project).

\end{acknowledgments}

\restartappendixnumbering
\appendix
\section{The spectra of very-small-pitch-angle synchrotron radiation} \label{sec:power}

\begin{figure*}
\centering
\includegraphics[scale=0.325]{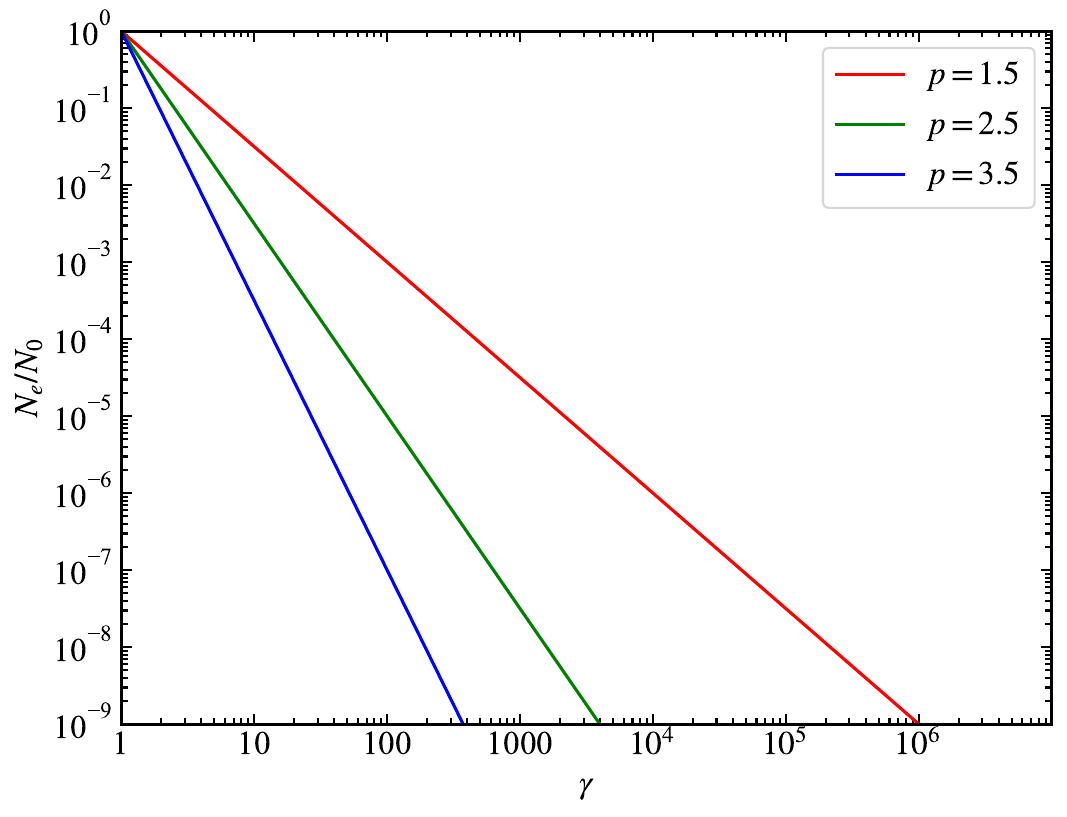}
\includegraphics[scale=0.325]{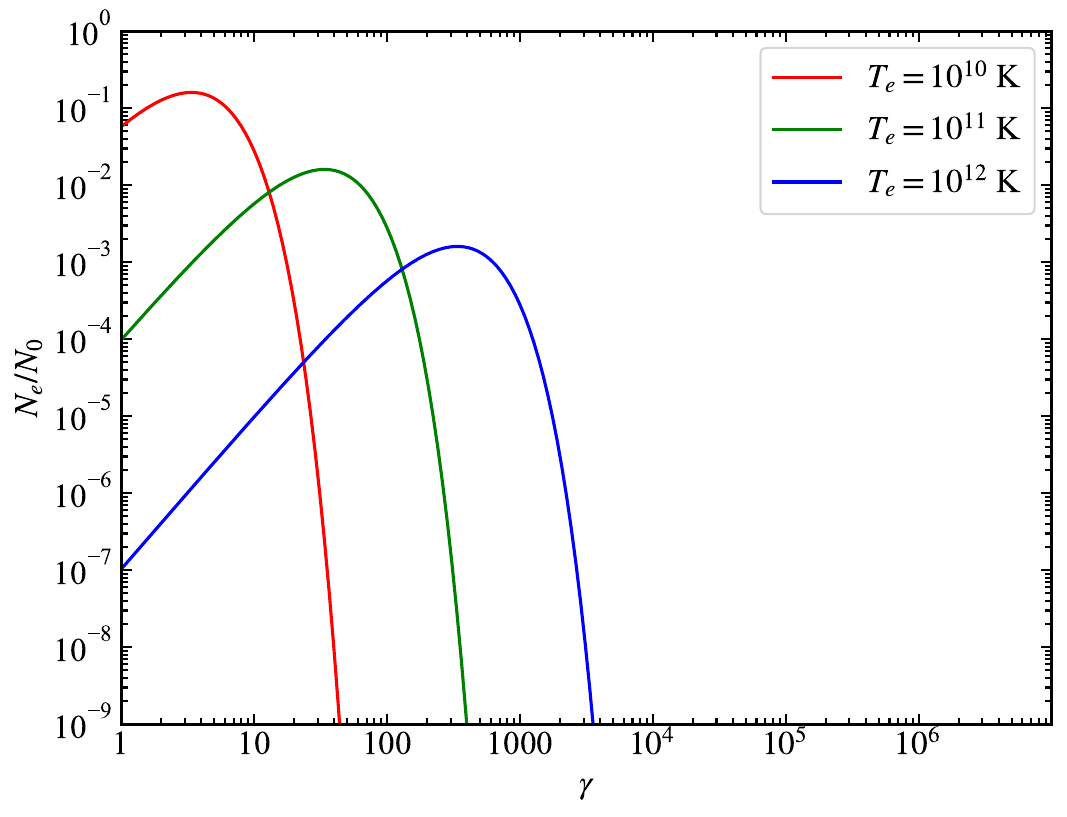}
\includegraphics[scale=0.325]{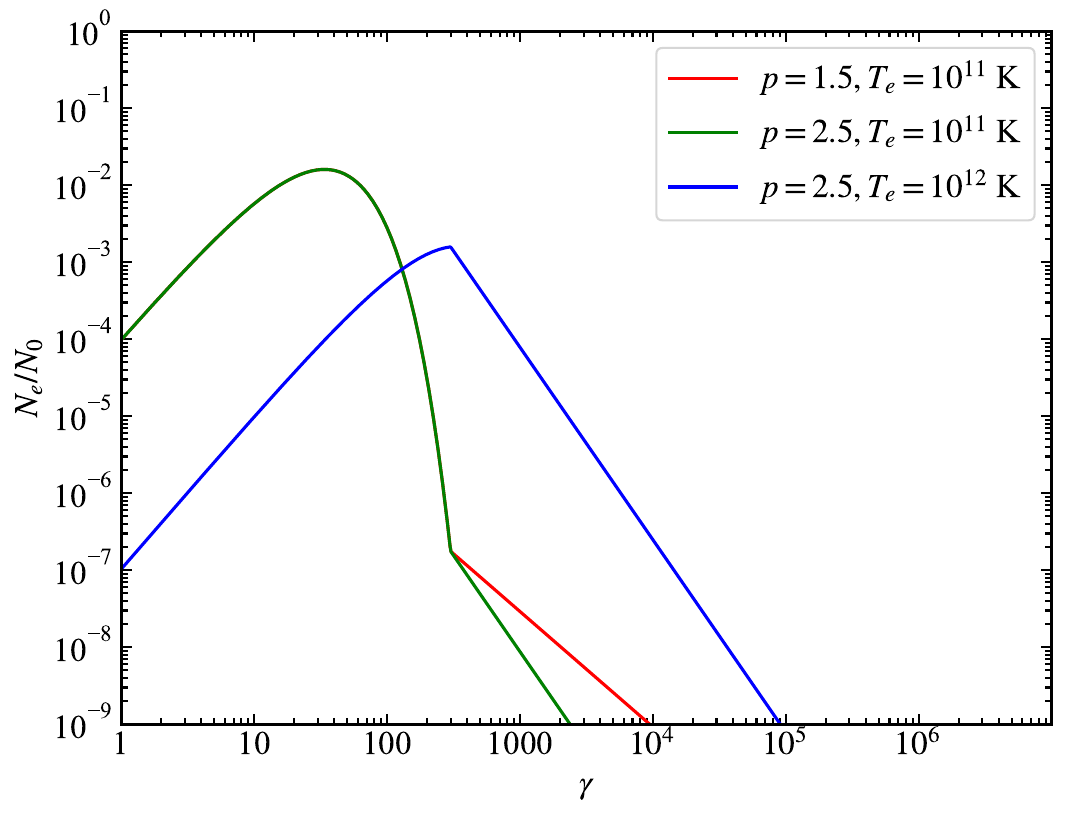}

\includegraphics[scale=0.325]{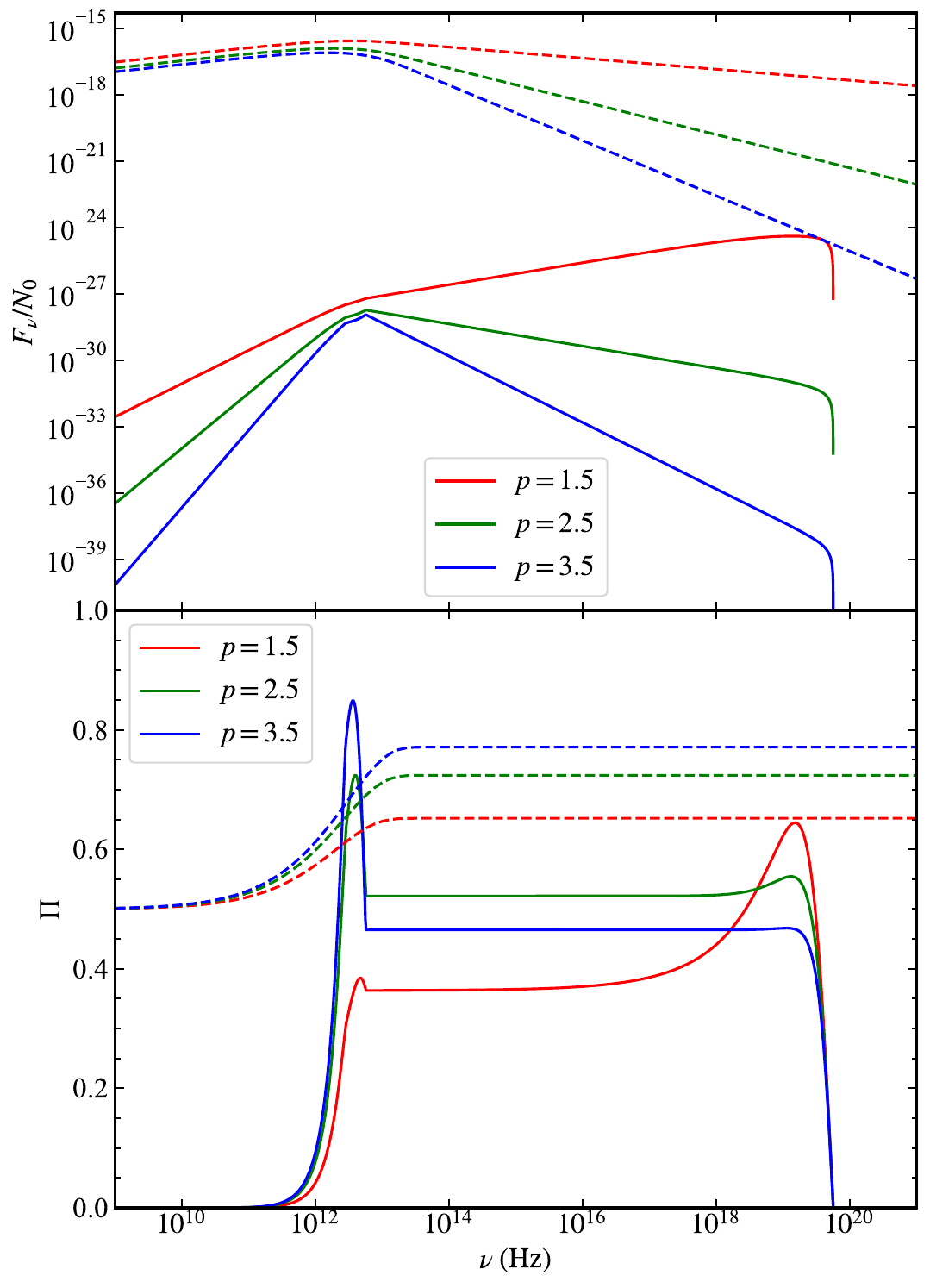}
\includegraphics[scale=0.325]{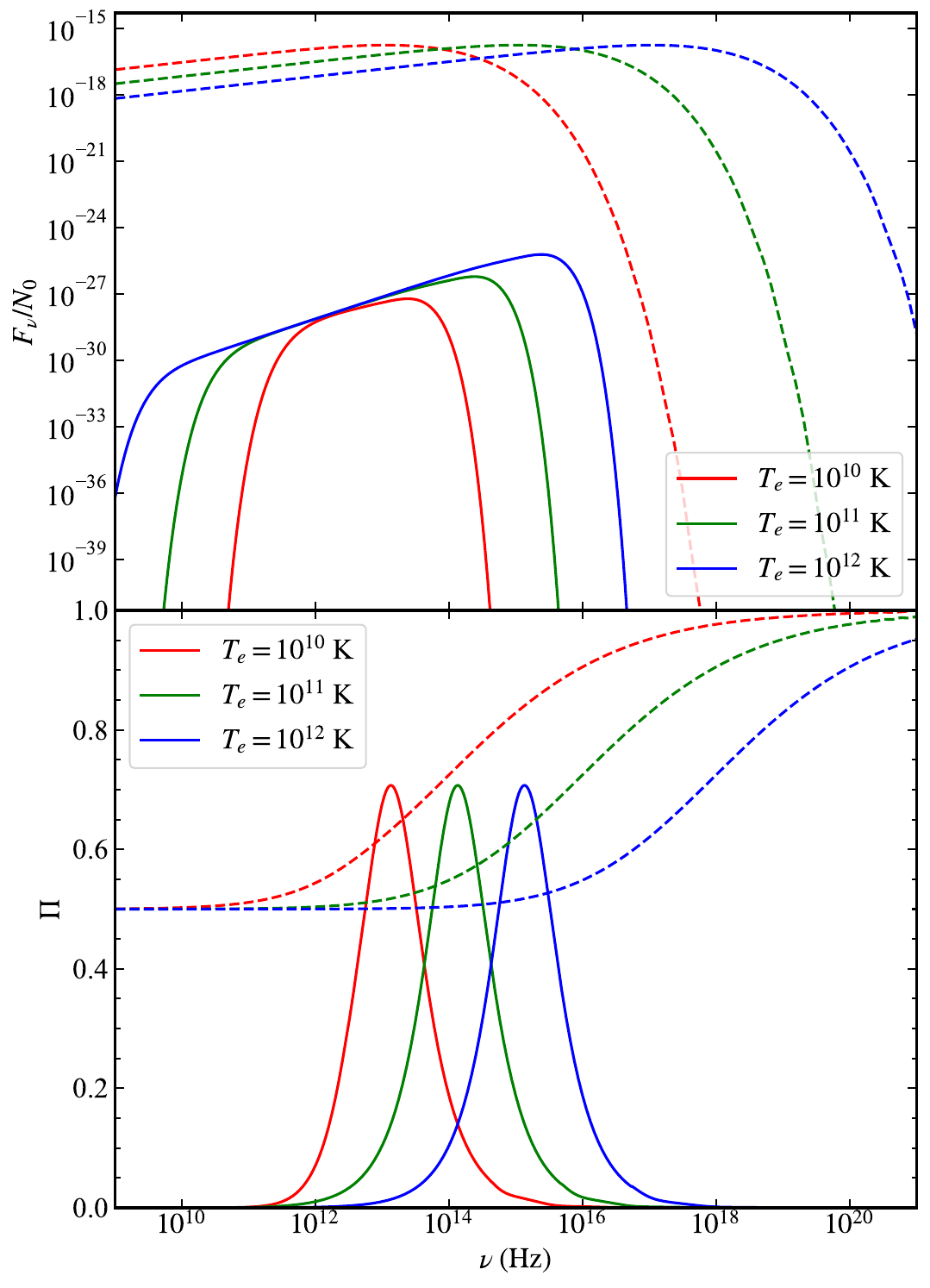}
\includegraphics[scale=0.325]{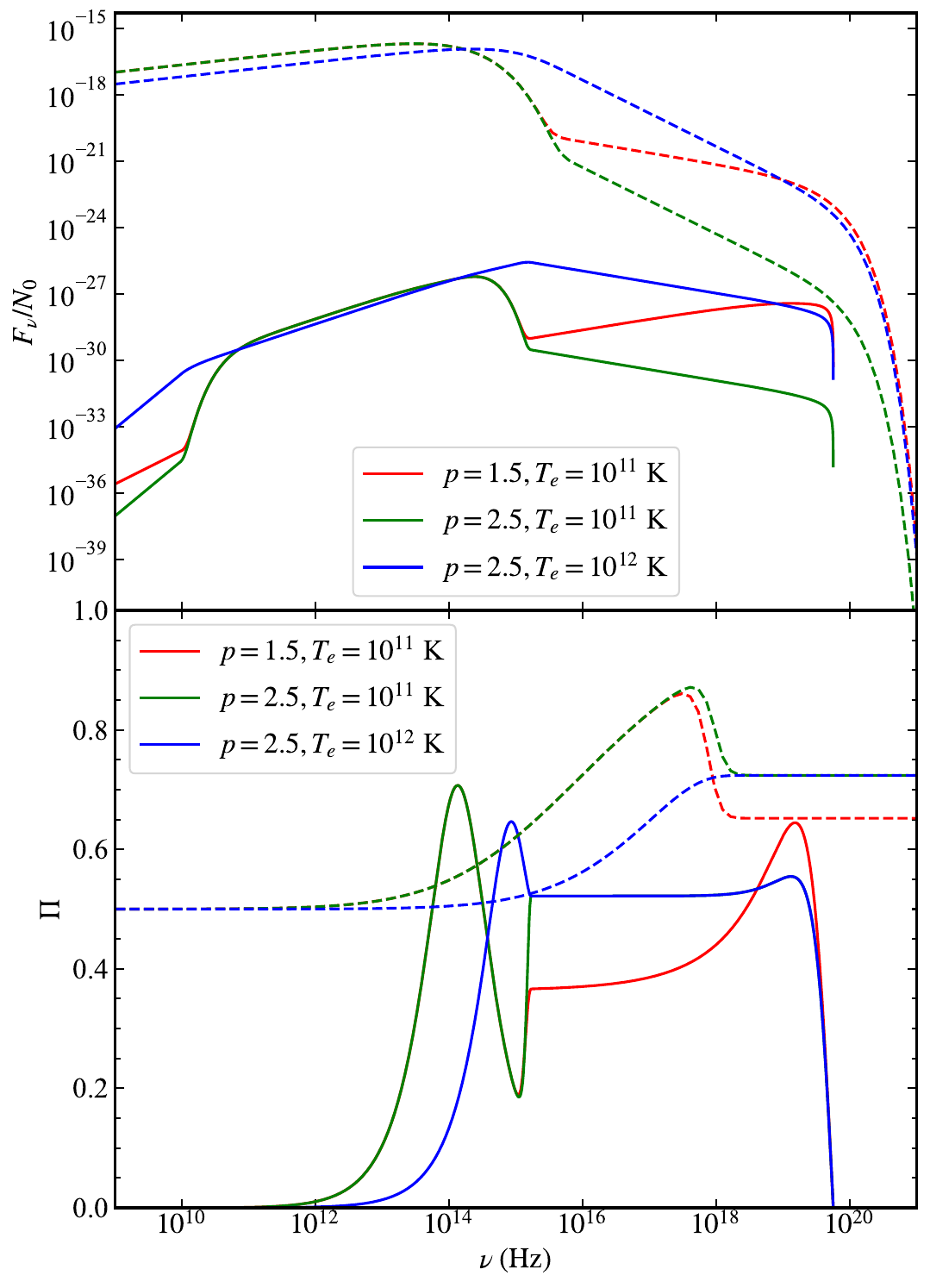}
\caption{The three columns show the very-small-pitch-angle and conventional synchrotron spectra and polarization degree of power-law, Maxwellian, and hybrid electron energy distributions. The solid and dashed lines represent the very-small-pitch-angle and conventional synchrotron radiation, respectively. The first row shows the electron energy distributions. We adopt $\gamma_{\rm th} = 300$ for the hybrid electron energy distribution. The second row shows the corresponding synchrotron spectra. We adopt $\tilde{\alpha} = 10^{-3}$ for the very-small-pitch-angle synchrotron spectra. The third row shows the linear polarization degree as a function of frequency. We adopt $B = 1.0 \times 10^{6}\, \mathrm{G}$ in the calculations of the spectra and linear polarization degree.
\label{CONV_VSPA_Spectrum}}
\end{figure*}

In this appendix, we illustrate the features of the spectra of the very-small-pitch-angle and conventional synchrotron radiation. By integrating Equations (\ref{power}) and (\ref{VSPA-PS}) over different electron energy distributions, we obtain the total power spectra of very-small-pitch-angle and conventional synchrotron radiation. Figure \ref{CONV_VSPA_Spectrum} shows the electron energy distribution, specific flux, and linear polarization degree. We adopt $\tilde{\alpha} = 10^{-3}$ in the calculations of the very-small-pitch-angle synchrotron radiation spectra. Meanwhile, in the calculations for the hybrid distribution, we adopt a conjunctive Lorentz factor $\gamma_{\rm th} = 300$.

We focus on the very-small-pitch-angle synchrotron radiation for power-law, Maxwellian, and hybrid electron energy distributions. The first row of Figure \ref{CONV_VSPA_Spectrum} shows the power-law, Maxwellian, and hybrid electron energy distributions. For the hybrid distribution, $\gamma_{\rm th}$ separates the electrons into thermal and nonthermal components. When $\gamma \leq \gamma_{\rm th}$, the electrons follow a Maxwellian distribution, whereas for $\gamma > \gamma_{\rm th}$, the electrons follow a power-law distribution.
The second row of Figure \ref{CONV_VSPA_Spectrum} shows the specific flux for power-law, Maxwellian, and hybrid electron energy distributions. For a power-law distribution, the radiation with a higher power-law index exhibits a weaker flux. In either the case of $p = 2.5$ or the case of $p = 3.5$, the flux increases with frequency in the low-frequency region, whereas it decreases with increasing frequency in the high-frequency region. However, in the case of $p = 1.5$, the flux in the high-frequency region increases with frequency. For a Maxwellian distribution, the radiation with a lower electron temperature exhibits a weaker flux. Moreover, when the thermal electrons have a lower temperature, the frequency range covered by the radiation becomes narrower. For a hybrid distribution, the spectrum asymptotes to a power law at both the low- and high-frequency ends, with a thermal component appearing at intermediate frequencies. The flux of the power-law components decreases with decreasing electron temperature. The thermal component is not affected by the power-law index. When the electron temperature decreases, the flux of the power-law components decreases significantly. 
The third row of Figure \ref{CONV_VSPA_Spectrum} shows the linear polarization degree corresponding to different electron distributions. The polarization degree results for power-law and Maxwellian electron distributions have been discussed in Sections \ref{3.1} and \ref{3.2}. For a hybrid electron distribution, the polarization degree at high frequencies is nearly independent of frequency over a broad frequency range.  At high frequencies, the cases with the same power-law index but different electron temperatures exhibit the same polarization degree. Therefore, the polarization degree depends only on the power-law index at high frequencies. In contrast, at low frequencies, the cases with the same electron temperature but different power-law indices exhibit the same polarization degree. Therefore, the polarization degree depends only on the electron temperature at low frequencies. Moreover, the polarization profile exhibits a peak at low frequencies, with the polarization degree decreasing rapidly on both sides of the peak. The peak of the polarization degree occurs at lower frequencies when the electron temperature is lower.

We find that for different electron energy distributions, the specific flux of very-small-pitch-angle synchrotron radiation is much lower than that of conventional synchrotron radiation. This is because the order of magnitude of the specific flux of very-small-pitch-angle synchrotron radiation is primarily determined by $\tilde{\alpha}$, and since $\tilde{\alpha} \ll 1$, the flux is extremely weak. In addition, for a power-law distribution, the spectral slope is $2-p$ for the very-small-pitch-angle synchrotron radiation in the high-frequency region, compared with the usual result of $(1-p)/2$ for the conventional synchrotron radiation. This is consistent with the Example 1 presented in \citet{Epstein1973a}. We also find that in the low-frequency region, the spectral slope is $p$ for the very-small-pitch-angle synchrotron radiation. For the Maxwellian distribution, the spectra of the very-small-pitch-angle synchrotron radiation follow the spectral shape of $F_\nu \propto \nu$. In addition, the flux drops rapidly at both the low- and high-frequency ends. For the hybrid distribution, the spectra of the very-small-pitch-angle synchrotron radiation asymptote to a power law at both the low- and high-frequency ends, with a thermal component in the intermediate-frequency range. In contrast, the spectra of conventional synchrotron radiation exhibit a thermal component at low frequencies and approach a power law at high frequencies.

\section{Analytical expression for the linear polarization Degree of very-small-pitch-angle synchrotron radiation with a power-law electron distribution}

The polarization profile of very-small-pitch-angle synchrotron radiation exhibits a prominent peak and a pronounced kink at low frequencies, corresponding to an abrupt transition from the peaked polarization to a constant value. To determine whether the structures in the polarization profile are intrinsic features of the mathematic process or numerical artifacts,  we analytically evaluate Equation (\ref{VSPA-PL}) for a power-law electron energy distribution. Changing variables of integration to $u = \nu/(\gamma\nu_L)$, we have
\begin{equation}\label{VSPA-PL-ANALY}
    \Pi_{L,{\rm VSPA}} =\frac{\int_{u_{\min}}^{u_{\max}}u^{p-2}\left(u-\frac{u^2}{2}\right)du}{\int_{u_{\min}}^{u_{\max}}u^{p-2}\left(1-u+\frac{u^2}{2}\right)du} = \frac{F_N(u_{\max})-F_N(u_{\min})}{F_D(u_{\max})-F_D(u_{\min})}
\end{equation}
where $u_{\rm min} = \nu/(\gamma_{\rm max} \nu_L)$, $u_{\rm max} = \nu/(\gamma_{\rm min} \nu_L)$, and $F_N(u)$ and $F_D(u)$ are given by
\begin{equation}
    F_N(u)=\frac{u^p}{p}-\frac{u^{p+1}}{2(p+1)},
\end{equation}
and
\begin{equation}
    F_D(u)=\frac{u^{p-1}}{p-1}-\frac{u^p}{p}+\frac{u^{p+1}}{2(p+1)}.
\end{equation}

\begin{figure*}
\centering
\includegraphics[scale=0.7]{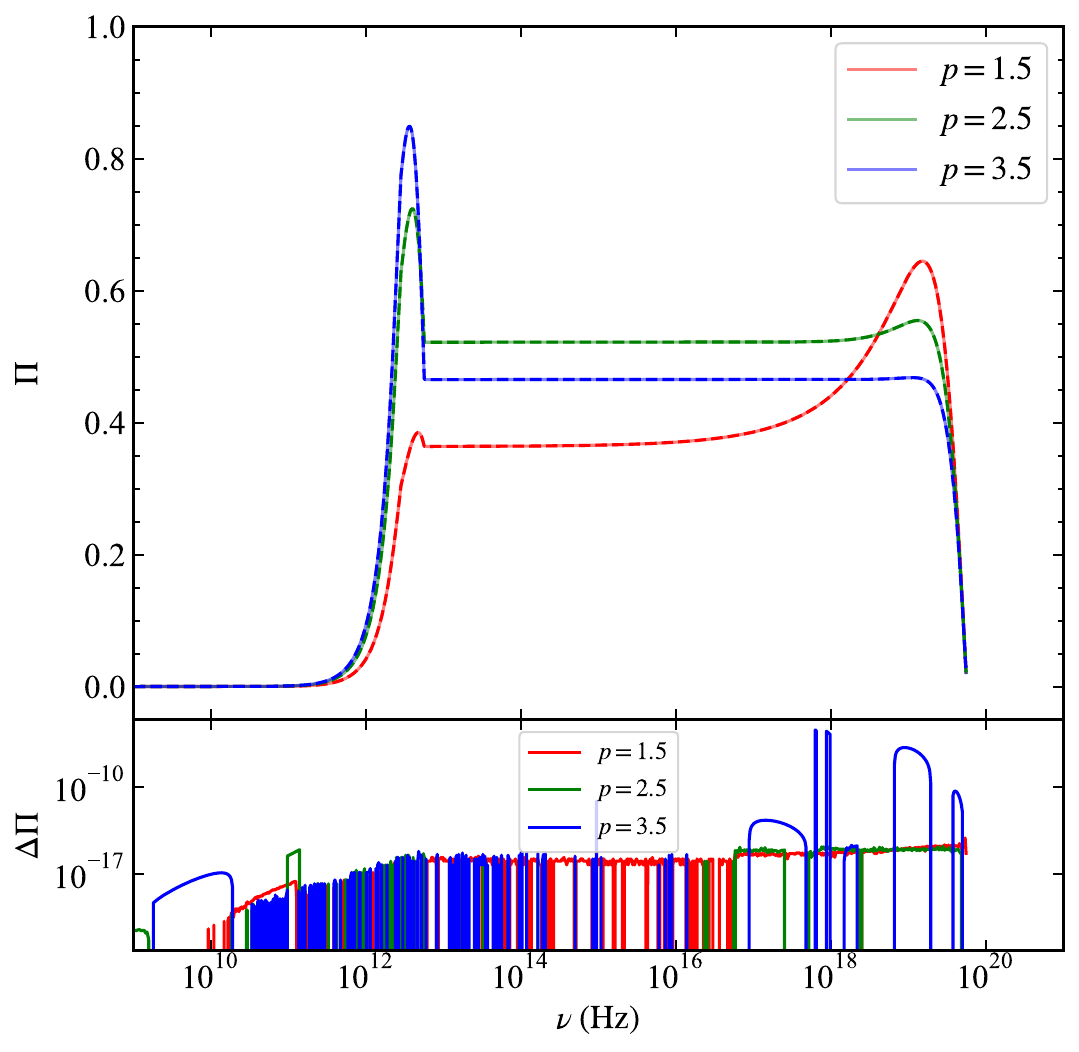}
\caption{Linear polarization degree of very-small-pitch-angle synchrotron radiation for power-law electron energy distributions. The solid and dashed lines represent the polarization degrees obtained from the analytical expression (Equation (\ref{VSPA-PL-ANALY})) and numerical integration of Equation (\ref{VSPA-PL}), respectively. $\Delta\Pi$ denotes the difference between the polarization degrees obtained using the two methods. In our calculations, we adopt a magnetic field strength of $B = 1.0 \times 10^{6}\,\mathrm{G}$.
\label{VSPA-ANALY}}
\end{figure*}

Figure \ref{VSPA-ANALY} shows the linear polarization degrees obtained from the analytical expression (Equation (\ref{VSPA-PL-ANALY})) and the numerical integration of Equation (\ref{VSPA-PL}), along with their difference $\Delta\Pi$. The difference between the two methods is $\Delta\Pi\sim10^{-16}$, demonstrating that the analytical and numerical results are nearly identical and reproduce the same structures in the polarization profile. Therefore, the structures in the polarization profile, including the peaks and kinks, are not numerical artifacts but genuine features from the mathematic process. 

The analytical expression allows us to further investigate the features of the polarization profile. From Equation (\ref{VSPA-PL-ANALY}), we note that the result of the integration depends sensitively on the integration limits of $u$. In our calculations, we consider $\gamma_{\rm min} = 1$ and $\gamma_{\rm max} = 10^7$. In addition, the electron Lorentz factor is required to satisfy $\gamma>\max(\nu_L/\nu,\nu/2\nu_L)$. Therefore, the minimum Lorentz factor can be written as $\gamma_{\min}=\max(1,\nu_L/\nu,\nu/2\nu_L)$. Consequently, we obtain
\begin{equation}\label{limit-umax}
    u_{\max}=
    \begin{cases}(\frac{\nu}{\nu_L})^2,&\nu<\nu_L,\\
        \frac{\nu}{\nu_L},&\nu_L\leq \nu\leq 2\nu_L,
        \\2,&\nu>2\nu_L.
    \end{cases}
\end{equation}
Equation (\ref{limit-umax}) shows that $u_{\max}$ changes with the observational frequency. Therefore, we can divide the frequency dependence of the polarization degree into three distinct regimes and derive an analytical approximation for each frequency range.

For $\nu < \nu_L$, where $u_{\max}=(\nu/\nu_L)^2$, neglecting the higher-order terms and taking $u_{\min}=10^{-7}\nu/\nu_L\approx0$, we obtain
\begin{equation}\label{ap1}
    \Pi_{L,{\rm VSPA}} \approx \frac{p-1}{p}\left(\frac{\nu}{\nu_L}\right)^2.
\end{equation}
We note that $\Pi_{L,{\rm VSPA}}\propto\nu^2$, which explains the steep increase of the polarization degree with frequency at low frequencies.

For $\nu_L \leq \nu \leq 2\nu_L$, $u_{\max}$ changes from $(\nu/\nu_L)^2$ to $\nu/\nu_L$, and the polarization degree can be written as
\begin{equation}\label{ap2}
    \Pi_{L,{\rm VSPA}} \simeq\frac{F_N(u_{\max})}{F_D(u_{\max})} = \frac{(\nu/\nu_L)(p-1)\left[2(p+1)-p(\nu/\nu_L)\right]}{2p(p+1)-2(p^2-1)(\nu/\nu_L)+p(p-1)(\nu/\nu_L)^2}.
\end{equation}
Differentiating Equation (\ref{ap2}) with respect to $\nu$ and setting $d\Pi_{L,{\rm VSPA}}/d\nu=0$, we find that the peak occurs at $\nu_{\mathrm{peak}}=(1+1/p)\nu_L$. The corresponding peak polarization degree is given by $(p^2-1)/(p^2+1)$. 

For $\nu > 2\nu_L$, $u_{\max}=2$ is independent of frequency. At relatively low frequencies, where $u_{\min}\approx0$, the expression can be further simplified to
\begin{equation}
    \Pi_{L,{\rm VSPA}}\approx\frac{F_N(2)}{F_D(2)} = \frac{2(p-1)}{p^2-p+2}.
\end{equation}
Thus, the polarization degree is independent of frequency in this regime, explaining its constant value over this frequency range. As $\nu$ increases, $u_{\min}$ increases, and its contribution to the polarization degree becomes nonnegligible at higher frequencies. Therefore, the polarization degree is then given by
\begin{equation} \label{high-ap}
    \Pi_{L,{\rm VSPA}}=\frac{F_N(2)-F_N(u_{\min})}{F_D(2)-F_D(u_{\min})}=\frac{2^{p+1}(p-1)-2(p^2-1)u_{\min}^p+p(p-1)u_{\min}^{p+1}}{2^p(p^2-p+2)-2p(p+1)u_{\min}^{p-1}+2(p^2-1)u_{\min}^p-p(p-1)u_{\min}^{p+1}}.
\end{equation}
where $u_{\min} = 10^{-7}\nu/\nu_L$. As $u_{\min}$ approaches 2, corresponding to $\nu$ approaching $2\times10^7\nu_L$, the integration range shrinks to zero, causing the polarization degree to rapidly decrease to zero.

\begin{figure*}
\centering
\includegraphics[scale=0.45]{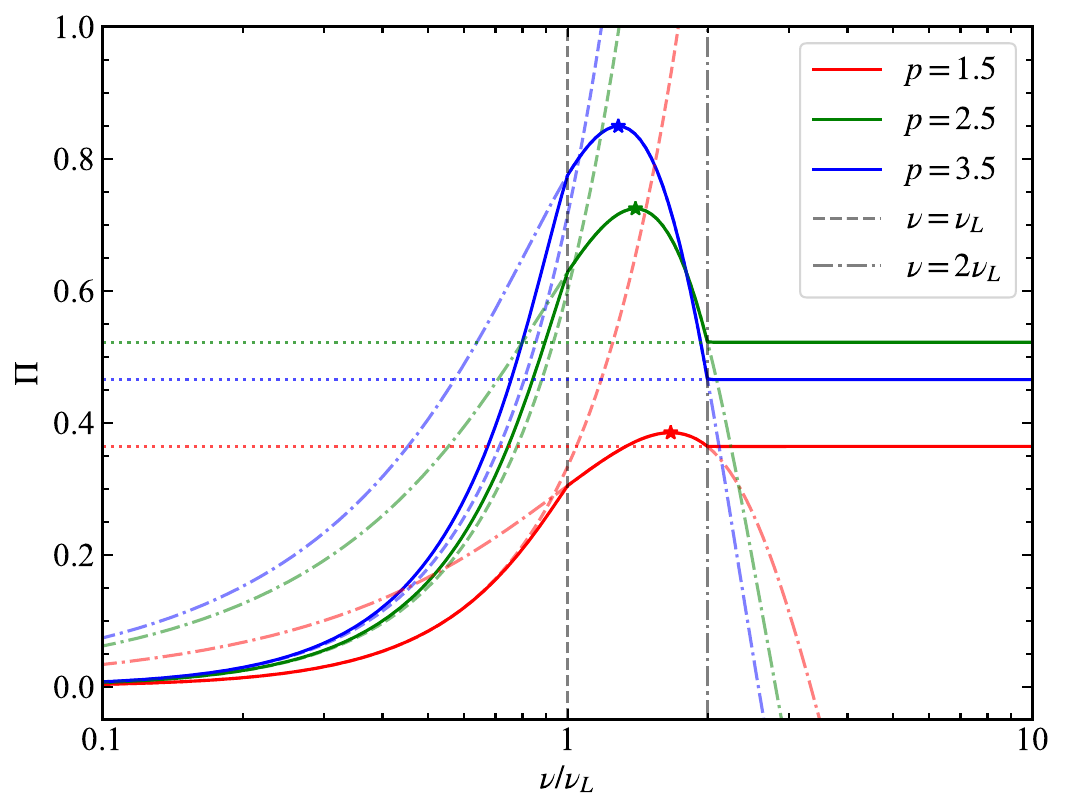}
\includegraphics[scale=0.45]{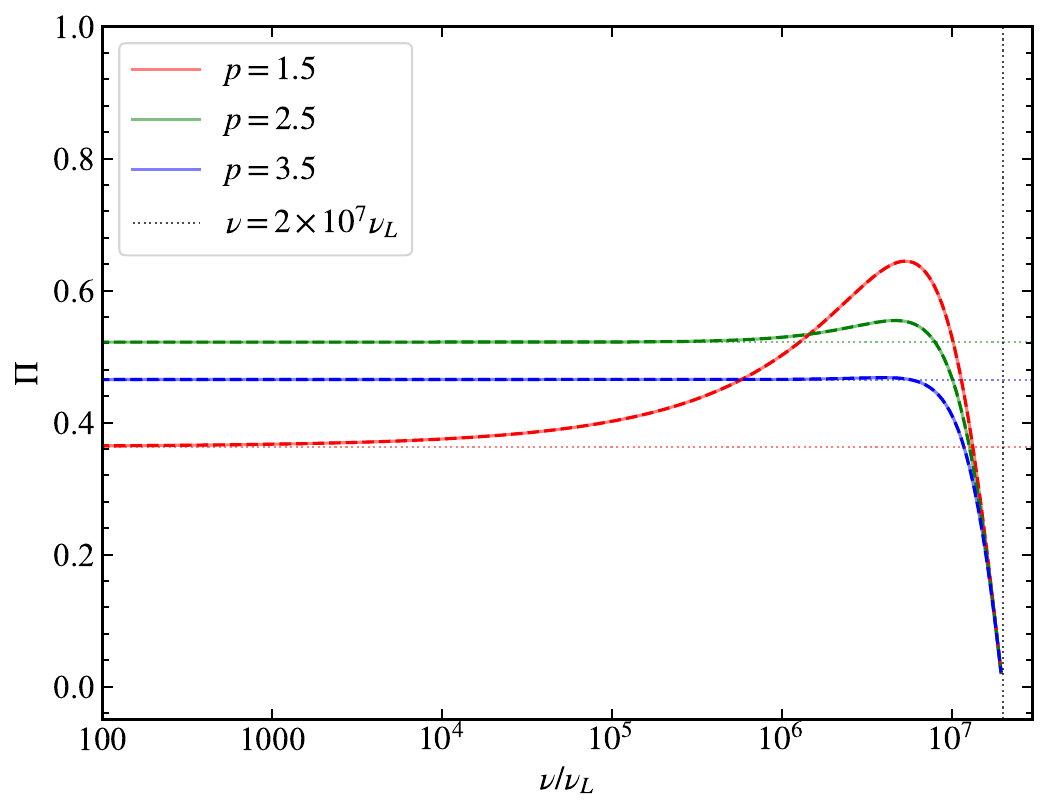}
\caption{Linear polarization degree of very-small-pitch-angle synchrotron radiation as a function of the frequency ratio $\nu/\nu_L$ for power-law electron distributions. Left panel: the solid lines show the numerical results, while the dashed and dashed-dotted lines denote the analytical approximations in the $\nu<\nu_L$ and $\nu_L\leq\nu\leq2\nu_L$ regimes, respectively. The gray dashed and dashed-dotted vertical lines indicate $\nu=\nu_L$ and $\nu=2\nu_L$, respectively. The horizontal dotted lines represent the analytical polarization degrees in the $\nu>2\nu_L$ regime. The stars mark the peak polarization degrees and their corresponding frequencies in the polarization profile. Right panel: the solid and dashed lines represent the numerical results and the high-frequency analytical approximations of the polarization degree, respectively. The gray dotted vertical line indicates the upper frequency cutoff, $\nu=2\times10^7\nu_L$. In our calculations, we adopt a magnetic field strength of $B=1.0\times10^6\,\mathrm{G}$.
\label{VSPA-approx}}
\end{figure*}

Figure \ref{VSPA-approx} presents the numerical results and analytical approximations of the linear polarization degree. We can see that the analytical approximations reproduce the numerical calculations well in their respective frequency regimes. The polarization profile exhibits two distinct kinks at $\nu=\nu_L$ and $\nu=2\nu_L$, respectively. These are precisely the frequencies at which the expression for $u_{\max}$ changes. At low frequencies, the polarization degree increases rapidly with frequency. When $\nu>\nu_L$, the upper integration limit $u_{\max}$ changes, and the analytical approximation for the polarization degree transitions from Equation (\ref{ap1}) to Equation (\ref{ap2}). Consequently, the rate of increase then becomes slower, producing the first kink, followed by a rapid decline after the polarization degree reaches its peak. When $\nu>2\nu_L$, $u_{\max}$ becomes constant, and the polarization degree transitions from a declining trend to a constant value, producing the second prominent kink in the polarization profile. This indicates that the kinks in the polarization profile arise from changes in the upper integration limit $u_{\max}$ induced by variations in frequency.
In addition, the high-frequency analytical expression (Equation \ref{high-ap}) successfully reproduces the high-frequency bump seen in the numerical polarization profile. This indicates that the high-frequency bump originates from the gradual increase of the lower integration limit $u_{\min}$, which becomes important for the polarization degree at sufficiently high frequencies. The effect of frequency on the polarization degree in the high-frequency regime is more pronounced for smaller values of the power-law index.
When $\nu=2\gamma_{\max}\nu_L$, with $\gamma_{\max}=10^7$ adopted in our calculations, the integration interval shrinks to zero, causing the polarization degree to rapidly decrease to zero. 

\bibliography{main}{}
\bibliographystyle{aasjournalv7}

\end{document}